\documentclass[twocolumn
]{aastex631}
\newcommand{\ie}{$i.e.,\;$}
\newcommand{\eg}{$e.g.,\;$}
\newcommand{\viz}{$viz.,\;$}
\shorttitle{Remnant radio galaxies in the XMM-LSS deep field}
\shortauthors{Dutta et al.}
\graphicspath{{./}{figures/}}

\begin{document}

\title{Search and characterization of remnant radio galaxies in the XMM$-$LSS deep field}

\author[0000-0002-6542-2939]{Sushant Dutta}
\affiliation{Physical Research Laboratory, Ahmedabad, 380009, Gujarat, India}
\affiliation{Indian Institue of Technology Gandhinagar, Palaj, Gandhinagar, 382355, Gujarat, India}
\author[0000-0002-6040-4993]{Veeresh Singh}
\affiliation{Physical Research Laboratory, Ahmedabad, 380009, Gujarat, India}
\author[0000-0001-5356-1221]{C.H. Ishwara Chandra}
\affiliation{National Centre for Radio Astrophysics, TIFR, Post Bag 3, Ganeshkhind,Pune 411007, India}
%
\author[0000-0002-1345-7371]{Yogesh Wadadekar}
\affiliation{National Centre for Radio Astrophysics, TIFR, Post Bag 3, Ganeshkhind,Pune 411007, India}
\author[0000-0001-9851-8243]{Abhijit Kayal}
\affiliation{Physical Research Laboratory, Ahmedabad, 380009, Gujarat, India}
\affiliation{Indian Institue of Technology Gandhinagar, Palaj, Gandhinagar, 382355, Gujarat, India}
\author[0000-0001-6864-5057]{Ian Heywood}
\affiliation{Astrophysics, Department of Physics, University of Oxford, Keble Road, Oxford, OX1 3RH, UK}
\affiliation{Centre for Radio Astronomy Techniques and Technologies, Department of Physics and Electronics, Rhodes University, PO Box 94, Makhanda 6140, South Africa}
\affiliation{South African Radio Astronomy Observatory, 2 Fir Street, Black River Park, Observatory 7925, South Africa}
%
%




\begin{abstract}
The remnant phase of a radio galaxy is characterized by the cessation 
of AGN activity resulting in the stoppage of jets supplying plasma to radio lobes. 
In this paper, we present the search and characterization of remnant candidates 
in 12.5 deg$^{2}$ of the {\em XMM$-$Newton} Large$-$Scale Structure (XMM$-$LSS) field by using deep 
radio observations at 325 MHz from the Giant Metrewave Radio Telescope (GMRT), 
at 150 MHz from the LOw Frequency ARray (LOFAR), at 1.4 GHz from the Jansky Very Large Array (JVLA), 
and at 3 GHz from the VLA Sky Survey (VLASS). 
By using both morphological criteria {\viz}undetected radio core as well as 
spectral criteria {\viz}high spectral curvature, and ultra$-$steep spectrum, 
we identify 21 remnant candidates that are found to reside mostly in non$-$cluster environments, and 
exhibit diverse properties in terms of morphology, spectral index ($\alpha_{\rm 150}^{\rm 1400}$ in 
the range of $-1.71$ to $-0.75$ with a median of $-1.10$), and linear radio size (ranging from 242 kpc to 1.3 Mpc with 
a median of 469 kpc). 
Our study attempts to identify remnant candidates down to the flux density limit of 6.0 mJy at 325 MHz, 
and yields an upper limit on the remnant fraction ($f_{\rm rem}$) to be around 5$\%$. 
The observed $f_{\rm rem}$ seems consistent with the predictions of an 
evolutionary model assuming power law distributions of the duration of active phase and jet kinetic power 
with index $-0.8$ to $-1.2$.  
\end{abstract}

%

\section{Introduction} \label{sec:intro}
Radio galaxies, a subclass of active galactic nuclei (AGN), emit copiously at radio wavelengths and exhibit well defined 
morphological structures that include a radio core, a pair of highly collimated bipolar jets emanating from the core, and radio lobes (cocoons of plasma) fed by the jets.  
It is well known that the AGN jet activity in radio galaxies is only a phase (active phase) 
lasting for several tens of Myr during which the radio size can grow 
upto a few hundred kpc and rarely, upto even a few Mpc \citep{Parma07,Machalski08}. 
After an active phase, AGN activity ceases or drops below a certain threshold level so that the outflowing jets are no longer supported and the radio lobes begin to fade away \citep{Parma99}. 
During the fading period termed as the `remnant phase' or `dying phase', the radio core and jets 
supposedly disappear but the radio lobes can still be detected for a period of 
a few tens of million years before they disappear due to radiative and dynamical losses \citep{Slee01,Murgia11}. 
The duration of remnant phase is arguably a small fraction of the duration of active phase, 
hence remnant sources represent a short$-$lived final phase of the radio galaxy's evolution \citep{Morganti17}.  
Remnant radio galaxies (hereafter `remnants') are deemed to be less abundant owing to the relatively short duration of the remnant phase. 
Due to their relative paucity remnants are poorly understood. 
Specifically, AGN duty cycle and its dependence on various factors such as the jet kinetic power, 
host galaxy, and large$-$scale cluster environment are still a matter of investigation. 
\par
To understand the characteristics of remnants there have been attempts to carry out systematic 
searches for them by exploiting mainly low$-$frequency ($<$1.4 GHz) radio surveys \citep[{\eg}][]{Parma07,Murgia11,Godfrey17,Brienza17,Mahatma18,Quici21}. 
Low$-$frequency radio surveys are advantageous to search for remnants owing to the steepening of 
radio spectrum of relic plasma in radio lobes. 
In fact, samples of Ultra$-$Steep Spectrum (USS) radio sources have been used to identify potential 
remnant candidates. For instance, \citet{Godfrey17} utilised shallow but wide$-$area 74 MHz 
VLA Low$-$frequency Sky Survey Redux (VLSSr, \citet{Lane14}) in combination with 
the 1.4 GHz Faint Images of the Radio Sky at Twenty$-$Centimeters survey (FIRST; \citet{Becker95}) 
and the NRAO VLA Sky Survey (NVSS; \citet{Condon98}) survey, 
and reported that fewer than 2$\%$ of Fanaroff$-$Riley type II (FR$-$II) radio galaxies 
with S$_{\rm 74~MHz}$ $>$1.5 Jy 
are candidate USS remnants (${\alpha}_{\rm 74}^{\rm 1400}$ $<$~$-1.2$, S$_{\nu}$ $\propto$ ${\nu}^{\alpha}$). 
The lobes of low-surface-brightness emission detected in the serendipitously discovered individual remnants 
{\eg}B2 0924+30 \citep{Cordey87,Jamrozy04,Shulevski17}, J1324$-$3138 \citep{Venturi98}, blob1 \citep{Brienza16}, NGC 1534 \citep{Duchesne19}, emphasizes the need of deep low$-$frequency observations. 
\par
The shallow radio surveys of previous generations are likely to miss a significant population of remnants possessing diffuse 
low$-$surface$-$brightness emission. 
Keeping this in view, sensitive low$-$frequency radio surveys have been carried out in deep extragalactic fields 
to search for and obtain large samples of remnants. 
For instance, using 150 MHz LOw Frequency ARray (LOFAR; \citet{vanHaarlem13}) observations in the Lockman Hole 
field \cite{Brienza17} identified 23 remnant 
candidates in a sample of 158 sources detected above a flux density cutoff limit of 40 mJy and 
radio size $\geq$40$^{\prime\prime}$ at 150 MHz. 
In a follow up study \cite{Jurlin21} performed deep (noise$-$rms $\sim$ 9$-$10 $\mu$Jy beam$^{-1}$) 
JVLA A$-$configuration observations of the remnant candidates reported in the Lockman Hole field, and 
identified only 13/23 candidates as genuine remnants.  
In a similar study \citet{Mahatma18} reported remnants in the 
Herschel$-$ATLAS field using 150 MHz LOFAR and 6.0 GHz JVLA observations and found only 
11 remnants in a sample of 127 sources with a flux density cutoff limit (S$_{\rm 150~MHz}$) $\geq$80 mJy 
and radio size $\geq$40$^{\prime\prime}$. The cutoff limits on the flux density and radio size were placed to 
examine morphological structures, attain completeness above a flux limit, 
and to find USS sources using the 1.4 GHz NVSS, a relatively shallow survey.   
\par
It is important to note that the aforementioned studies have demonstrated the potential of the combination of 
deep low$-$frequency and deep high$-$frequency observations to identify remnants. However, a high flux density 
cutoff limit hinders the identification of remnants and their nature 
at lower flux densities. Also, a high flux density cutoff limit is likely to result in a smaller sample  
that can be biased towards the remnants of powerful radio galaxies. Considering the 
limitations introduced by the high flux density cutoff limit we attempt to search 
and characterize remnants down to a fainter flux density limit {\ie}8.0 mJy at 325 MHz that 
corresponds to 13.7 mJy at 150 MHz, assuming a typical spectral index of $-0.7$. 
The expectation is that a search for remnants over a wide range of flux densities reaching down to faint levels would allow us to obtain a larger sample, a more robust determination of remnant fraction, and its dependence 
on the flux density and luminosity. 
Also, a larger sample would enable us to use a more robust statistical approach in constraining the evolutionary models of remnant phase.
In this work, we attempt to carry out a systematic search for remnants in the {\em XMM$-$Newton} Large$-$Scale Structure (XMM$-$LSS) field by using sensitive 325 MHz Giant Meterwave Radio Telescope (GMRT) radio observations 
(5$\sigma$ $\sim$ 0.75 mJy beam$^{-1}$) and existing ancillary multi$-$frequency radio data. 
We note that our study deals with only remnants and excludes active sources with episodic or intermittent AGN activity.  
\par 
This paper is organized as follows. In Section~\ref{sec:data} we provide the details of multi$-$frequency radio data and optical data available in the XMM$-$LSS field. 
Section~\ref{sec:criteria} describes the selection criteria adopted to identify remnant candidates. 
In Section~\ref{sec:discussion} we discuss various characteristic properties of remnant candidates and their comparison 
to active sources. In Section~\ref{sec:frac} we determine remnant fraction and discuss its dependence 
on the flux density. 
In Section~\ref{sec:conclusions} we list the conclusions of our study.  
\\ 
In this paper, we use the cosmological parameters H$_{\rm 0}$ = 70 km s$^{-1}$ Mpc$^{-1}$, ${\Omega}_{\rm m}$ = 0.3, 
and ${\Omega}_{\Lambda}$ = 0.7. 
\section{Multi$-$wavelength data in the XMM$-$LSS field}
\label{sec:data}
The XMM$-$LSS field has been observed across nearly entire electromagnetic spectrum 
spanning over radio \citep{Tasse07,Hale19,Heywood20}, near$-$IR \citep{Jarvis13}, mid$-$IR \citep{Lonsdale03,Mauduit12}, 
far$-$IR \citep{Oliver12}, optical \citep{Erben13,Aihara18}, to X$-$ray \citep{Pierre04,Chen18}, including optical 
spectroscopic data \citep{LeFevre13,LeFevre15,Davies18,Scodeggio18}. 
To search for remnants, we exploit deep low$-$frequency 325 MHz GMRT observations, 150 MHz LOFAR observations and 1.4 GHz JVLA observations available in the XMM-LSS field. In addition to the deep multi$-$frequency radio observations 
we also utilize data from large$-$area surveys such as 150 MHz TIFR GMRT Sky Survey Alternative Data Release 1 \citep[TGSS;][]{Intema17}, 1.4 GHz NVSS as well as 
FIRST and 3.0 GHz Very Large Array Sky Survey \citep[VLASS;][]{Lacy20}.    
Table~\ref{tab:RadioData} lists the basic parameters of deep field radio observations. 
Figure~\ref{fig:Footprints} shows the footprints of 325 MHz GMRT, 150 MHz LOFAR and 1.4 GHz JVLA observations. 
We emphasize that the low$-$frequency 325 MHz GMRT and 150 MHz LOFAR observations are better 
suited to detect extended diffuse emission associated with radio lobes. 
While, 3 GHz VLASS and 1.4 GHz JVLA observations of relatively higher resolution (2$^{\prime\prime}$.5$-$4$^{\prime\prime}$.5) can allow us to detect radio cores. 
To estimate the spectral indices of sources detected at 325 MHz and 150 MHz we mainly 
use 1.4 GHz NVSS data owing to 
its large beam$-$size (45$^{\prime\prime}$) and short$-$spacing uv coverage in the VLA D$-$configuration.
We note that the coverage of 1.4 GHz JVLA observations (5.0~deg$^2$) is much smaller than that of 325 MHz 
GMRT observations (12.5~deg$^2$) (see Figure~\ref{fig:Footprints}). 
Hence, outside the JVLA region, we use relatively less sensitive 3.0 GHz VLASS data, 
whenever high resolution images are required. 
To account for the impact of differing sensitivities of high$-$frequency ($\geq$1.4~GHz) observations we divide 12.5 deg$^{2}$ area of 325 MHz GMRT into two sub$-$regions : (i) 5.0~deg$^2$ area covered by the JVLA observations named as the XMM$-$LSS$-$JVLA, and (ii) 7.5~deg$^2$ area outside the JVLA observations named as 
the XMM$-$LSS$-$Out.
In the following sub$-$sections, we provide a brief description of the radio and optical data used in our study. 
\subsection{Radio data}
\label{sec:radio}
\subsubsection{325 MHz GMRT observations}
The 325 MHz GMRT observations centered at RA = 02$^{h}$ 21$^{m}$ 00$^{s}$ (J2000) and DEC = $-04^{\circ}$ 30$^{\prime}$ 00$^{\prime\prime}$ (J2000) covers an area of 12.5 deg$^2$ with 16 pointings in the XMM$-$LSS field.  
These observations were carried out with the legacy GMRT equipped with the software correlator and instantaneous bandwidth of 32 MHz. 
To optimize the uv coverage, observations were performed in semi$-$snapshot mode with each scan of 6$-$17 minutes duration. 
With an average exposure time of 2.5 hours per pointing 
the final 325 MHz GMRT mosaiced image has nearly uniform sensitivity with an average noise$-$rms of 150 $\mu$Jy beam$^{-1}$. The final map with a synthesized beam$-$size of 10$^{\prime\prime}$.2 $\times$ 7$^{\prime\prime}$.9 
yields the detection of 3739 individual radio sources above the flux limit of 5$\sigma$ in the deeper 
regions (noise$-$rms $<$200 $\mu$Jy beam$^{-1}$) 
and 6$\sigma$ in the relatively shallower regions (noise$-$rms $>$200 $\mu$Jy beam$^{-1}$). 
More details on these observations can be found in \citet{Singh14} who used it to study the population of 
USS radio sources. These observations have also been used to study the radio$-$FIR correlation in blue$-$cloud galaxies \citep{Basu15}, 
a giant remnant radio galaxy J021659$-$044920 \citep{Tamhane15}, the population of infrared$-$faint radio sources \citep{Singh17} and remnant candidates of small angular sizes \citep{Singh21}.
\begin{table*}
\begin{minipage}{160mm}
\caption{Summary of deep multi$-$frequency radio observations}
\label{tab:RadioData}
\begin{tabular}{cccccccc} 
\hline
Frequency & Telescope & Area      & 5$\sigma$ limit & Beam$-$size & PA & No. of total  & Reference \\
          &           & (deg$^2$) & (mJy~beam$^{-1}$) & ($\prime\prime \times \prime\prime$) & (deg) & sources  &     \\
\hline
150 MHz & LOFAR & 27.0 & 1.4$-$1.97 & 7.5 $\times$ 8.5 &  106    & 3044  & \citet{Hale19}  \\ 
325 MHz & GMRT & 12.5 & 0.75 & 10.2 $\times$ 7.9 & 74.6     & 3739  & \citet{Singh14} \\
1.4 GHz & JVLA & 5.0 & 0.08  & 4.5 & 0.0   & 5760 & \citet{Heywood20}  \\
\hline
\end{tabular}
\end{minipage}
\end{table*}
\subsubsection{150 MHz LOFAR observations}
\begin{figure}
\centering
\includegraphics[angle=0,width=8.5cm,trim={0.5cm 7.5cm 0.0cm 4.5cm},clip]{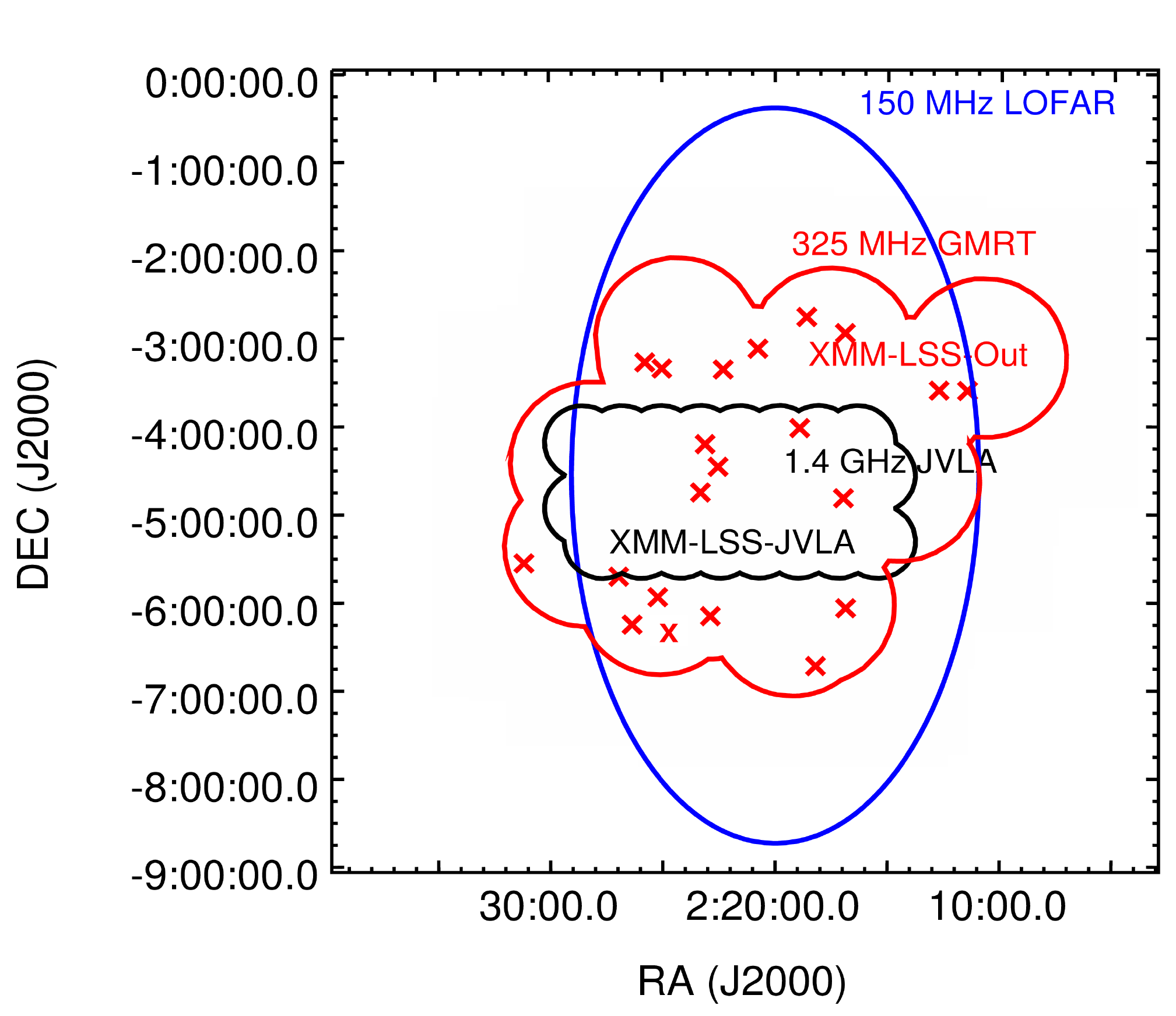}
\caption{The footprints of 325 MHz GMRT observations (in red), 150 MHz LOFAR observations (in blue) and 1.4 GHz JVLA observations (in black) in the XMM$-$LSS field. The HSC$-$SSP wide component covers full region of the 
XMM$-$LSS. The 5.0 deg$^{2}$ XMM$-$LSS$-$JVLA region covered with deep 1.4 GHz JVLA observations and 
the XMM$-$LSS$-$Out region outside to JVLA region, are marked. The positions of our remnant candidates are 
marked by red cross symbols.}
\label{fig:Footprints} 
\end{figure}

The XMM$-$LSS field has been observed with the LOFAR using 
the High Band Antenna (HBA; 110$-$240 MHz) having central frequency at 144 MHz \citep{Hale19}. 
With three four$-$hour pointings LOFAR observations cover 27 deg$^{2}$ sky area, centered at 
RA = 02$^{h}$ 20$^{m}$ 00$^{s}$ (J2000) and DEC = -04$^{\circ}$ 30$^{\prime}$ 00$^{\prime\prime}$ (J2000), with elliptical footprints aligned in the north$-$south direction (see Figure~\ref{fig:Footprints}).  The final mosaiced map has a median noise$-$rms of 0.40 mJy beam$^{-1}$, while noise$-$rms 
reaches down to 0.28 mJy beam$^{-1}$ in the central region. 
With the synthesized beam$-$size of 7$^{\prime\prime}$.5 $\times$ 8$^{\prime\prime}$.5 
these observations detect a total of 3044 individual radio sources.  
We note that the angular resolution and sensitivity of 150 MHz LOFAR observations are 
comparable to that of 325 MHz GMRT observations.
Average 5$\sigma$ depth of 1.97 mJy beam$^{-1}$  in the 150 MHz LOFAR observations scales 
to $\sim$ 1.15 mJy beam$^{-1}$ at 325 MHz, assuming a typical spectral index value of $-0.7$.
Hence, combination of sensitive 150 MHz LOFAR observations and 325 MHz GMRT observations
will enhance the chance of remnant detection. 
We use 150 MHz LOFAR source catalogue as radio images are not publicly available. 
For multi$-$component sources we sum the flux densities of individual components where source components 
are identified by over$-$plotting 150 MHz source positions on to the 325 MHz GMRT image.  
It is worth noting that most of the 325 MHz GMRT survey area is covered by the 150 MHz LOFAR observations 
(see Figure~\ref{fig:Footprints}). However, for a small area with no LOFAR coverage we use 
less sensitive 150 MHz TGSS data. 
The 150 MHz TGSS radio survey covers the entire sky north of $-53^{\circ}$ DEC and provides continuum images with 
a median noise$-$rms of 3.5 mJy beam$^{-1}$ and angular resolution of 25$^{\prime\prime}$ $\times$ 25$^{\prime\prime}$.
\subsubsection{1.4 GHz JVLA observations}
Recently, \citet{Heywood20} carried out 1.4 GHz wide$-$band (0.994$-$2.018 GHz) JVLA radio observations of 5.0 deg$^{2}$ sky area over the near$-$IR VIDEO region \citep{Jarvis13} in the XMM$-$LSS field. 
With 32 pointings the final mosaiced map has a median noise$-$rms of 16 $\mu$Jy beam$^{-1}$ and 
angular resolution of 4$^{\prime\prime}$.5. These are one of the deepest 1.4 GHz observations 
over the largest area (5.0 deg$^{2}$) in the XMM$-$LSS field,
and detect a total of 5762 radio sources above $5\sigma$ flux limit. 
The JVLA observations performed with the B$-$configuration in the wide$-$band continuum mode 
allow sub$-$band imaging, and provide in$-$band spectral indices 
for 3458 radio sources. 
\subsubsection{3 GHz Very Large Array Sky Survey (VLASS)} 
In our work, we also use data from the 3 GHz VLASS which covers the entire sky north of $-40^{\circ}$ 
declination. The VLASS is being carried out with 2$-$4 GHz wide$-$band of JVLA in B/BnA$-$configuration offering 
angular resolution of 2$^{\prime\prime}$.5 and noise$-$rms of 0.12 mJy beam$^{-1}$ in one epoch observations 
\citep{Lacy20}. The VLASS is designed to observe each field over three 
epochs with the separation of 32 months, allowing it to study variable radio sources. 
The coadded images of three epochs are expected to achieve noise$-$rms of 0.07 mJy beam$^{-1}$. 
We use the VLASS Epoch 1 Quick Look images and source catalog available at 
the Canadian Initiative for Radio Astronomy Data Analysis (CIRADA) website\footnote{http://cutouts.cirada.ca/}.  
The high resolution deep VLASS images are useful in detecting the presence of a radio core.    
\subsubsection{Auxiliary radio observations}
In addition to the aforementioned radio observations, the XMM$-$LSS field is covered with 
shallow 74 MHz VLA observations having noise$-$rms 32 mJy beam$^{-1}$ \citep{Tasse06}.
Due to lack of sensitivity similar to the 325 MHz GMRT observations 
these data are not very useful to perform a systematic search for remnants. 
We used 74 MHz flux densities, whenever available, for the SED fitting of 
our selected remnant candidates. 
Further, the XMM$-$LSS field is also surveyed at 240 MHz and 610 MHz with the GMRT. 
The 240 MHz GMRT observations achieved noise$-$rms of 2.5 mJy beam$^{-1}$ and detected 466 sources 
in 18.0 deg$^{2}$ area \citep{Tasse07}. 
\cite{Smolcic18} present 610 MHz observations covering an area of 30.4 deg$^2$ with a non$-$uniform noise$-$rms (200 $\mu$Jy beam$^{-1}$ in the inner area of 11.9 deg$^{2}$, and 45 $\mu$Jy beam$^{-1}$ in the outer area of 18.5 deg$^{2}$) and the resolution of 6$^{\prime\prime}$.5 yields a catalog of 5434 radio sources at $\geq$7$\sigma$ level. 
In principle, deep 610 MHz observations can be utilised to search for remnants but we find that, 
both 240 MHz and 610 MHz observations suffer from underestimated or missing flux density issue. The 610 MHz observations are also not ideal for the core detection due to moderate angular resolution and sensitivity. 
Therefore, we choose not to use 240 MHz and 610 MHz GMRT observations for our study.

\subsection{Optical Data}
\label{sec:optical}
Deep optical data in the XMM$-$LSS field are available from the Hyper Suprime$-$Cam Subaru Strategic Program 
(HSC$-$SSP\footnote{https://hsc-release.mtk.nao.ac.jp/doc/}) which is a three$-$tiered (wide, deep and ultra$-$deep), 
multi$-$band ($g$, $r$, $i$, $z$, $y$ and four narrow$-$band filters) imaging survey carried out with a wide$-$field camera named Hyper Suprime$-$Cam installed at the 8.2$-$m Subaru telescope \citep{Aihara18}. 
The HSC$-$SSP survey's wide component reaches to a depth of 26.2$_{\rm -0.3}^{\rm +0.2}$ mag at 5$\sigma$ level 
for point sources in the $i$ band and covers nearly 1200 deg$^{2}$ sky area mostly around the celestial equator. 
Deep component is nearly one magnitude deeper with 5$\sigma$ detection limit of 26.9$_{\rm -0.3}^{\rm +0.2}$ mag in  
$i$ band and covers 7.0 deg$^{2}$ in each of the four separate fields (XMM$-$LSS, E$-$COSMOS, ELAIS$-$N1, DEEP2$-$F3). 
The ultra$-$deep survey is nearly 0.8 magnitude deeper than the deep component but it covers 
only 1.7 deg$^{2}$ sky area 
in each of the two sub$-$fields centered at the COSMOS and Subaru XMM$-$Newton Deep field Survey (SXDS). 
The 5$\sigma$ limiting magnitudes are AB magnitudes \citep{Oke83} measured within 2$^{\prime\prime}$.0 diameter apertures 
and include effects of source confusion. 
We note that the 325 MHz GMRT observed region in the XMM$-$LSS field is completely covered with the HSC$-$SSP wide 
survey and 7.0 deg$^{2}$ area is covered with the deep component. 
We obtain optical data from the HSC$-$SSP third public data release (PDR3) that provides source catalogues and 
images with the median seeing of 0$^{\prime\prime}$.6 in $i$ band. 
We note that the HSC$-$SSP photometric data are more than one magnitude deeper than 
the previously available optical photometric data from the Canada$-$France$-$Hawaii Telescope Legacy Survey (CFHTLS) that 
reached down only to $i^{\prime}$ = 24.5 mag limiting magnitude at 5$\sigma$ level \citep{Ilbert06}. 
Hence, we prefer to use the HSC$-$SSP instead of the CFHTLS.    
\par
In addition to photometric data, HSC$-$SSP PDR3 also provides a collection of publicly available spectroscopic redshifts 
(spec-$z$ table in the data access website\footnote{https://hsc-release.mtk.nao.ac.jp/doc/index.php/available-data\_pdr3/}). 
The spectroscopic redshifts are gleaned from the existing spectroscopic surveys in the field such as 
Sloan Digital Sky Survey (SDSS) DR16 \citep{Ahumada20}, SDSS IV QSO catalog \citep{Paris18}, VIMOS$-$VLT Deep Survey (VVDS; \cite{LeFevre13}), 
PRIsm MUlti$-$object Survey (PRIMUS) DR1 \citep{Cool13}, 
VIMOS Public Extragalactic Redshift Survey (VIPERS) PDR1 \citep{Garilli14}) in the XMM$-$LSS field. 
The spectroscopic objects are matched by the position within 1$^{\prime\prime}$.0 to the HSC$-$SSP objects 
and the closest match is considered. In case, a HSC$-$SSP source is found to have more than one spectroscopic redshifts the secure redshift value with the smallest error is considered. 
We note that the VIPER survey pre$-$selected redshift range of 0.5 $<$ $z$ $<$ 1.0 based on the optical and IR color$-$color plots, and spec$-z$ 
measurements are limited to relatively bright optical sources ($i^{\prime}_{\rm AB}$ $<$ 22.5).  
The PRIMUS also measured spec$-z$ of galaxies only upto $z$ $=$ 1.0.  
%
In case of unavailability of spectroscopic redshift we use photometric redshifts 
available from the HSC$-$SSP PDR2 \citep{Aihara19}. 
We opted for photo$-z$ estimates derived by using a convolutional neural network (CNN) method that uses galaxy images in contrast to 
the earlier methods that used only integrated photometry of galaxies \citep[see][]{Schuldt21}. 
The CNN based photo$-z$ estimates gives a precision of $\sigma$ = 0.12 (68$\%$ confidence interval) for 
$|$~z~$_{\rm pred}$ $-$ z$_{\rm ref}$~$|$ in the redshift range of 0 to 4.0 for the full HSC$-$SSP photometric sample.
\section{Selection criteria and identification of remnants}
\label{sec:criteria}
In the literature, remnant sources have been identified by using both morphological as well as spectral criteria.
For instance, \cite{Mahatma18} selected remnant sources based on the morphological criterion {\ie} 
lack of radio core in sensitive high$-$frequency radio observations, 
while \cite{Godfrey17} selected remnant sources using the steepness of the radio spectrum. 
We note that the non$-$detection of radio core, a seemingly robust criterion, is limited by the sensitivity 
of observations performed to examine the existence of core. 
Using both morphological and spectral criteria in their search of remnants \cite{Jurlin21} argued that a weak core representing a dying or rejuvenated AGN activity may still be present in a remnant source. 
However, it is difficult to distinguish a dying radio core from a rejuvenated radio core.    
Also, there are suggestions that the AGN activity particularly in massive galaxies may always be present at some level, irrespective of their evolutionary stage \citep{Sabater19}. 
Hence, in addition to absent core criterion, we also consider spectral criteria to obtain an unbiased sample of remnants.  
In the following subsections we provide the details of our selection criteria for remnants, 
and resultant remnant sub$-$samples.   
\subsection{Morphological criterion}
\label{sec:morph}
A typical remnant is expected to show relaxed radio morphology 
without compact components like core, hotspots, or jets \citep{Saripalli12}. 
Although, a young remnant with recently switched off jets can still show hotspots in the lobes. 
As time elapses, radio lobes can become amorphous due to expansion, if lobes are over$-$pressured at the end 
of its life \citep{WangKaiser08}.
More importantly, absence of radio core in deep high$-$resolution observations can be considered a 
manifestation of switched$-$off AGN activity. Hence, in principle, a remnant is expected to lack the radio core at all frequencies even in the deepest possible radio images \citep[see][]{Mahatma18,Quici21}.     
Therefore, we use the absence of radio core in deep radio images as a criterion to identify remnants. 
\par
To search for remnants we began with a sample of 268 extended radio sources detected in 
the 325 MHz GMRT observations. Our initial sample is gleaned from the catalogue of 3739 sources 
detected in the 325 MHz GMRT radio observations by applying two conditions : (i) sources are detected with the Signal$-$to$-$Noise Ratio (SNR) $\geq$10 to avoid contamination from spurious sources, and 
(ii) radio sources are extended with the largest angular size (LAS) $\geq$30$^{\prime\prime}$, nearly three times of 
the synthesized beam of GMRT at 325 MHz, to enable us to perform visual inspection of radio morphology. 
With size $\geq$30$^{\prime\prime}$ we ensure a minimum of six 1.4~GHz JVLA/FIRST synthesized beams 
and twelve 3.0~GHz VLASS synthesized beams spread out across the source.
This cutoff limit on the size allows us to minimise the blending of radio core with radio lobes, 
and is required for a clear interpretation of radio morphology. 
\par
To identify radio galaxies with absent radio core we visually inspected 
3 GHz VLASS and 1.4 GHz FIRST cutouts for our all 268 sample sources. 
We attempted to identify the core component down to 3$\sigma$ level by overlaying 
3 GHz VLASS and 1.4 GHz FIRST radio contours onto the 325 MHz grey$-$scale images.   
We note that 3 GHz VLASS images with the resolution of 2$^{\prime\prime}$.5 can identify radio core 
at 3$\sigma$ limit of  0.36 mJy beam$^{-1}$ flux density level. While, 1.4 GHz FIRST survey with the resolution of 5$^{\prime\prime}$.0 can detect core component only upto 3$\sigma$ = 0.6 mJy beam$^{-1}$. 
The FIRST 3$\sigma$ flux density limit of 0.6 mJy at 1.4 GHz corresponds to 0.35 mJy at 3 GHz, if a typical spectral index of -0.7 is assumed. 
Thus, the VLASS is more effective in detecting a flat spectrum ($\alpha$ $\geq$-0.7) radio core, while the FIRST 
can be useful, if core exhibits steeper spectral index ($\alpha$ $<$~$-0.7$). 
Hence, the FIRST and VLASS can be used in a complementary fashion to detect the core component. 
Using both 3 GHz VLASS and 1.4 GHz FIRST image cutouts we find only 20/268 sources with no detected radio core, 
but distinctly detected radio lobes in our 325 MHz image. 
\par
We note that 1.4 GHz JVLA observations available in 5.0 deg$^{2}$ XMM$-$LSS$-$JVLA region can confirm the presence or absence of radio core at a much fainter level (3$\sigma$ = 0.06~mJy beam$^{-1}$). There are 10/20 sources falling within the XMM$-$LSS$-$JVLA region. The inspection of 1.4 GHz JVLA images reveals the presence of radio core in 5/10 sources, 
and confirms them to be active. Hence, we find only five sources with no detected radio core in the XMM$-$LSS$-$JVLA region, and 10 sources with no detected radio core in the XMM$-$LSS$-$Out region.    
Thus, using absent$-$core criterion we identify a total of 15/268 ($\sim$ 5.6$\%$) sources as the remnant candidates. 
\par
It is important to note that our criterion based on the non$-$detection of radio core selects only candidate 
remnants, and we cannot rule out the possibility of existence of a faint but undetected radio core. 
This is vindicated by the fact that, within the XMM$-$LSS$-$JVLA region, five out of ten sources with no detected core in the VLASS and FIRST images showed a faint core in deep 1.4 GHz JVLA images. 
We also point out that several of our sample sources, in particular those of smaller angular sizes 
($\leq$60$^{\prime\prime}$), tend to pose difficulty in clearly separating out core from radio lobes. 
Thus, our remnant candidates selected by using morphological criteria are biased towards the 
sources of larger angular sizes. We note that all but two of our remnant candidates identified by using absent$-$core  criterion show LAS $\geq$59$^{\prime\prime}$. 
Therefore, we use spectral criteria to identify remnant candidates that may have been missed by 
the morphological criterion owing to the limited sensitivity and angular resolution of the images used.
\begin{table*}
\begin{minipage}{160mm}
\caption{Selection criteria and number of remnant candidates}
\begin{tabular}{lrrr} \hline 
Sample                &    Criteria      &  Sample Size    &   Fraction \\ \hline
Extended sources with AGN activity & SNR $\geq$10, size $>$30$^{\prime\prime}$ & 268 &  \\
Sources above flux density cutoff limit &     S$_{\rm 325~MHz}$ $\geq$8.0 mJy  & 263 &     \\
Remnant candidates  & Morphology       &     15 (6 SPC or USS) &   5.7$\%$      \\
Remnant candidates  &   SPC            &     9 (4 USS or Morph)   & 3.4$\%$ \\
Remnant candidates  &  USS             &     6 (6 SPC or Morph)   & 2.3$\%$ \\
Total no. of remnant candidates & Morphology or SPC or USS  &  21/263     &  8.0 $\%$  \\ \hline
\end{tabular}
\\
{\bf Notes}$-$ SNR and radio size are measured at 325~MHz. The flux density cutoff limit is imposed for completeness. 
Remnant candidates are identified from the subsample of 263 sources. 
  
\end{minipage}
\end{table*} 
\subsection{Spectral curvature criterion}
\label{sec:spec}
According to the spectral ageing models integrated radio spectrum of an active radio source can be 
characterized with a broken power law, with a typical spectral index in the range of $-0.5$ to $-0.7$ 
below a break frequency ${\nu}_{\rm break}$, and a steeper spectral 
index ${\alpha}$ = ${\alpha}_{\rm inj}$ $-0.5$ above ${\nu}_{\rm break}$ \citep{Jaffe73,Carilli91}. 
The appearance of spectral break can be explained by the fact that 
the high energy relativistic electrons in the lobes lose their energy faster with shorter radiative lifetimes 
than that for the low energy electrons, and it results in the steeping of spectrum at the 
high$-$frequency regime beyond a break frequency ${\nu}_{\rm break}$ \citep{Komissarov94}. 
In remnant sources, the break frequency (${\nu}_{\rm break}$) 
progressively shifts to lower frequencies with time due to radiative cooling and lack of any supply of new plasma to the lobes. 
Hence, radio spectrum of a remnant source can be characterised with a steeper 
spectral index $\alpha$ $\sim$ ${\alpha}_{\rm inj}$ $-$ 0.5 above ${\nu}_{\rm break}$, and a flatter 
spectral index similar to the injection spectral index ${\alpha}_{\rm inj}$ $\sim$ $-0.5$ $-$ $-0.7$ below ${\nu}_{\rm break}$ \citep{Blandford78,Murgia11}. 
To search for remnants we exploit the spectral curvature characteristic of remnant sources.  
We define spectral curvature as (SPC) = ${\alpha}_{\rm high}$ $-$ ${\alpha}_{\rm low}$ in 
the frequency range of 150 MHz to 1.4 GHz; where 
${\alpha}_{\rm high}$ = ${\alpha}_{\rm 325}^{\rm 1400}$ and ${\alpha}_{\rm low}$ = ${\alpha}_{\rm 150}^{\rm 325}$. We use SPC $\leq$~$-0.5$ to identify remnant sources in our sample, 
although its value depends on the evolutionary stage of the sources \citep[{see}][]{Murgia11}. 
The error on the spectral index (${\alpha}_{\rm err}$) is estimated as per the following expression. 
\begin{equation}
\label{eq:1}
{\alpha}_{\rm err} = \frac{1}{ln({\nu}_{1}/{\nu}_{2})} \sqrt{\Bigg(\frac{S_{\rm 1,err}}{S_{1}}\Bigg)^{2} + \Bigg(\frac{S_{\rm 2,err}}{S_{2}}\Bigg)^{2}}
\end{equation}
Where $S_{\rm 1,err}$ and $S_{\rm 2,err}$ are errors in the flux densities $S_{\rm 1}$ and $S_{\rm 2}$, measured at 
the frequencies ${\nu}_{\rm 1}$ and ${\nu}_{\rm 2}$, respectively. 
\par
We estimate the two$-$point spectral index between 325 MHz and 1.4 GHz (${\alpha}_{\rm 325}^{\rm 1400}$) 
using the total flux densities at the respective frequencies. 
The 1.4 GHz flux densities are mainly taken from the NVSS considering the fact that larger NVSS 
synthesized beam (45$^{\prime\prime}$) would be advantageous in capturing faint diffuse emission, while radio observations of higher resolution {\ie} FIRST and JVLA are likely to miss the detection of diffuse extended emission. 
In our sample of 268 sources we find 1.4 GHz NVSS counterparts for 245 sources. 
Out of remaining 23 sources, nine sources fall within the XMM$-$LSS$-$JVLA region and we obtain 1.4 GHz flux density from 
the JVLA observations as these sources show complete detection of extended emission, 
and flux densities are derived from the JVLA image convolved with 45$^{\prime\prime}$ Gaussian beam equivalent to 
the NVSS beam$-$size.  
Thus, we estimate two$-$point spectral index ${\alpha}_{\rm 325}^{\rm 1400}$ 
for 254/268 sources and place an upper limit on ${\alpha}_{\rm 325}^{\rm 1400}$ for remaining 14 sources 
by using the NVSS 5$\sigma$ flux density limit of 2.5 mJy for point sources. For extended sources, the upper limit on 1.4 GHz flux density 
is obtained by using the number of NVSS beams (45$^{\prime\prime}$) required to fully cover the radio size measured at 325 MHz.   
\par
To obtain spectral index between 150 MHz and 325 MHz (${\alpha}_{\rm 150}^{\rm 325}$) we search for 150 MHz counterparts of our sample sources mainly using 150 MHz LOFAR catalogue. 
Due to high sensitivity of the LOFAR observations (5$\sigma$ $\sim$ 1.4 mJy) all 231/268 sources falling within the LOFAR region are detected at 150 MHz. 
We ensure that all source components are consistently matched at both the frequencies 
by overlaying 150 MHz catalogue source positions onto the 325 MHz GMRT images. 
We note that the similar resolution and depth of 150 MHz LOFAR and 325 MHz GMRT enable 
us to detect the extended emission at both the frequencies.
For remaining 37 sources we check their detection in the 150 MHz TGSS survey. 
22/37 sources show their detection in the TGSS catalog with 7$\sigma$ flux limit 
cutoff (S$_{\rm 150~MHz}$ $<$24.5 mJy).
For remaining 15 sources we place an upper limit on their 150 MHz flux density 
(S$_{\rm 150~MHz}$ $<$24.5 mJy) based on their non$-$detection in the TGSS.
Thus, we estimate two$-$point spectral index 
${\alpha}_{\rm 150}^{\rm 325}$ for 253/268 sources and place an upper limit for remaining 15 sources.  
There is only one source that lacks the detection at both 150 MHz and 1.4 GHz frequencies 
and is excluded from our spectral analysis as both ${\alpha}_{\rm low}$ and ${\alpha}_{\rm high}$ remain unconstrained.  
\par
\begin{figure}
\centering
\includegraphics[angle=0,width=8.5cm,trim={0.0cm 7.1cm 0.0cm 7.5cm},clip]{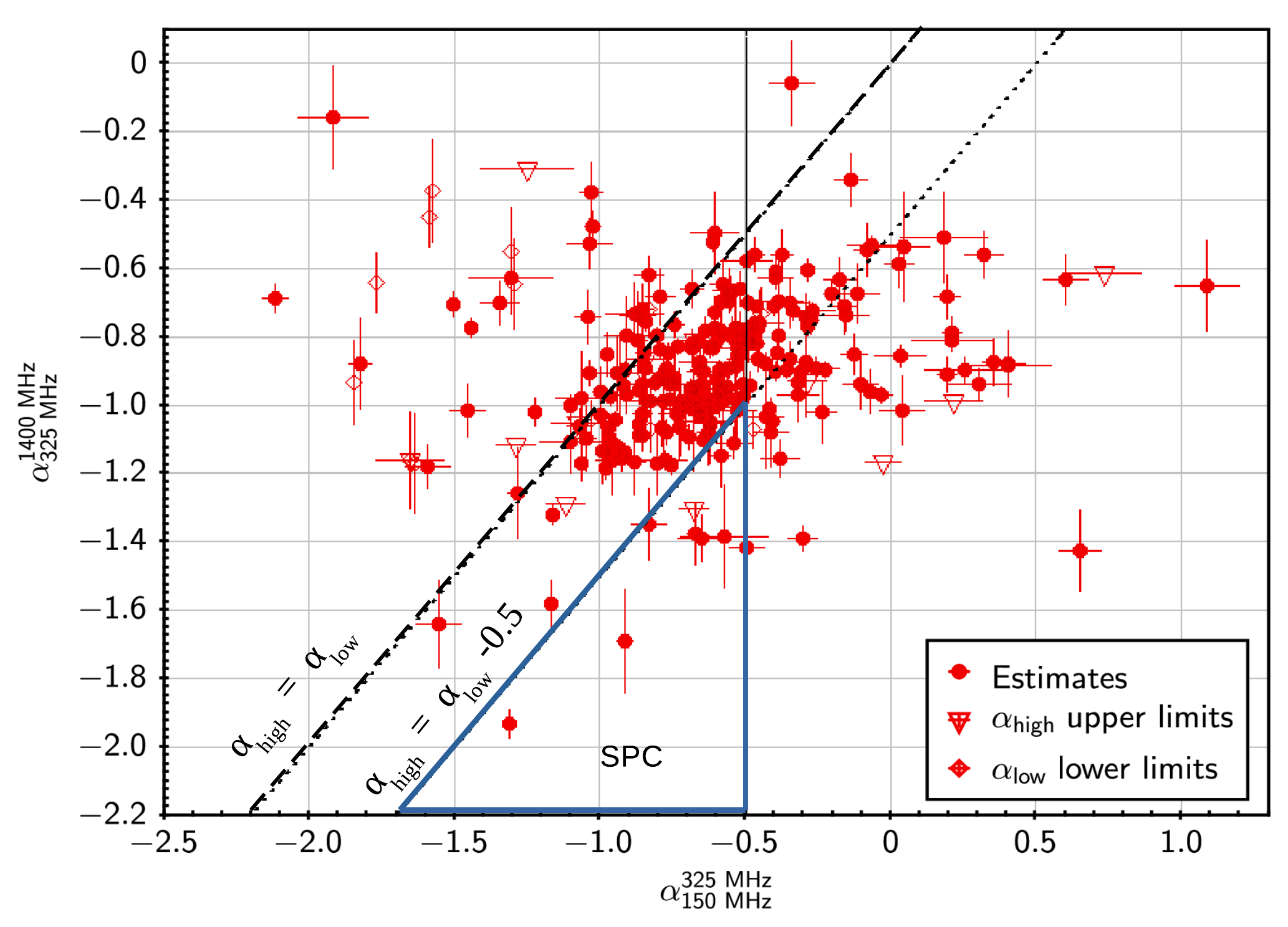}
\caption{Low frequency spectral index (${\alpha}_{\rm 150}^{\rm 325}$) versus high frequency spectral index 
(${\alpha}_{\rm 325}^{\rm 1400}$) plot. Sources showing strong spectral curvature 
(${\Delta}{\alpha}$ $\leq$~$-0.5$) fall in the blue triangle region bounded by lines 
${\alpha}_{\rm 325}^{\rm 1400}$ = ${\alpha}_{\rm 150}^{\rm 325}$ $-$ 0.5 and 
${\alpha}_{\rm 150}^{\rm 325}$ = -0.5. The sources undetected at 1.4 GHz have upper limits on 
${\alpha}_{\rm high}$ (${\alpha}_{\rm 325}^{\rm 1400}$) and they are shown by open downward triangle symbols. 
The sources undetected at 150 MHz have lower limits on ${\alpha}_{\rm low}$ (${\alpha}_{\rm 150}^{\rm 325}$) and they are 
shown by open rhombus symbols}.
\label{fig:SPC} 
\end{figure}

 
Figure~\ref{fig:SPC} shows the diagnostic plot of ${\alpha}_{\rm 150}^{\rm 325}$ versus 
${\alpha}_{\rm 325}^{\rm 1400}$ that allows us to identify sources with strong spectral 
curvature (SPC $\leq$~$-0.5$) in the frequency range of 150 MHz to 1.4 GHz. 
We note that while identifying remnant sources using spectral curvature diagnostic plot we 
avoid sources with ${\alpha}_{\rm 150}^{\rm 325}$ $\geq$~$-0.5$ {\ie}sources exhibiting 
flat or turnover spectrum in the low$-$frequency regime (below 325 MHz). 
Radio sources with peak frequency at $\leq$325 MHz can represent mega$-$hertz (MHz) peaked spectrum 
radio sources comprising mainly active young radio galaxies \citep[see][]{Callingham17}.
Further, sources showing upturn spectrum {\ie}${\alpha}_{\rm 150}^{\rm 325}$ steeper than 
${\alpha}_{\rm 325}^{\rm 1400}$ are also avoided. Sources showing upturn spectrum can 
possibly indicate episodic AGN activity wherein both remnant as well as young radio plasma co$-$exist,  
and a source appears much brighter at low$-$frequency due to substantial contribution from the remnant plasma 
\citep[see][]{Parma07}. However, we caution that the sources suffering with missing flux density at 325 MHz 
but not at 150 MHz can also give rise an artificially inverted spectrum. In fact, 
by using morphological criteria we found three remnant candidates showing inverted spectrum 
(see Table~\ref{tab:RRGSample}). In the spectral diagnostic plot, we 
identify sources with spectral curvature $\leq$~$-0.5$ that are found in a region bounded by the lines 
${\alpha}_{\rm high}$ = ${\alpha}_{\rm low}$ $-$ 0.5 and ${\alpha}_{\rm 150}^{\rm 325}$ = $-0.5$ 
(see Figure~~\ref{fig:SPC}). 
We find only nine sources with SPC $\leq$~$-0.5$ that are selected as remnant candidates. 
Of these, there are 3/9 remnant candidates that have already been identified by using the 
morphological criterion. In principle, the remaining six remnant candidates can also be identified by 
the absent$-$core criterion, but were deemed to be uncertain due to core$-$lobe deblending issue owing to 
their relatively smaller angular size. Therefore, the morphological and spectral selection criteria complement each other.
\subsection{Ultra$-$Steep Spectral (USS) index criterion}
\label{sec:USS}
We note that, with our limited frequency coverage,  
the spectral curvature criterion (SPC = ${\alpha}_{\rm high}$ - ${\alpha}_{\rm low}$ $\leq$-0.5) 
would miss remnant sources in which spectral break falls below 325 MHz. 
According to the evolutionary models break frequency (${\nu}_{\rm b}$) progressively shifts 
towards lower frequencies as source ages. The relation between total source age ($t_{\rm s}$) 
and ${\nu}_{\rm b}$ can be expressed with the following equation \citep[see][]{Komissarov94,Slee01}. 
\begin{equation}
\label{eq:2}
t_{\rm s} = 1590 \Bigg[~\frac{B^{\rm 0.5}_{\rm eq}}{(B_{\rm eq}^{2} + B_{\rm CMB}^{2})\sqrt{{\nu}_{\rm b}(1 + z)}}~\Bigg]~{\rm Myr}
\end{equation}
Where $B_{\rm eq}$ and $B_{\rm CMB}$ $=$ 3.25(1+$z$)$^2$ are equipartition magnetic field and inverse Compton equivalent 
magnetic field, respectively, in the unit of $\mu$G, $\nu_{\rm b}$ is break frequency in GHz and $z$ is the redshift of 
radio source. 
For a remnant source with ${\nu}_{\rm b}$ = 325 MHz, we find $t_{\rm s}$ = 41 Myr by 
assuming a typical value for equipartition magnetic field ($B_{\rm eq}$) 3.0~$\mu$G and redshift ($z$) 0.67, 
the median redshift of our sample. Considering the same parameters but ${\nu}_{\rm b}$ = 150 MHz would result 
$t_{\rm s}$ = 60 Myr. We note that our  estimates for the total source ages are similar to the remnants reported 
in the literature \citep[see][]{Parma07,Brienza16,Shulevski17}. 
Although, we mention that the source ages estimated here are based on various assumptions, and hence, 
these need to be considered as only characteristic timescales.       
\par 
In our sample, remnant sources with ${\nu}_{\rm b}$ $<$ 325 MHz would appear as the 
USS sources. Hence, USS radio sources can be regarded as potential remnant 
candidates. Although, USS criterion selects only a fraction of remnant population consisting of 
mainly old remnants \citep[see][]{Godfrey17}. 
We search for USS sources (${\alpha}_{\rm 150}^{\rm 1400}$ $\leq$~$-1.2$) based on 
the spectral index measured between 150 MHz and 1.4 GHz {\viz}the widest spectral window available. 
We caution that the limit placed on the spectral index for defining an USS source is somewhat arbitrary,  
and is based on the fact that many remnant sources show steep spectral index ($\alpha$) $<$~$-1.2$ 
at low frequencies \citep{Godfrey17}. 
\par
\begin{figure}
\centering
\includegraphics[angle=0,width=8.5cm,trim={0.0cm 0.0cm 0.0cm 0.0cm},clip]{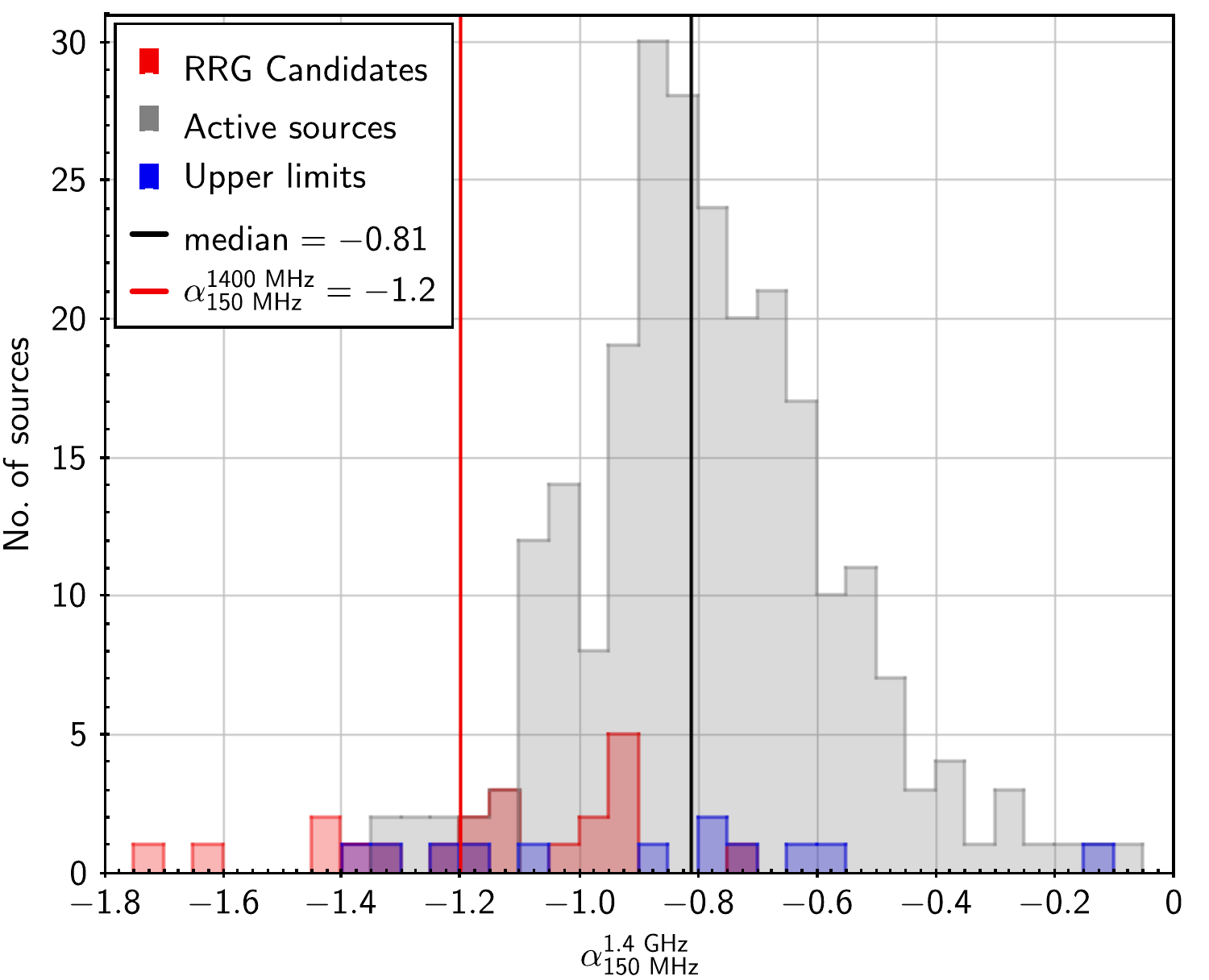}
\caption{The histogram of spectral index between 150~MHz and 1.4~GHz (${\alpha}_{\rm 150}^{\rm 1400}$).
The vertical black line marks a median value of $-0.81$ and the vertical red line segregates USS sources 
having ${\alpha}_{\rm 150}^{\rm 1400}$ $\leq$~$-1.2$. Our remnant candidates selected from all the criteria are shown by red bins.} 
\label{fig:USSHist} 
\end{figure}

In our sample, 242/268 sources have estimates of 150 MHz $-$ 1.4 GHz spectral index (${\alpha}_{\rm 150}^{\rm 1400}$) from their detection at both 150 MHz and 1.4 GHz frequencies, while 11 sources have 
only upper limits on ${\alpha}_{\rm 150}^{\rm 1400}$ due to their non$-$detection at 1.4 GHz. 
The 1.4 GHz flux densities are mainly from the NVSS and the upper limits on  ${\alpha}_{\rm 150}^{\rm 1400}$ are derived by using the NVSS flux density limit of 2.5 mJy (see Section~\ref{sec:spec}).  
The remaining 15 sources without 150 MHz flux density estimates have a lower limit on 
${\alpha}_{\rm 150}^{\rm 1400}$, and are likely to be sources with flatter index. 
These 15 sources are excluded from our analysis considering the fact that we are searching for USS sources. 
Figure~\ref{fig:USSHist} shows the distribution of ${\alpha}_{\rm 150}^{\rm 1400}$ ranging from 
$-1.72$ to $-0.05$ with a median value of $-0.81{\pm}0.02$. We find that only 10/253 ($\sim$3.9$\%$) sources can 
be identified as USS sources. There are 2/10 USS sources with an upper limit on ${\alpha}_{\rm 150}^{\rm 1400}$. Accounting for the errors on spectral indices 
the fraction of USS sources can be in the range of 2.6$\%$~$-$5.2$\%$. 
We note that the fraction of USS sources found in our sample is similar to that reported in some previous 
studies \citep[{\eg}][]{Brienza17}. 
\par
We caution that our USS sample can be contaminated by other types of radio sources with same spectral 
characteristic such as high$-z$ radio galaxies (HzRGs, \citet{Singh14}), some active FR~II radio galaxies  \citep{Harwood15} and cluster relics and halos \citep{vanWeeren09}. 
Radio relics and halos can be identified from their radio morphologies 
showing elongated filamentary structure and circular halo$-$like emission in contrast to double$-$lobe radio galaxies 
\citep{vanWeeren11}. We find that none among our ten USS sources shows morphology similar to a relic or a radio halo.  
We note that 2/10 of our USS sources are identified as active radio galaxies with core detected in the 1.4 GHz JVLA images. While, another 2/10 USS sources are identified as potential H$z$RGs due to their high 
surface$-$brightness and the non$-$detection of optical host.
Thus, after discarding potential contaminants the USS criterion yields a sample of only six remnant candidates. Unsurprisingly, all but one of our USS remnant candidates are already identified by the morphological criterion, while remaining one USS remnant candidate is selected by the SPC criterion. 
Therefore, again we find that the morphological criterion {\ie}absent$-$core is effective in 
identifying majority of our remnant candidates, except 
sources with small angular sizes where deblending of core from lobes is difficult using 
images of 2$^{\prime\prime}$.5$-$5$^{\prime\prime}$.0 resolution. 
Further, identification of remnant candidates from both spectral as well as morphological criteria strengthens 
their remnant status.   
Using both morphological and spectral criteria we find a total of 21 remnant candidates that are listed in Table~\ref{tab:RRGSample}. Radio contours overlaid on the corresponding optical images are presented in Figure~\ref{fig:RRGImages}.

\section{Characteristics of remnant candidates} 
\label{sec:discussion}
With an aim to decipher the nature of our remnant candidates, we discuss their characteristic properties {\ie}radio spectra, surface$-$brightness, core$-$prominence upper limits, redshifts and luminosities, 
and compare them with the remnants reported in the literature.    
\subsection{Spectral properties}
\label{sec:specprop}
\subsubsection{Spectral index distribution}
Figure~\ref{fig:USSHist} shows the distributions of two$-$point spectral index between 150 MHz and 1.4 GHz 
(${\alpha}_{\rm 150}^{\rm 1400}$) for our remnant candidates and active sources. 
We find that our 21 remnant candidates have ${\alpha}_{\rm 150}^{\rm 1400}$ in the range of $-1.72$ to $-0.75$ with a median value of $-1.02{\pm}0.05$. While, active sources 
have ${\alpha}_{\rm 150}^{\rm 1400}$ distributed in the range of $-1.32$ to $-0.05$ with 
a median value of $-0.79{\pm}0.02$. 
Therefore, in comparison to active sources, our remnant candidates show systematically steeper 
spectral indices even when majority (15/21) of them can be identified by using only morphological criterion. 
The two sample Kolmogorov$-$Smirnov (KS) test suggests that the spectral index distributions of active and remnant candidates are not similar {\ie}probability that the two samples come from the same distribution is only 0.54 (see Table~\ref{tab:RRGComp}). 
We caution that the small number of remnant candidates can affect the statistical result. 
The steep spectral indices (${\alpha}_{\rm 150}^{\rm 1400}$ $<$~$-1.0$) of our remnant 
candidates can be understood as the spectral evolution of radio lobes suffering energy 
losses during the remnant phase. However, the remnant candidates with less steep spectral indices 
(${\alpha}_{\rm 150}^{\rm 1400}$ $>$$-1.0$) need to be examined more critically. We note that our remnant candidates with ${\alpha}_{\rm 150}^{\rm 1400}$ $>$$-1.0$ tend to exhibit spectral curvature and the 
spectral indices between 325 MHz and 1.4 GHz (${\alpha}_{\rm 325}^{\rm 1400}$) are steeper than $-1.0$. 
\par
From Figure~\ref{fig:USSHist} it is evident that there is no sharp cutoff limit on the spectral index 
distribution that can enable us to differentiate between active sources and remnant candidates. 
In fact, active sources continue to remain present even if only USS sources are 
considered (see Section~\ref{sec:USS}). Also, we note that only 12/21 (57$\%$) of our remnant candidates 
are identified by using spectral criteria {\ie}spectral curvature as well as USS criterion, implying that a 
substantial fraction of remnants would be missed if search is limited only to the spectral criteria. 
Hence, we conclude that the morphological criteria are important for identifying the full population of 
remnant sources. This conclusion is similar to that found in some recent studies \citep[{\eg}][]{Mahatma18,Jurlin21}.   

%
\begin{figure*}
\centering
\includegraphics[angle=0,width=6.0cm,trim={0.5cm 6.0cm 0.0cm 4.0cm},clip]{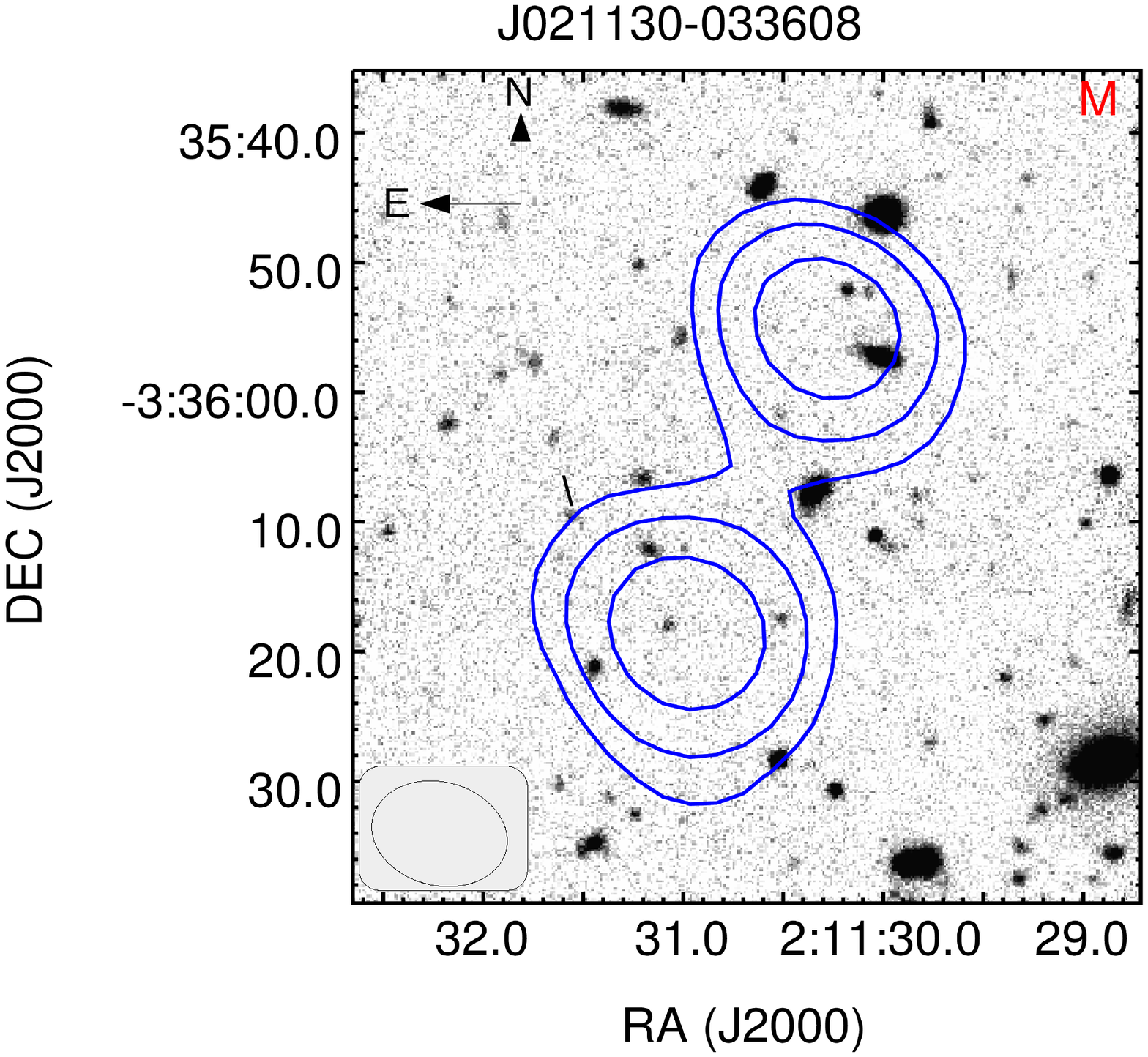}
\includegraphics[angle=0,width=5.8cm,trim={0.0cm 4.0cm 0.0cm 3.5cm},clip]{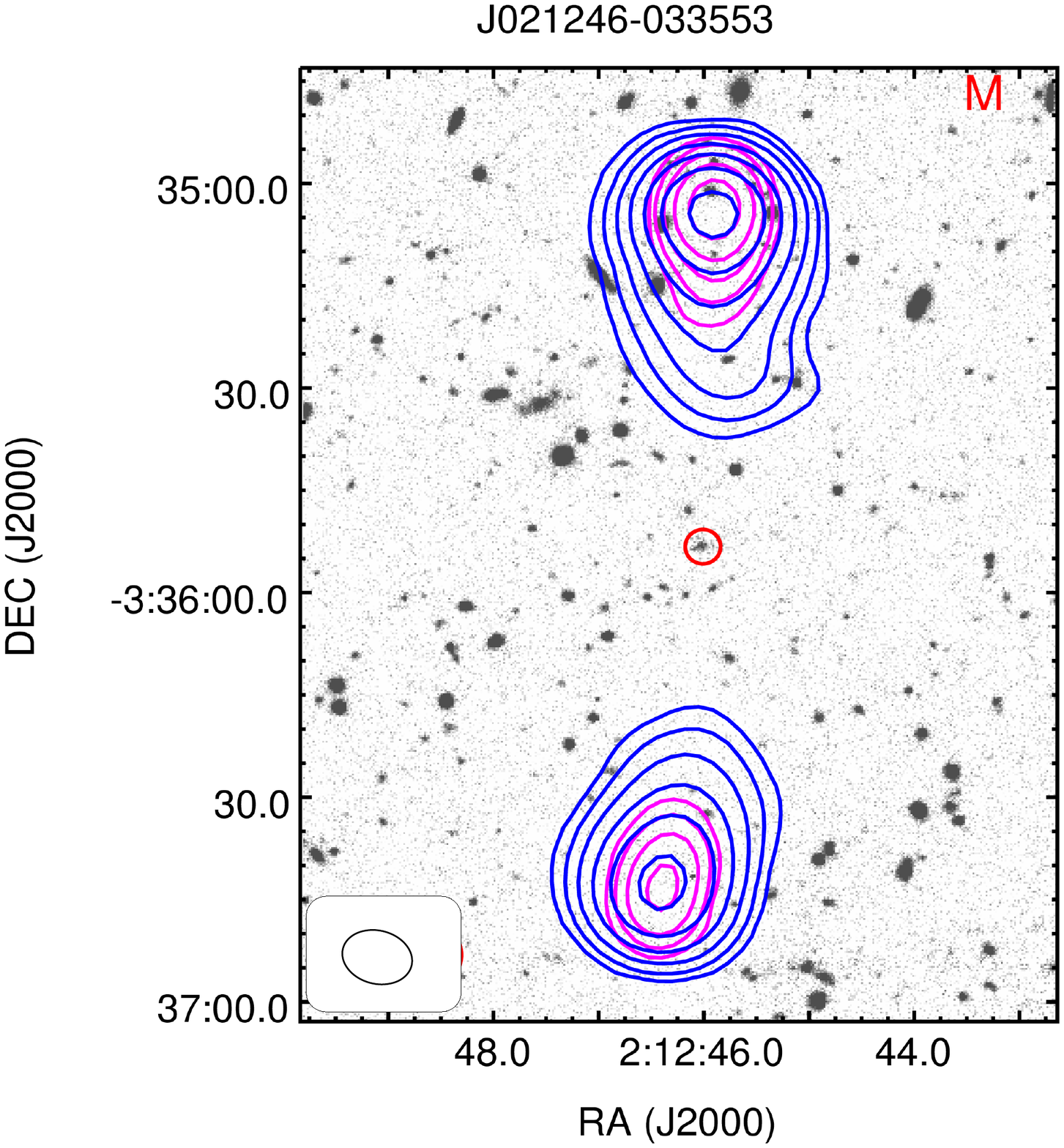}
\includegraphics[angle=0,width=6.0cm,trim={0.0cm 6.0cm 0.0cm 6.0cm},clip]{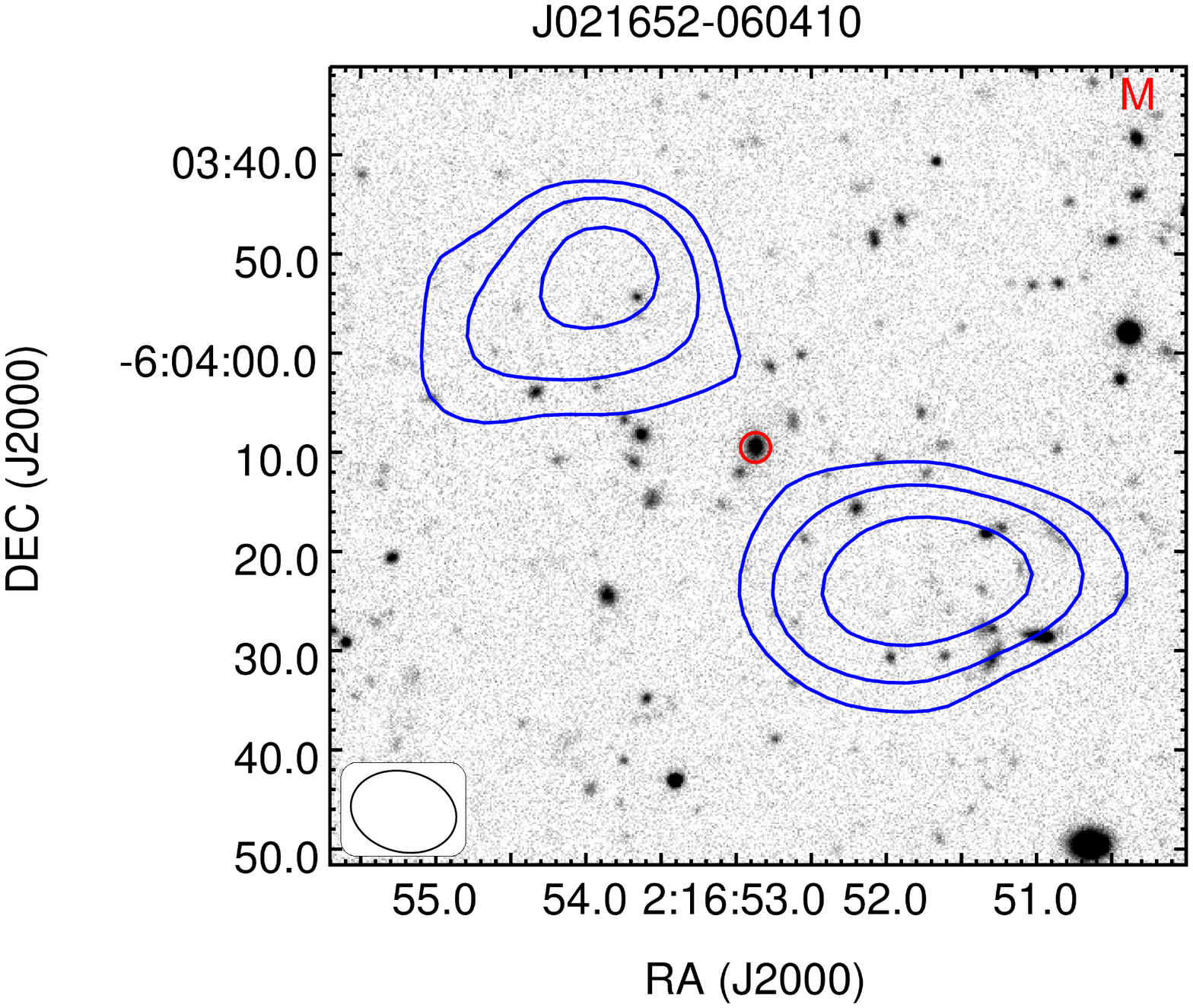}
\includegraphics[angle=0,width=6.0cm,trim={0.0cm 3.5cm 0.0cm 6.2cm},clip]{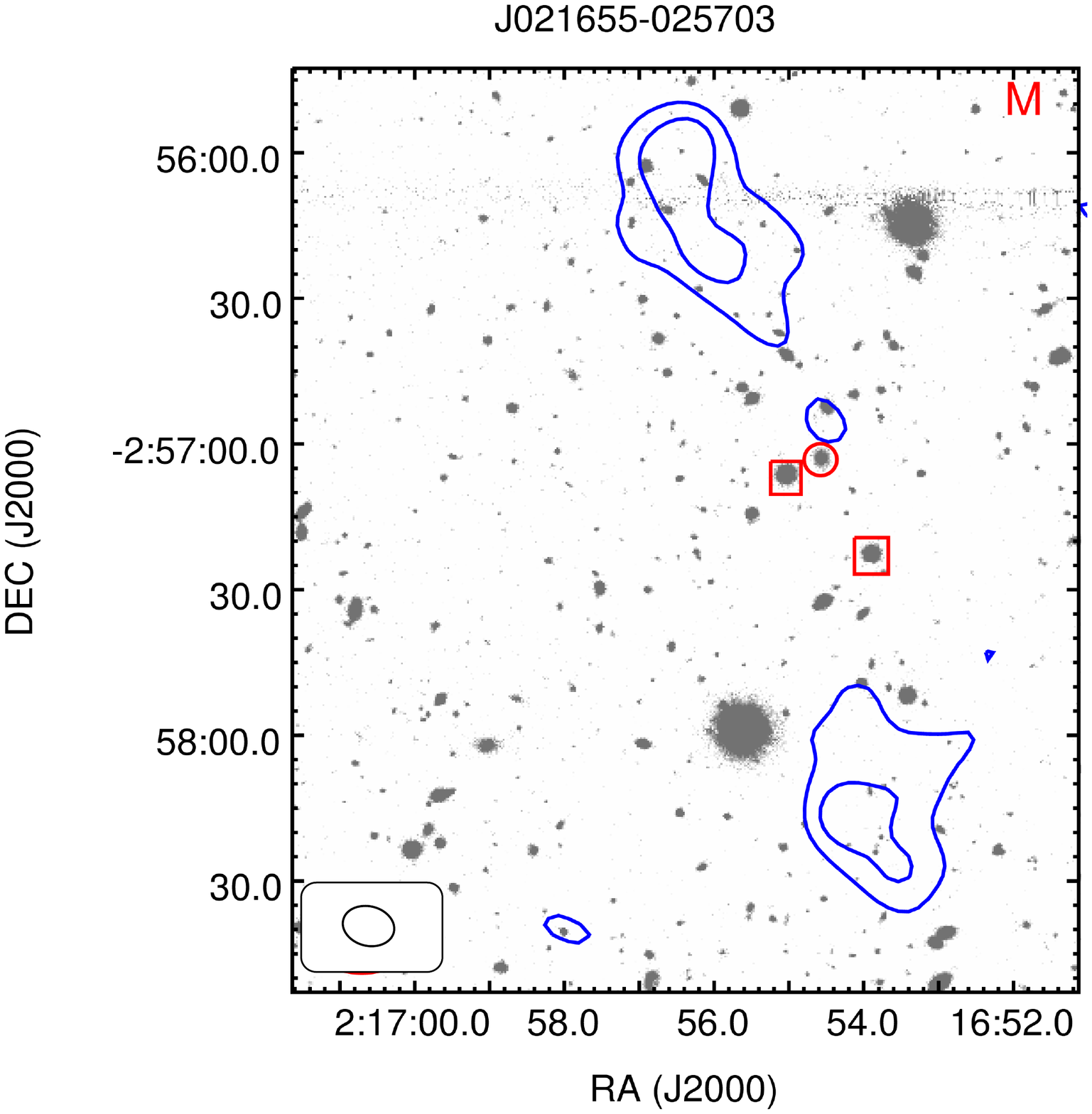}
\includegraphics[angle=0,width=5.8cm,trim={0.5cm 4.0cm 0.5cm 7.0cm},clip]{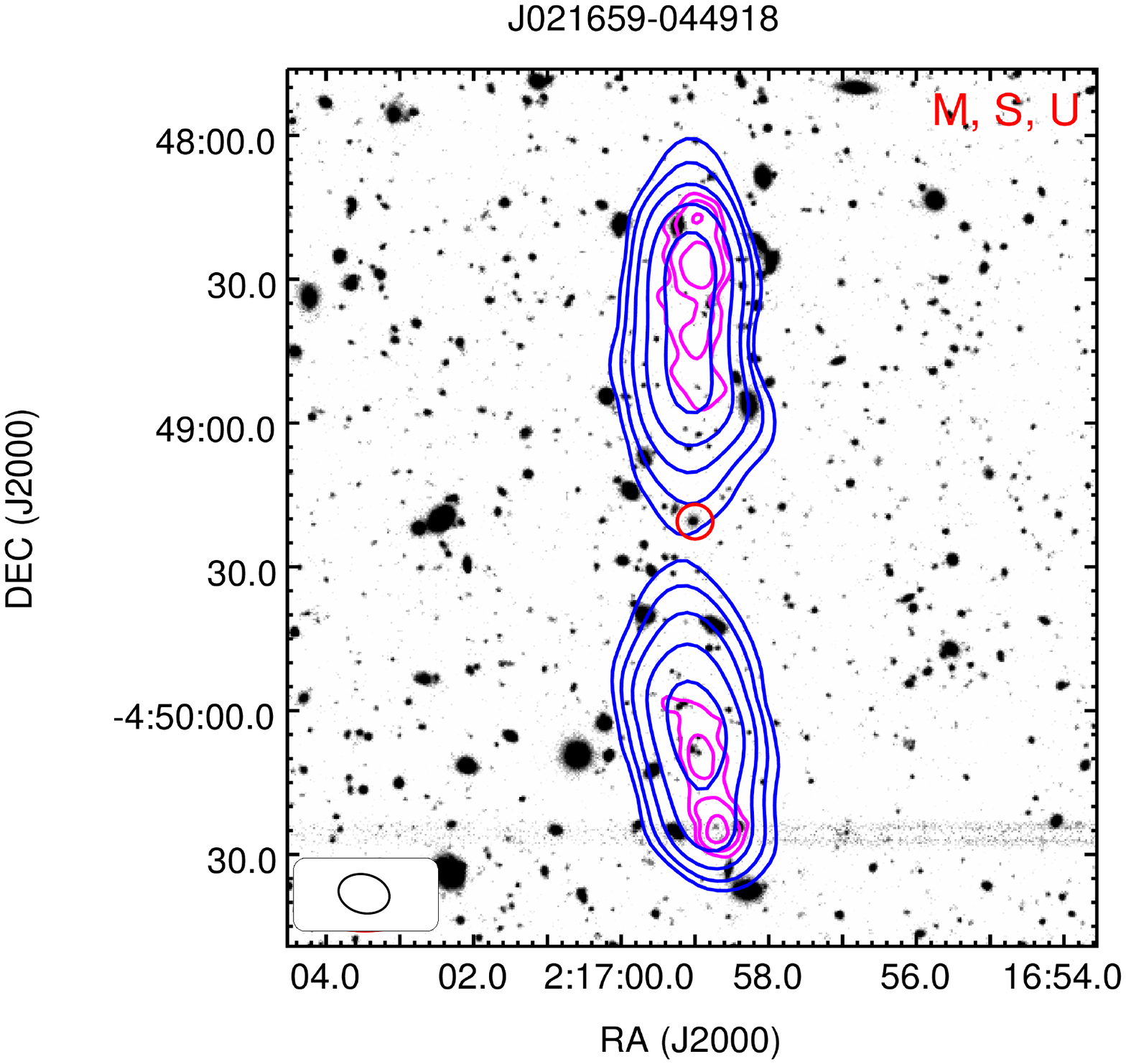}
\includegraphics[angle=0,width=5.8cm,trim={0.5cm 4.5cm 0.5cm 4.5cm},clip]{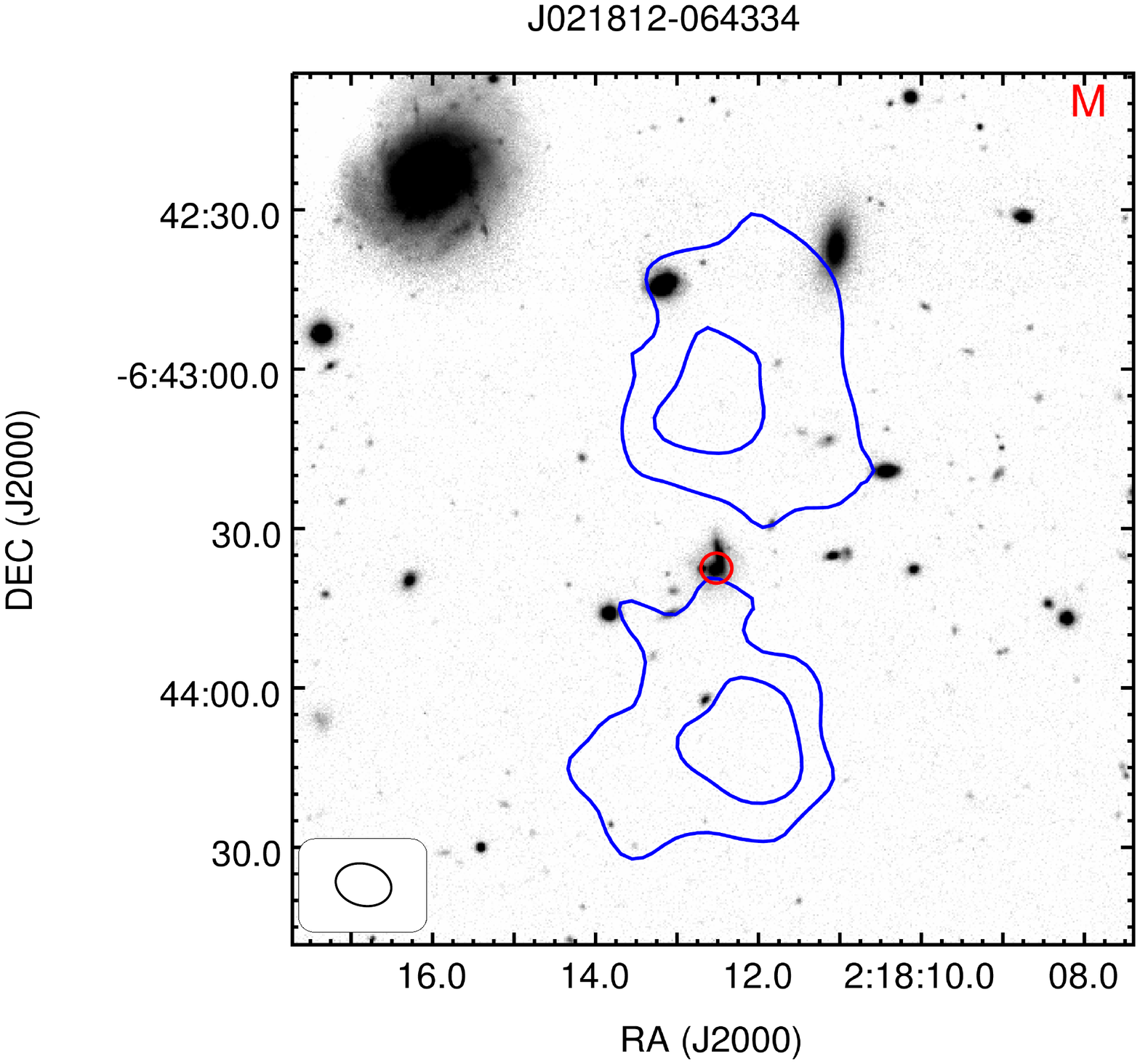}
\includegraphics[angle=0,width=5.8cm,trim={0.5cm 3.0cm 0.5cm 4.0cm},clip]{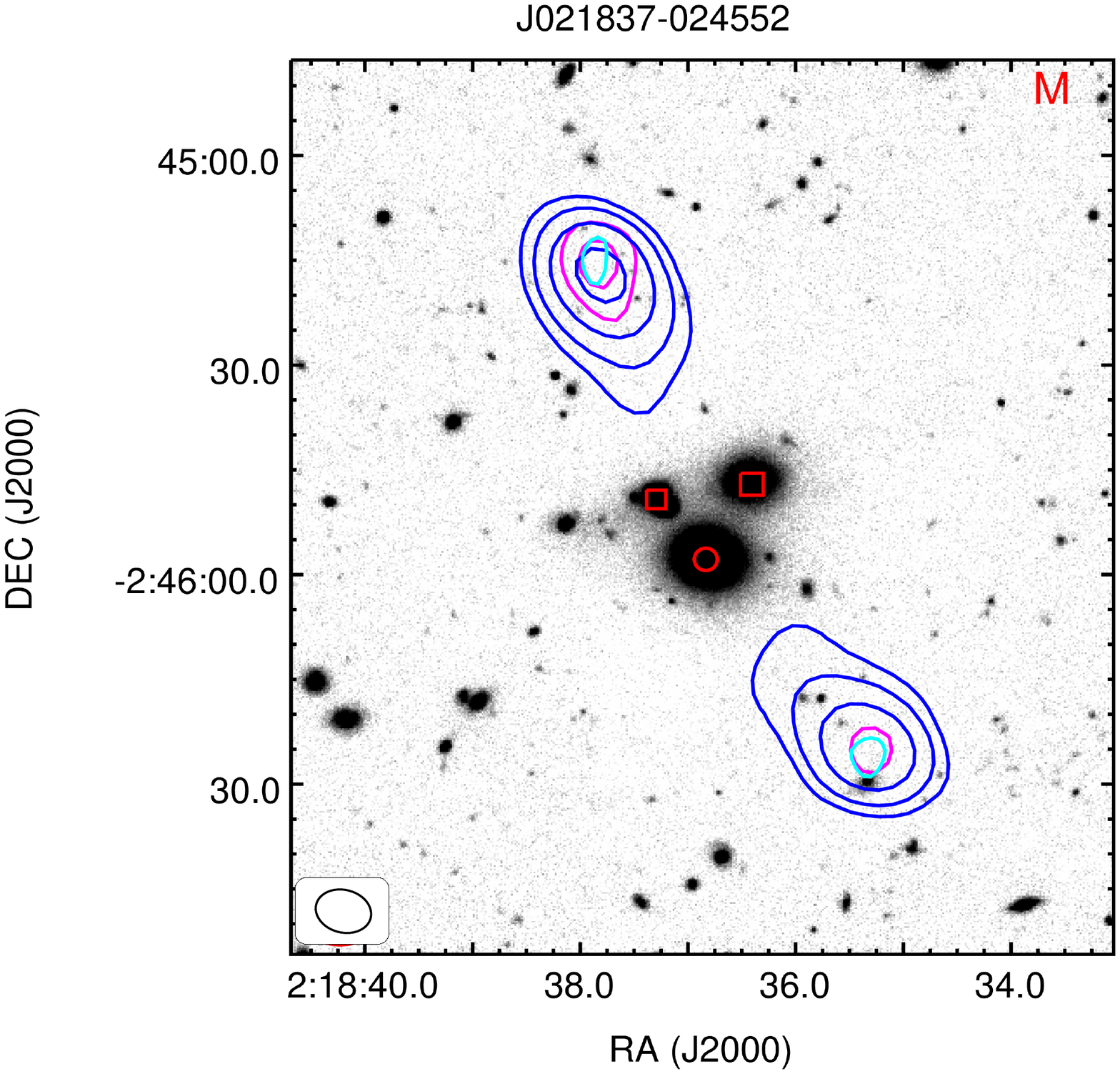}
\includegraphics[angle=0,width=5.8cm,trim={0.0cm 4.0cm 0.5cm 5.0cm},clip]{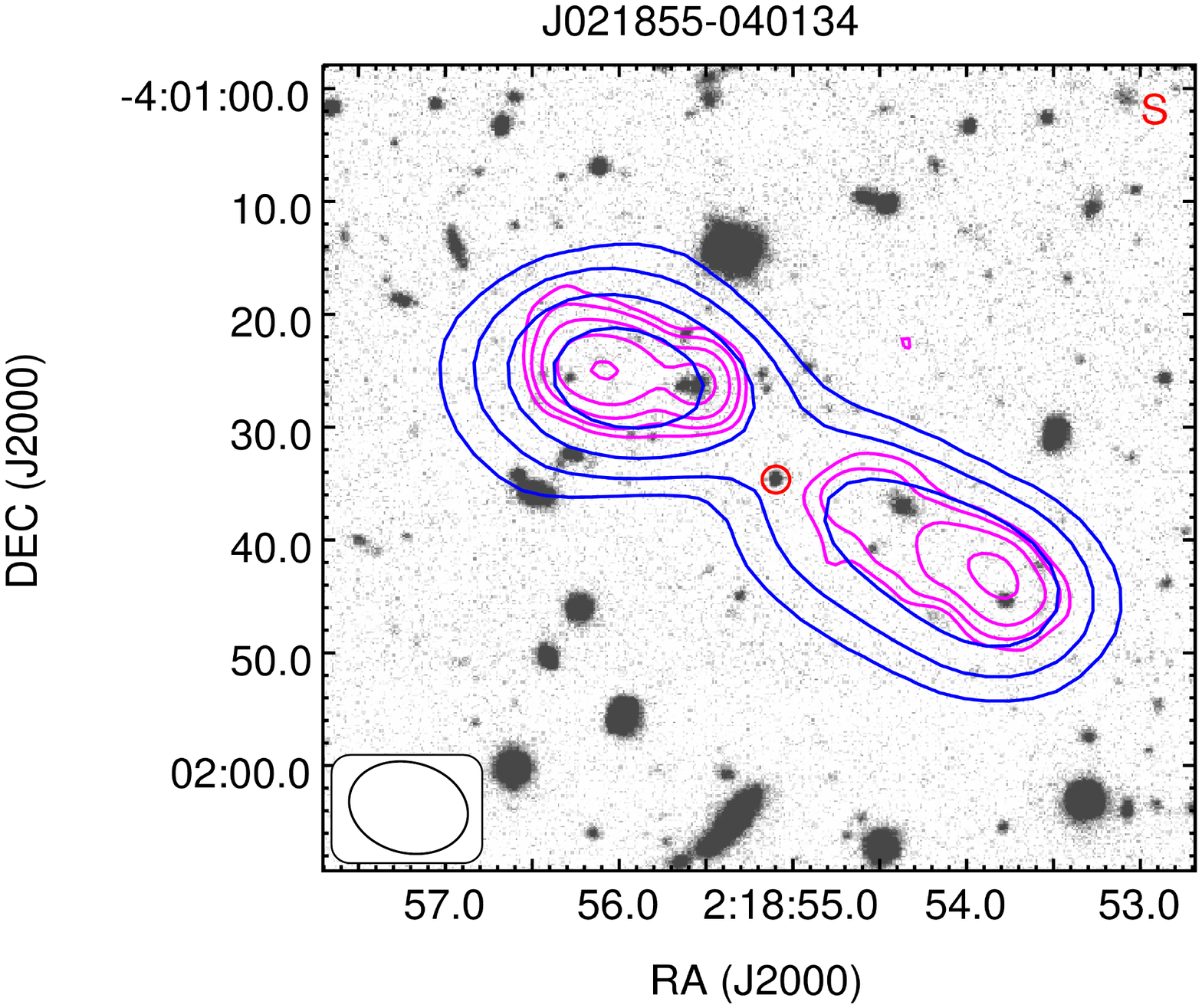}
\includegraphics[angle=0,width=5.8cm,trim={0.0cm 4.0cm 0.0cm 5.0cm},clip]{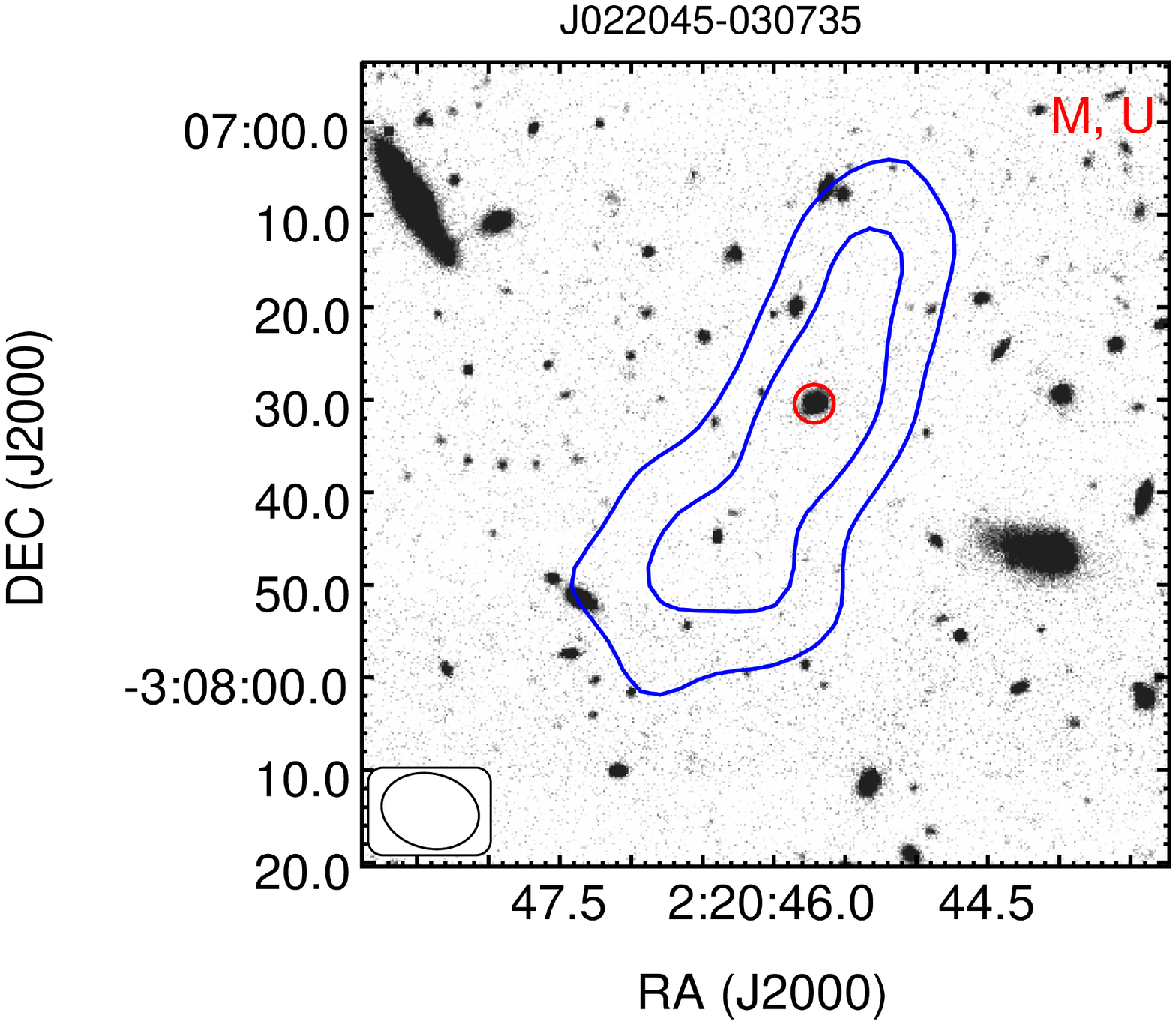}
\caption{The images of our remnant candidates showing the radio contours from 325 MHz GMRT (in Blue), 1.4 GHz JVLA/FIRST (in Magenta) and 3.0 GHz VLASS (in Cyan) overlaid onto the corresponding HSC$-$SSP $i$ band optical image. 
The radio contour levels are at 3$\sigma$ $\times$ (1, 2, 4, 8, 16 ......) and 
the corresponding optical image is logarithmically scaled. The likely host galaxy is marked with a 
red circle on it. Other potential host candidates are marked by red boxes. 
The GMRT 325 MHz synthesized beam of 10$^{\prime\prime}$.2 $\times$ 7$^{\prime\prime}$.9 
is shown in the bottom left corner. The arrows indicating the north and east directions are shown in 
the first image.}  
\label{fig:RRGImages} 
\end{figure*}
\addtocounter{figure}{-1}
\begin{figure*}
\centering
\includegraphics[angle=0,width=5.8cm,trim={0.0cm 3.5cm 0.0cm 6.0cm},clip]{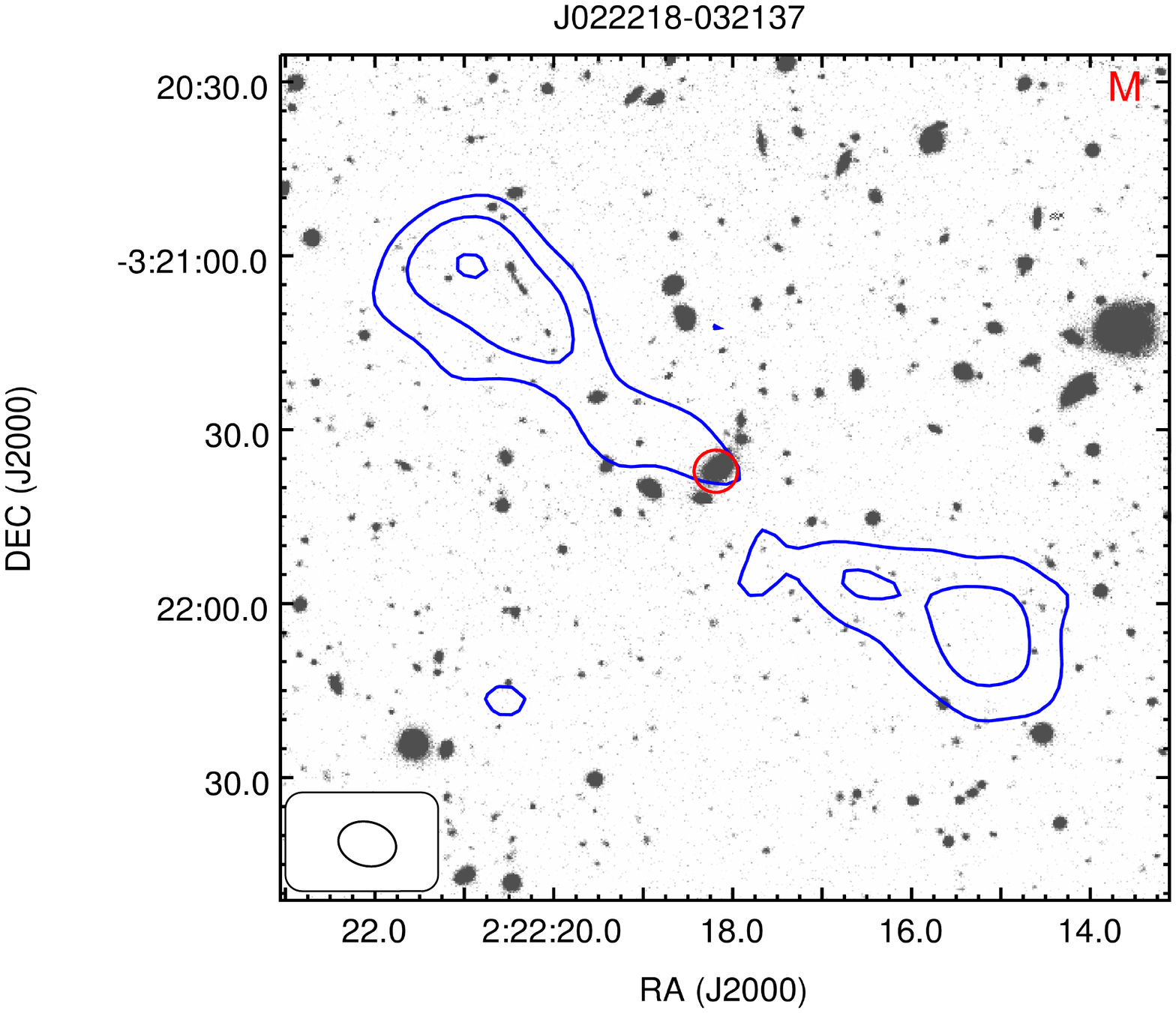}
\includegraphics[angle=0,width=5.8cm,trim={0.5cm 5.0cm 0.5cm 6.0cm},clip]{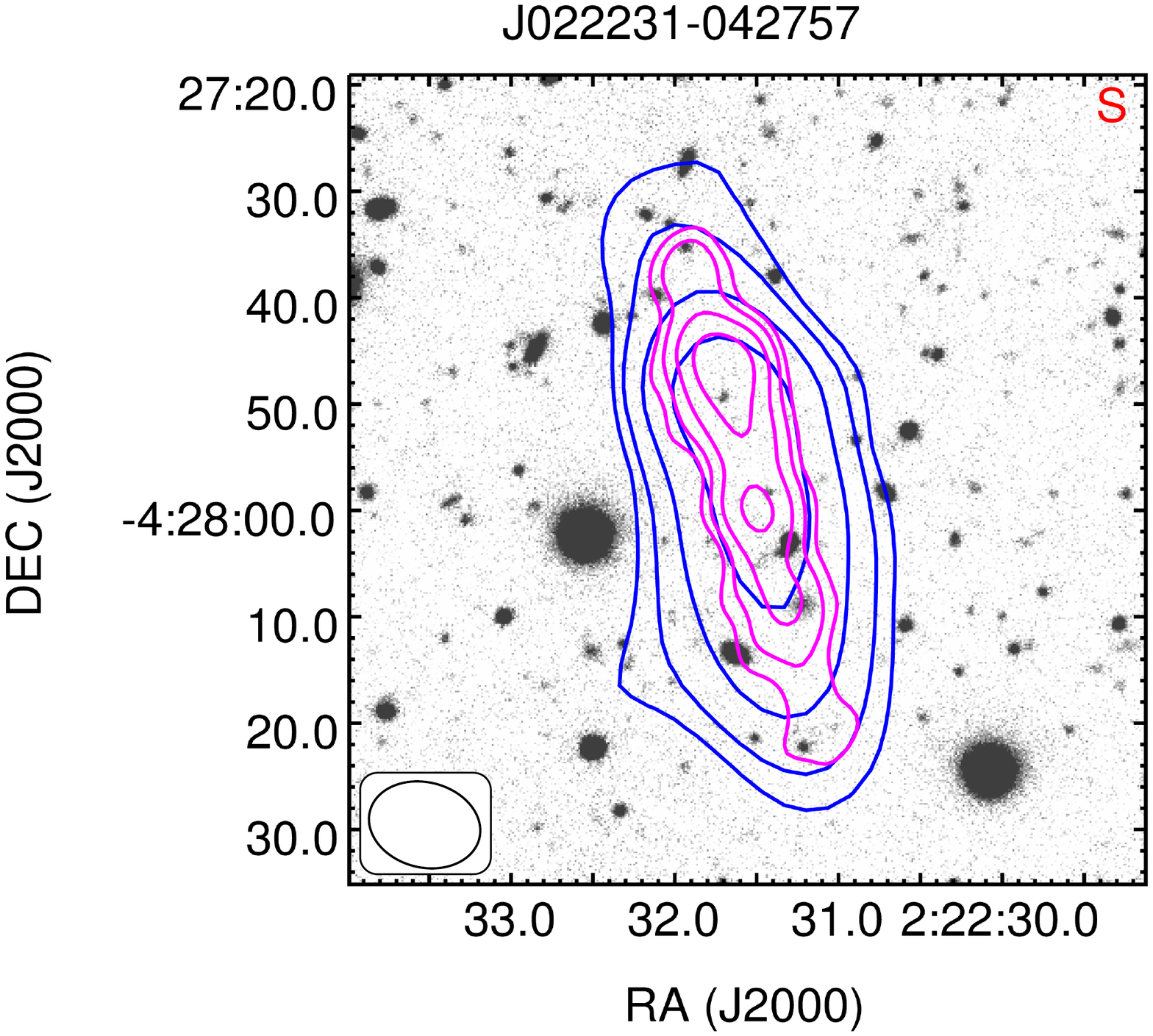}
\includegraphics[angle=0,width=6.0cm,trim={0.5cm 4.5cm 0.5cm 6.0cm},clip]{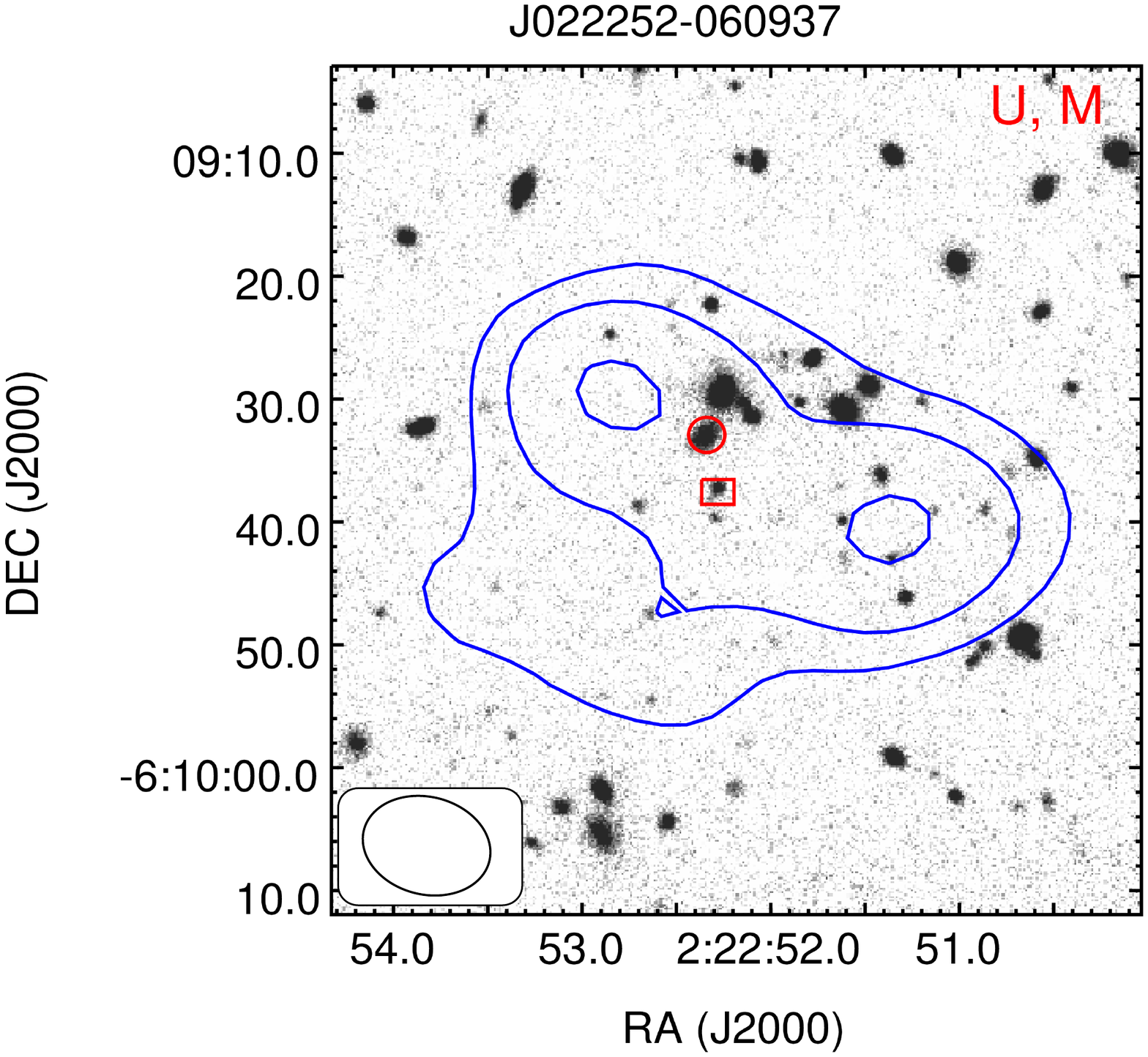}
\includegraphics[angle=0,width=5.8cm,trim={0.5cm 4.0cm 0.5cm 6.0cm},clip]{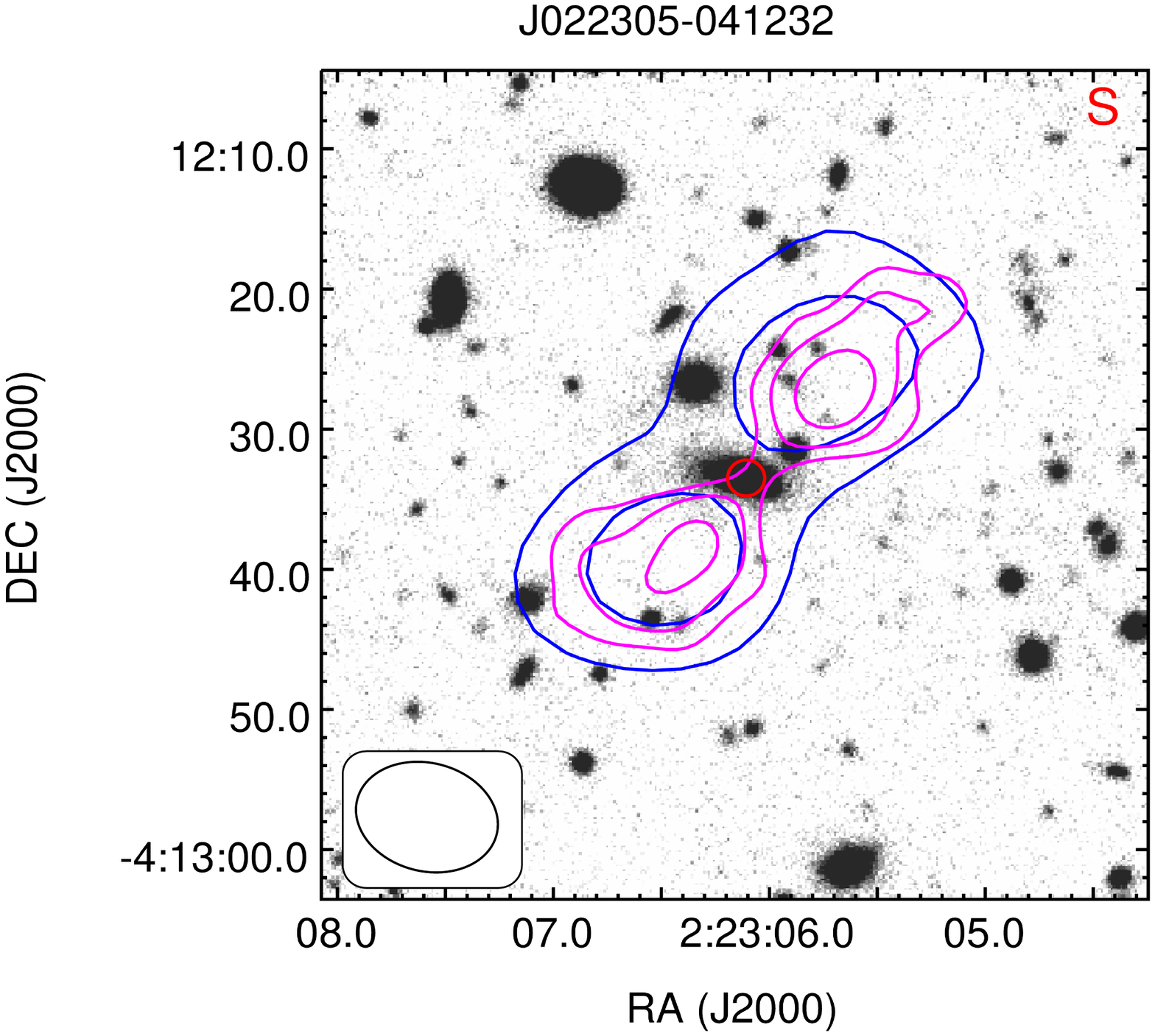}
\includegraphics[angle=0,width=5.8cm,trim={0.5cm 4.0cm 0.5cm 6.0cm},clip]{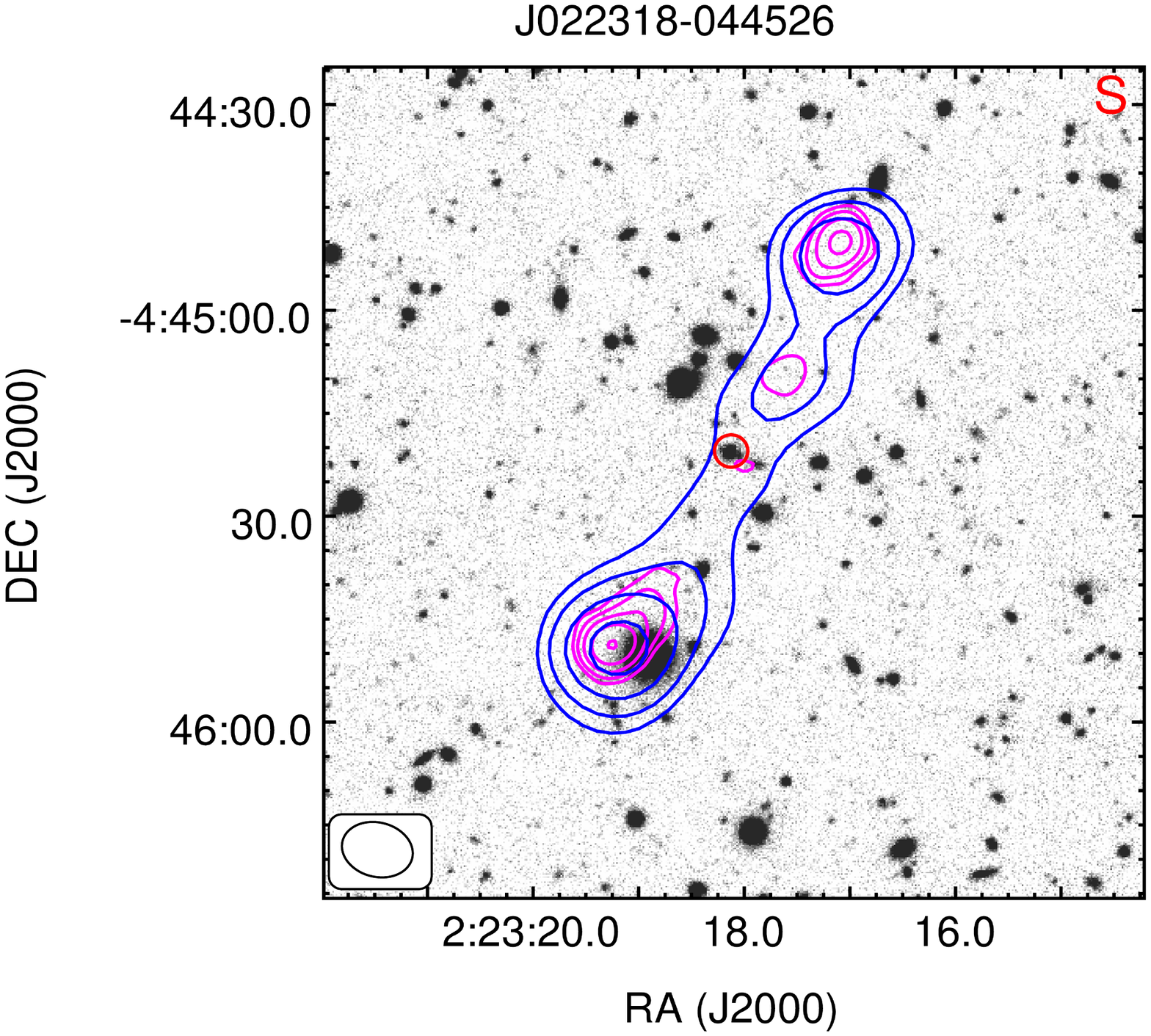}
\includegraphics[angle=0,width=5.8cm,trim={0.0cm 4.0cm 0.5cm 6.0cm},clip]{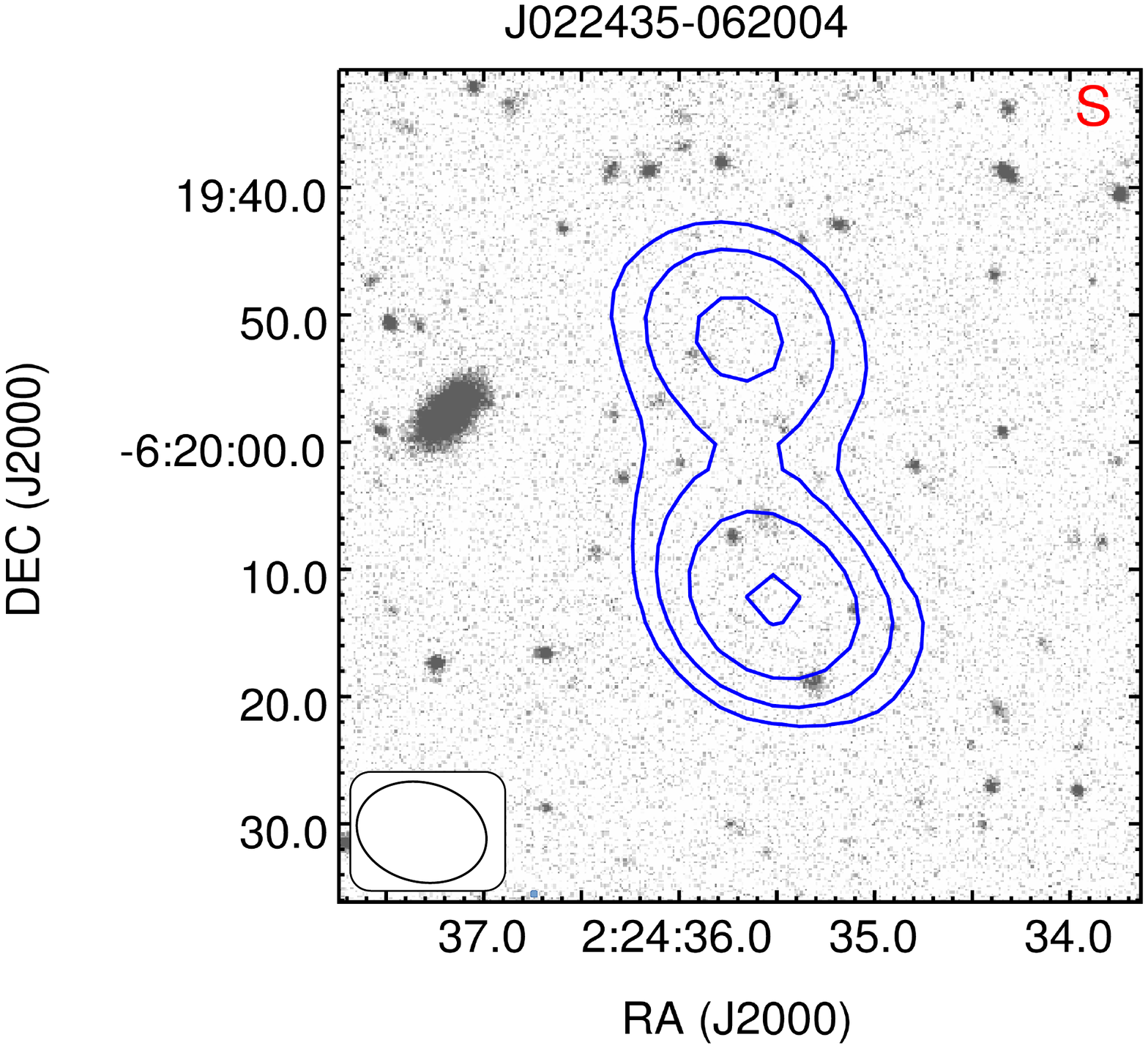}
\includegraphics[angle=0,width=5.8cm,trim={0.0cm 4.0cm 0.5cm 6.0cm},clip]{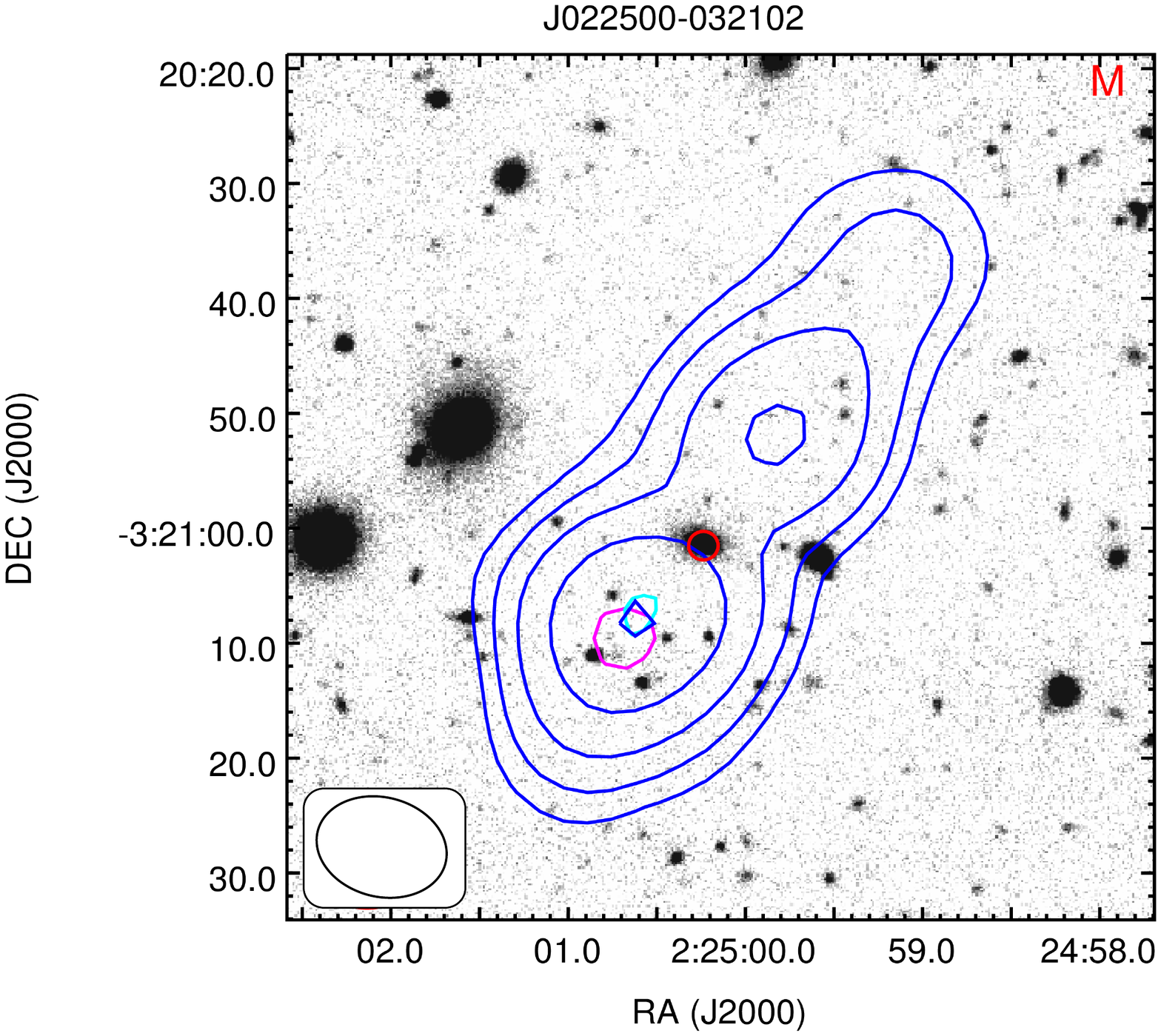}
\includegraphics[angle=0,width=5.8cm,trim={0.5cm 4.0cm 0.5cm 6.0cm},clip]{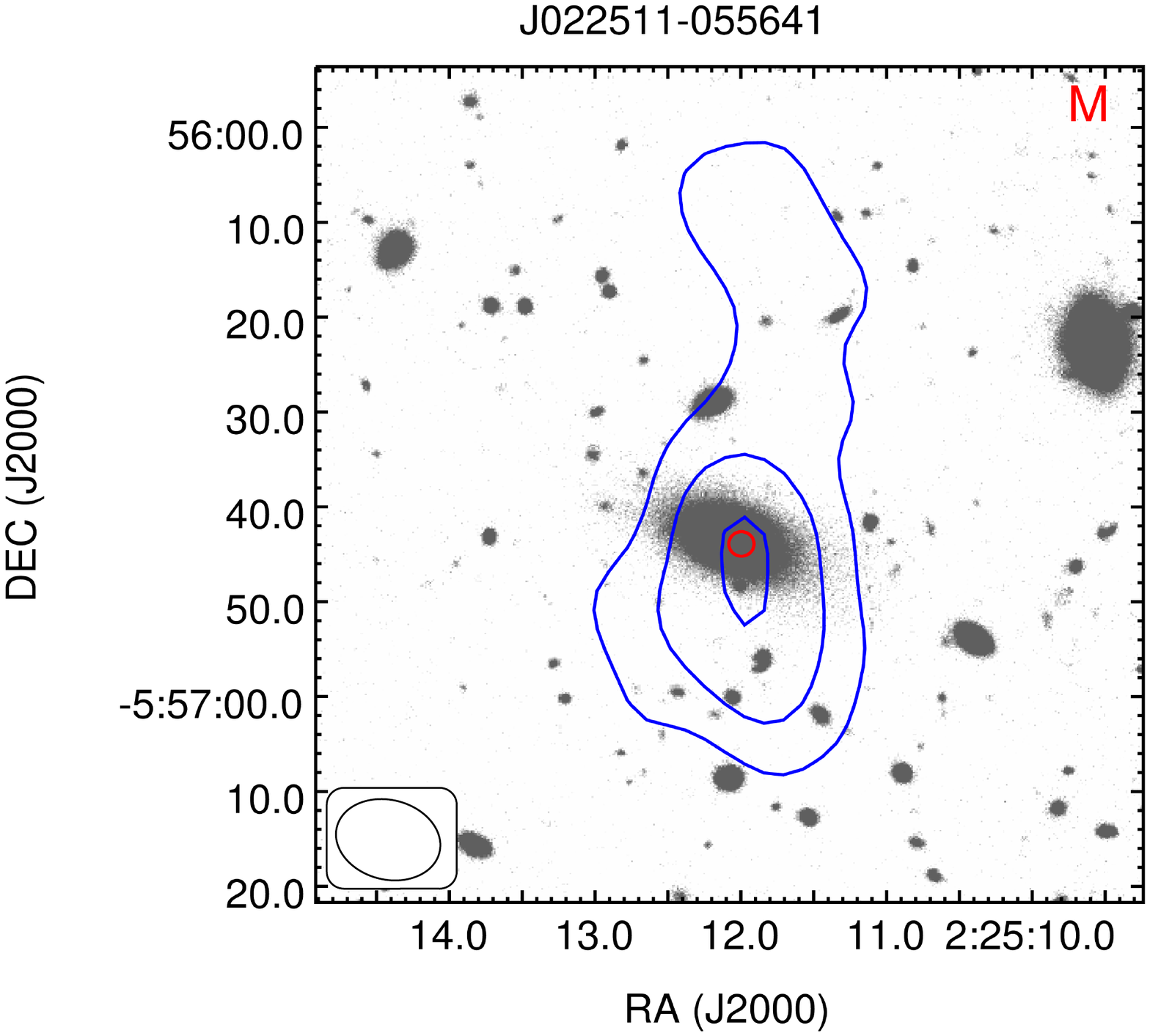}
\includegraphics[angle=0,width=5.8cm,trim={0.5cm 4.0cm 0.0cm 6.0cm},clip]{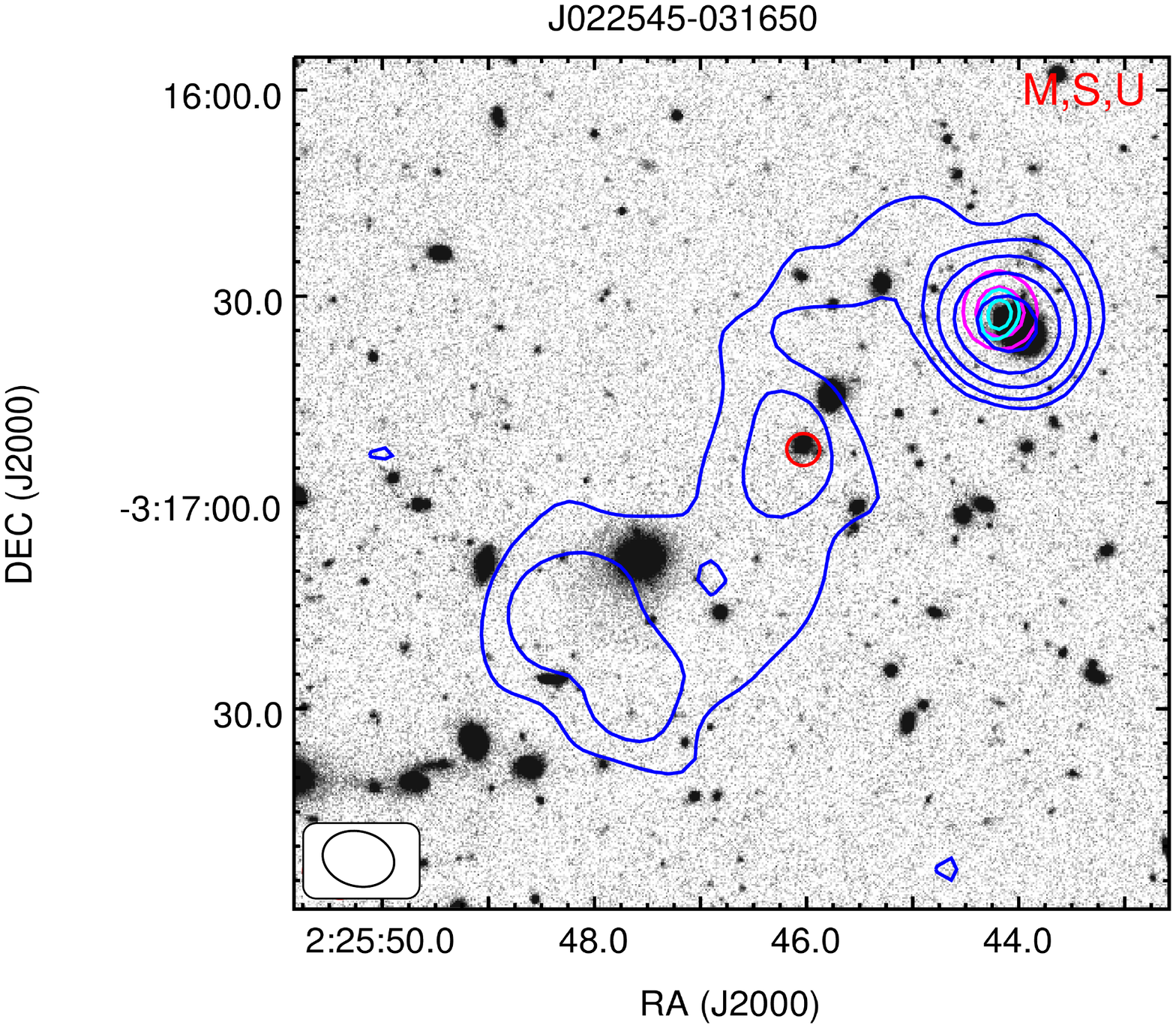}
\includegraphics[angle=0,width=5.8cm,trim={0.5cm 4.0cm 0.0cm 6.0cm},clip]{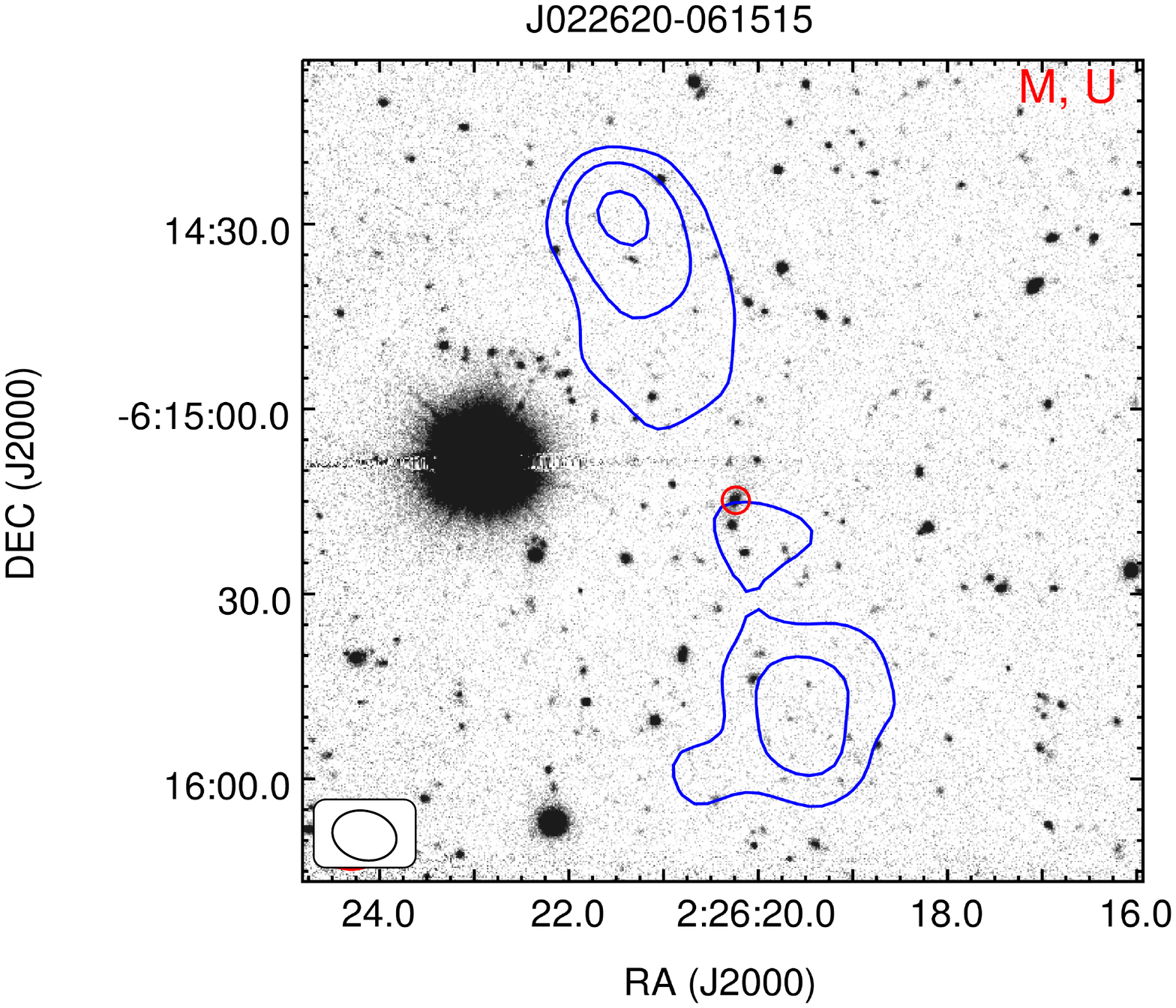}
\includegraphics[angle=0,width=5.8cm,trim={0.5cm 4.0cm 0.0cm 6.0cm},clip]{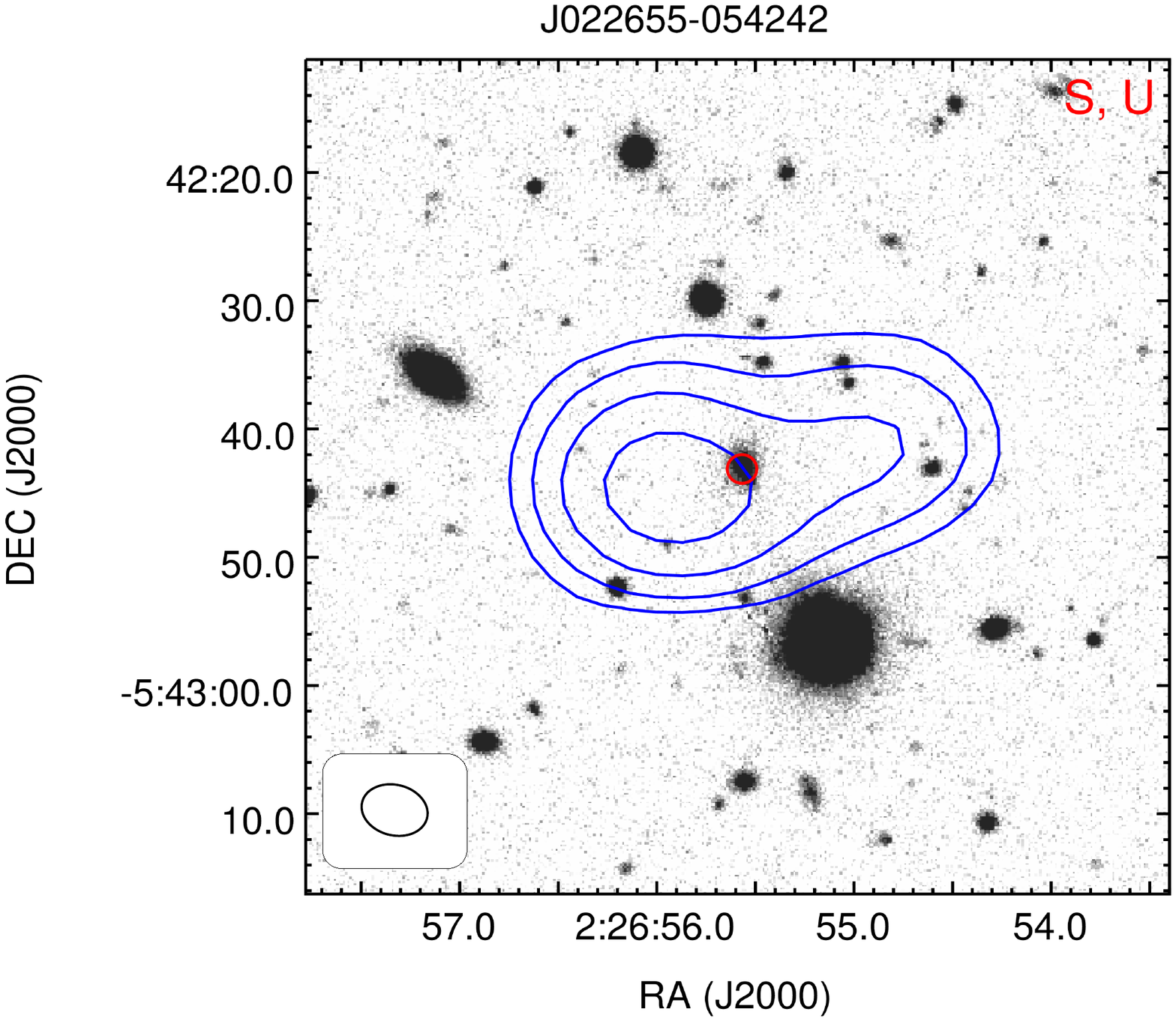}
\includegraphics[angle=0,width=5.8cm,trim={0.5cm 4.0cm 0.0cm 6.0cm},clip]{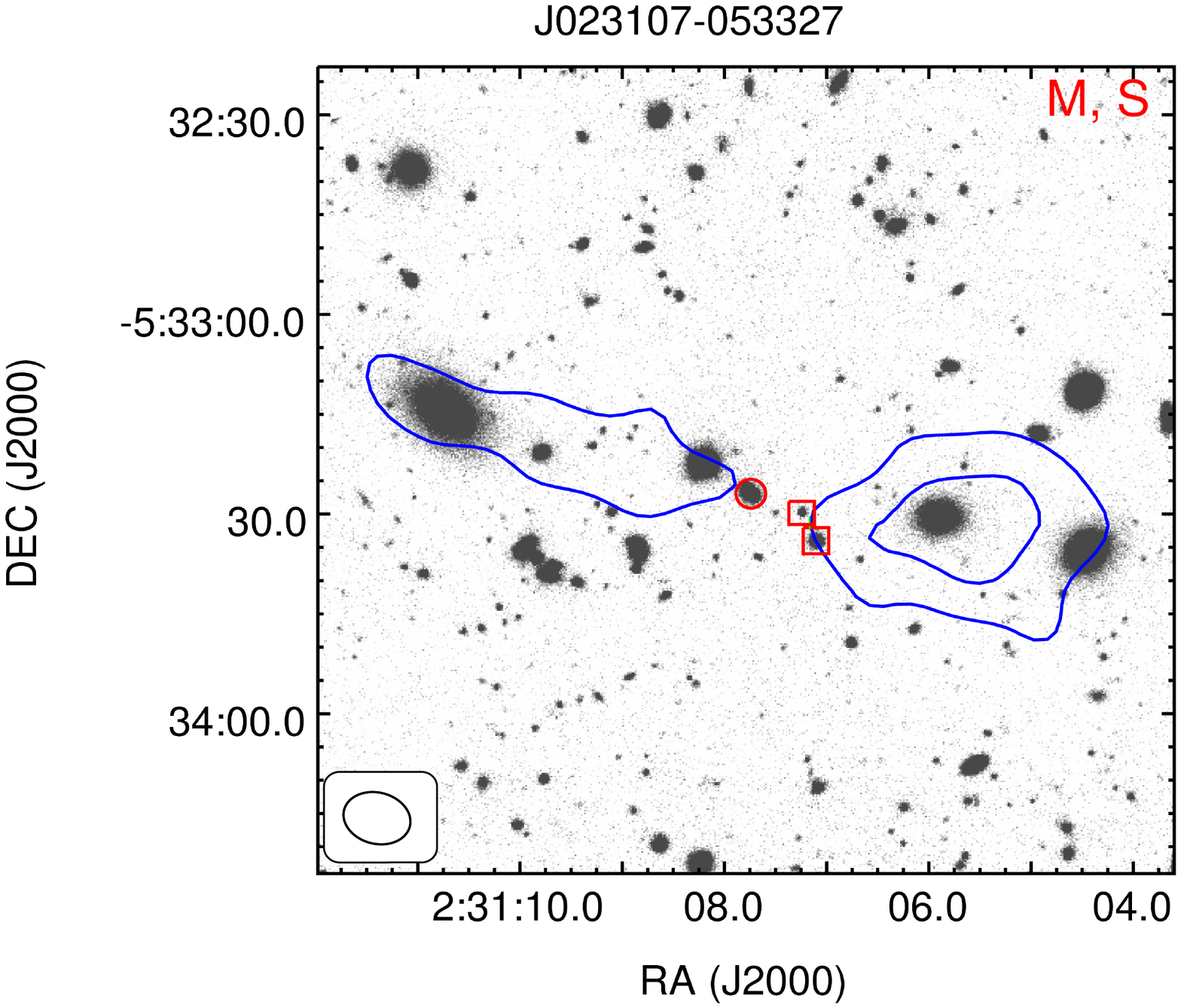}
\caption{{\it -continued}} 
\label{fig:RRGImages} 
\end{figure*}
%
\begin{table*}
\begin{minipage}{180mm}
\caption{The sample of remnant candidates}
\label{tab:RRGSample}
\hskip-3.0cm
\scalebox{0.9}{
\begin{tabular}{lcccccccccc}
\hline
Source  & $S_{\rm 74~MHz}$ & $S_{\rm 150~MHz}$ &  $S_{\rm 325~MHz}$ &  $S_{\rm 1400~MHz}$ & ${\alpha}_{\rm 150}^{\rm 1400}$ & ${\alpha}_{\rm 150}^{\rm 325}$ &  ${\alpha}_{\rm 325}^{\rm 1400}$  & $CP_{\rm 150~MHz}$ & $SB_{\rm 325~MHz}$ & Selection  \\
Name        &    (mJy)  &    (mJy)       &   (mJy)       &    (mJy)      &       &       &    & ($\times$ 10$^{-3}$) &  & Criteria  \\  \hline
J021130$-$033608 &     & $<$ 24.5   & 13.7$\pm$0.7  & $<$2.5      &     &  $>$$-0.76$ & $<$$-1.16$   &        & 47  &  M   \\
J021246$-$033553 & 790$\pm$78 & 617.3$\pm$2.6  & 316.2$\pm$3.7 & 64.6$\pm$1.4  & $-1.01\pm0.01$ & $-0.87\pm0.01$ & $-1.09\pm0.02$  &$<$0.58 & 264 &  M      \\
J021652$-$060410 &      & 56.1$\pm$1.5   & 46.6$\pm$2.7  & 10.5$\pm$1.3  & $-0.75\pm0.06$ & $-0.24\pm0.05$ & $-1.02\pm0.09$  &$<$6.4  & 65  &  M       \\
J021655$-$025703 &     & 48.7$\pm$0.8   & 23.7$\pm$3.7  & 6.3$\pm$0.5   & $-0.92\pm0.04$ & $-0.93\pm0.12$ & $-0.91\pm0.12$ &$<$7.4  & 22 &  M      \\
J021659$-$044918$^{\rm J}$ & 920$\pm$120 & 428.2$\pm$1.8  & 155.3$\pm$2.5 & 9.2$\pm$0.5   & $-1.72\pm0.02$ & $-1.31\pm0.01$ & $-1.94\pm0.04$  &$<$0.14 & 105 &  M, S, U  \\
J021812$-$064334 &     & 220.5$\pm$0.8  & 78.0$\pm$6.5  & 28.0$\pm$1.3  & $-0.92\pm0.02$ & $-1.34\pm0.06$ & $-0.70\pm0.07$  &$<$1.6  & 39  &  M     \\
J021837$-$024552 &     & 79.6$\pm$1.3   & 41.6$\pm$3.1  & 9.8$\pm$0.5   & $-0.94\pm0.02$ & $-0.84\pm0.06$ & $-0.99\pm0.06$ &$<$4.5  & 68 &  M        \\
J021855$-$040134$^{\rm J}$ &     & 38.5$\pm$0.8   & 24.5$\pm$0.8  & 4.6$\pm$0.6   & $-0.95\pm0.06$ & $-0.58\pm0.03$ & $-1.15\pm0.09$   &$<$1.5  & 29   &  S       \\
J022045$-$030735 &     & 39.2$\pm$0.6   & 16.5$\pm$1.5  &$<$2.5         &$<$$-1.23$       & $-1.12\pm0.07$ & $<$$-$1.29 &$<$9.1  & 64  & M, U     \\
J022218$-$032137 &      & 85.5$\pm$0.9   & 27.8$\pm$2.3  & 6.3$\pm$0.5   & $-1.17\pm0.04$ & $-1.45\pm0.06$ & $-1.02\pm0.08$  &$<$4.2  & 28 &  M       \\
J022231$-$042757$^{\rm J}$ &     & 48.7$\pm$0.6   & 29.0$\pm$0.9  & 3.9$\pm$0.5   & $-1.13\pm0.06$ & $-0.67\pm0.03$ & $-1.37\pm0.09$  &$<$1.2  & 160 &  S        \\
J022252$-$060937 &     & 104.3$\pm$0.9  & 42.3$\pm$1.2  & 4.2$\pm$0.4   & $-1.44\pm0.04$ & $-1.17\pm0.02$ & $-1.58\pm0.07$   &$<$3.5  & 170 &  U, M   \\
J022305$-$041232$^{\rm J}$ &     & 13.8$\pm$0.6   & 9.1$\pm$0.6   & 1.8$\pm$0.1   & $-0.91\pm0.02$ & $-0.54\pm0.06$ & $-1.11\pm0.05$  &$<$4.4  & 50 &  S       \\
J022318$-$044526$^{\rm J}$ &     & 33.9$\pm$1.0   & 20.6$\pm$2.1  & 2.7$\pm$0.1   & $-1.13\pm0.01$ & $-0.65\pm0.08$ & $-1.39\pm0.07$    &$<$1.8  & 63  &  S      \\
J022435$-$062004 &     & 41.0$\pm$1.4   & 21.6$\pm$1.5  & 3.0$\pm$0.4   & $-1.17\pm0.06$ & $-0.83\pm0.06$ & $-1.35\pm0.10$ &$<$8.8  & 132 &  S       \\
J022500$-$032102 & 200$\pm$50 & 109.5$\pm$1.1  & 59.5$\pm$1.2  & 12.5$\pm$1.0  & $-0.97\pm0.04$ & $-0.79\pm0.02$ & $-1.07\pm0.06$ &$<$3.3  & 197 &  M       \\
J022511$-$055641 &     & 28.2$\pm$0.3   & 16.0$\pm$1.5  & 3.6$\pm$0.6   & $-0.92\pm0.07$ & $-0.73\pm0.07$ & $-1.02\pm0.13$   &$<$12.8 & 63  &  M       \\
J022545$-$031650 &     & 56.6$\pm$0.5   & 33.5$\pm$2.2  &$<$2.5    &$<-1.39$     & $-0.68\pm0.05$ & $<-1.78$ &$<$6.4 & 37  &  M, S, U  \\
J022620$-$061515 &     & 138.6$\pm$1.6  & 41.7$\pm$4.2  & 3.8$\pm$0.6   & $-1.61\pm0.07$ & $-1.56\pm0.08$ & $-1.64\pm0.13$  &$<$2.6  & 56  &  M, U    \\
J022655$-$054242 &     & 55.1$\pm$1.1   & 27.2$\pm$0.8  & 2.3$\pm$0.5   & $-1.42\pm0.10$ & $-0.91\pm0.03$ & $-1.69\pm0.15$  &$<$6.5  & 198 & S, U     \\
J023107$-$053327 &     & 46.0$\pm$6.7   & 29.6$\pm$3.7  & 3.9$\pm$0.7   & $-1.10\pm0.10$ & $-0.57\pm0.15$ & $-1.39\pm0.15$  &$<$7.8  & 55  &  M, S   \\
\hline
\end{tabular}}
\\
{\bf Notes}$-$ Column 1 : Source name based on the 325 MHz position in J2000 coordinates; Column 2 : 74 MHz flux density from the VLA observations; 
Column 3 : 150 MHz flux density from the LOFAR  observations; Column 4 : 325 MHz flux density from the GMRT observations; 
Column 5 : 1.4 GHz flux density from the NVSS, whenever available, otherwise from the JVLA observations; Column 6 : global spectral index between 150 MHz and 1.4 GHz; 
Column 7 : spectral index between 150 MHz and 325 MHz; Column 8 : spectral index between 325 MHz and 1.4 GHz; 
Column 9 : Upper limit on the core prominence (CP) where 3$\sigma$ upper limit on the non$-$detected core is taken 
from 3.0 GHz VLASS or 1.4 GHz JVLA observations and total flux density is from 150 MHz LOFAR observations; 
Column 10 : surface brightness (SB) in units of mJy~arcsec$^{-2}$ based on the 325 MHz GMRT observations; 
Column 11 : selection criteria used to identify remnant candidate (M = morphology, S = spectral curvature, U = USS). 
Sources falling within the JVLA observations region are marked with $^{J}$ in Column 1. 
\end{minipage}
\end{table*} 	
\begin{table*}
\begin{minipage}{180mm}
\caption{Redshifts, radio luminosities and sizes of remnant candidates}
\label{tab:RRGPD}
\hskip-3.0cm
\scalebox{0.8}{
\begin{tabular}{lcccccccccccc}
\hline
Source           &  $z$   &  Reference  &  log$L_{\rm 150~MHz}$ & log$L_{\rm 1.4~MHz}$ & LAS &  
Size & \multicolumn{2}{c}{power law}  &  \multicolumn{3}{c}{Curved power law}  & SED  \\  \cline{8-9} \cline{10-12} 
Name             &        &         &  (W~Hz$^{-1}$)           & (W~Hz$^{-1}$)       & ($^{\prime\prime}$) & (kpc) &  ${\alpha}_{\rm fit}^{\rm PL}$ &  ${\chi}^{2}$ &  ${\alpha}_{\rm fit}^{\rm CPL}$ & q    &  ${\chi}^{2}$ &   \\ \hline
 J021130$-$033608 & $>$1.0             &             & $>$26.15 & $>$25.17  & 35.5  & $>$284.3 & $<$$-$1.00$\pm$0.10 & 0.52   & -1.30$\pm$0.17 & $<$-0.18 &  1.17  & CPL \\
 J021246$-$033553 & 1.16$\pm$0.11 & HSS$-$SSP (p) & 27.71  & 26.73  & 121.0 & 998.8 & $-0.87\pm0.10$ & 3.10 & $-1.28\pm0.09$ & $-0.17\pm0.03$ & 1.75    & CPL \\
 J021652$-$060410 & 0.98$\pm$0.10 & HSS$-$SSP (p) & 26.46  & 25.74  & 63.0  & 502.1 & $-0.77\pm0.17$ & 2.25 & $-1.30\pm0.30$ & $-0.35\pm0.08$  & 2.09 &   CPL \\
 J021655$-$025703 & 0.65$\pm$0.04 & HSS$-$SSP (p) & 25.92  & 25.03  & 166.0 & 1149.1 & $-0.91\pm0.01$ & 1.21 & $-0.90\pm0.06$  & $0.01\pm0.001$ & 2.89  & PL \\
 J021659$-$044918 & 1.3247$\pm$0.0011  & SXDF (s) & 27.97  & 26.30  & 155.0 & 1301.0 & $-1.60\pm0.12$ & 7.20 & $-1.40\pm0.07$  & $-0.24\pm0.02$  & 2.90 & CPL \\
 J021812$-$064334 & 0.26$\pm$0.05 & HSS$-$SSP (p) & 25.61  & 24.71  & 117.0 & 469.1 & $-0.88\pm0.19$ & 3.12 & $-0.50\pm0.05$  & 0.29$\pm$0.03 & 0.70 &    CPL \\
 J021837$-$024552 & 0.14$\pm$0.01  & HSS$-$SSP (p) & 24.61  & 23.70  & 98.5  & 241.6 & $-0.94\pm0.04$ & 1.08 & $-1.00\pm0.08$ & $-0.07\pm0.01$ & 0.32 &  PL \\
 J021855$-$040134 & 0.63$\pm$0.19  & HSS$-$SSP (p) & 25.84  & 24.92  & 60.7  & 413.8 & $-0.91\pm0.01$ & 0.56 & $-0.90\pm0.06$  & $-0.25\pm0.02$ & 1.37 & CPL \\
 J022045$-$030735 & 0.76$\pm$0.07  & HSS$-$SSP (p) & 26.10  & 24.91  & 59.3  & 437.1 & $-1.20\pm0.05$ & 1.04  & $-1.40\pm0.09$ & $-0.08\pm0.01$ & 0.21 &    PL \\
 J022218$-$032137 & 0.58$\pm$0.05  & HSS$-$SSP (p) & 26.07  & 24.94  & 128.0 & 845.1 & $-1.10\pm0.12$ &  2.23 & $-0.86\pm0.03$  & $0.20\pm0.01$ & 0.58   &  CPL \\
 J022231$-$042757 & $>$1.0   &             & $>$26.52  & $>$25.43  & 50.1  & $>$401.2 & $-1.20\pm0.15$ & 1.95 & $-1.60\pm0.07$   & $-0.31\pm0.01$ & 0.89 & CPL \\
 J022252$-$060937 & 0.73$\pm$0.23  & HSS$-$SSP (p) & 26.54  & 25.14  & 44.3  & 321.2 & $-1.50\pm0.10$ & 2.29 & $-1.70\pm0.08$ & $-0.19\pm0.01$ &  0.59  &    CPL  \\
 J022305$-$041232 & 0.6314$\pm$0.0001  & PRIMUS (s)  & 25.40  & 24.52  & 36.5  & 249.7 & $-0.93\pm0.13$ & 0.35 & $-1.30\pm0.04$ & $-0.26\pm0.01$  & 1.06 & CPL \\
 J022318$-$044526 & 1.15$\pm$0.17  & HSS$-$SSP (p) & 26.54  & 25.44  & 78.0  & 642.1 & $-1.20\pm0.16$ & 1.91 & $-1.70\pm0.05$  & $-0.33\pm0.01$ & 0.65   & CPL  \\
 J022435$-$062004 & 1.89$\pm$0.16  & HSS$-$SSP (p) & 27.13  & 25.99  & 40.0  & 336.6 & $-1.20\pm0.12$ & 0.39 & $-1.50\pm0.06$  & $-0.23\pm0.01$  & 0.72 & CPL  \\
 J022500$-$032102 & 0.61$\pm$0.06  & HSS$-$SSP (p) & 25.76  & 24.82  & 64.0  & 429.5 & $-0.94\pm0.05$ & 0.30 & $-1.10\pm0.07$  & $-0.07\pm0.02$ & 0.50  &  CPL \\
 J022511$-$055641 & 0.3193$\pm$0.0001  &  SDSS (s)   & 24.98  & 24.09  & 65.0  & 302.1 & $-0.93\pm0.07$ & 0.14 & $-1.10\pm0.04$  & $-0.13\pm0.004$  & 0.48   &  CPL  \\
 J022545$-$031650 & 0.36$\pm$0.07 & HSS$-$SSP (p) & 25.55  & 24.19  & 94.3  & 478.3 & $-1.40\pm0.22$ & 2.17 & $-1.20\pm0.08$  & $-0.49\pm0.03$ &  0.52  &  CPL \\
 J022620$-$061515 & 0.94$\pm$0.09 & HSS$-$SSP (p) & 26.99  & 25.43  & 107.0 & 844.2 & $-1.60\pm0.02$ & 1.23 & $-1.70\pm0.13$  & $-0.04\pm0.003$  & 3.46   & PL  \\
 J022655$-$054242 & 1.0255$\pm$0.0002  &  VIPER (s)  & 26.72  & 25.34  & 37.8  & 304.5 & $-1.40\pm0.17$ & 2.29 & $-1.30\pm0.07$  & $-0.35\pm0.02$  & 1.91   &  CPL  \\
 J023107$-$053327 & 0.65$\pm$0.09 & HSS$-$SSP (p) & 26.04  & 24.97  & 110.0 & 764.2 & $-1.10\pm0.18$ & 1.68 & $-1.30\pm0.11$  & $-0.37\pm0.03$  & 1.39  &  CPL \\
\hline
\end{tabular}}
\\
{\bf Notes}$-$ Spectroscopic and photometric redshifts are indicated by `s' and `p' within brackets in column 3. 
The linear angular size is measured from the 325 MHz GMRT image and represents end-to-end linear 
distance along the direction of maximum elongation. The ${\alpha}_{\rm fit}^{\rm PL}$ and ${\alpha}_{\rm fit}^{\rm CPL}$ represent spectral indices in power law and curved power law fits, respectively. 
The $q$ value defines curvature parameter obtained from the curved power law model fit. 
The nature of SED is determined from the goodness-of-fit indicated by the reduced ${\chi}^{2}$ value. The 150 MHz and 1.4~GHz luminosities are k-corrected assuming curved power law SEDs. 
\end{minipage}
\end{table*} 	
\subsubsection{Radio SEDs}
\label{sec:SED}
To understand the spectral characteristics of our remnant candidates we model their radio spectral energy distributions (SEDs) 
over the widest possible range between 74 MHz to 1.4 GHz, subject to the availability of flux densities.  
In principle, flux densities measured at different frequencies should be matched in resolution and 
uv$-$coverage. However, we point out that the beam$-$sizes at 150 MHz (7$^{\prime\prime}$.5$\times$8$^{\prime\prime}$.5) 
and 325 MHz (10$^{\prime\prime}$.2$\times$7$^{\prime\prime}$.9) are comparable, and much larger beam$-$sizes at 74 MHz (30$^{\prime\prime}$ in the VLA) and at 1.4 GHz (45$^{\prime\prime}$ in the NVSS) 
would ensure the detection of low$-$surface$-$brightness emission. Although, unlike low-frequency observations, 
1.4 GHz VLA observations have sparse uv-coverage that can possibly miss diffuse emission of very 
low-surface-brightness, which in turn would result flux density lower than the actual. 
Hence, we caution that the missing flux issue, if significant, would make spectral curvature artificially more stronger. 
We mention that 15/21 (71.4 per cent) of our remnant candidates have been identified by using 
the morphological criteria. While, remaining six remnant candidates identified with spectral curvature criterion 
would mostly continue to exhibit strong curvature (${\Delta}{\alpha}$ $<$$-$0.5), even if 1.4 GHz flux density increases by a margin of 10 percent or so. Thus, a marginal missing flux density upto 10 per cent at 1.4~GHz is unlikely impact our results.
\par
Since remnant sources can show power law or curved spectra we attempt to fit the radio SED of 
each remnant candidate with a power law model as well as with a curved power law model. 
The best fit is determined on the basis of goodness$-$of$-$fit defined by the reduced ${\chi}^{2}$. 
The standard power law model is described as S$_{\nu}$ = S$_{\rm 0}$ ${({\nu}/{\nu}_{\rm 0})}^{\alpha}$, 
where spectral index $\alpha$ and flux normalization S$_{\rm 0}$ are constrained by the fitting, and ${\nu}_{\rm 0}$ is the frequency 
at which S$_{\rm 0}$ is evaluated. 
Curved power law is defined as 
S$_{\nu}$ = S$_{\rm 0}$ ${({\nu}/{\nu}_{\rm 0})}^{\alpha}$ e$^{\rm q(ln{\nu})^2}$, 
where $q$ parameterises the curvature of spectrum such that $q~<0$ represents a convex spectrum \citep[{see}][]{Quici21}. 
For optically$-$thin synchrotron emission arising from radio lobes, typical value of $q$ ranges over -0.2$\leq q \leq$0. 
We note that the $q$ parameter is not physically motivated and describes curvature only mathematically, although \citet{Duffy12} 
suggested that $q$ can be related to the physical quantities such as energy and magnetic field strength in radio lobes.
\par
We note that the radio SEDs of all but three of our remnant candidates are limited within 150~MHz to 1.4 GHz spectral window, and only three sources have 74 MHz flux density measurements available. 
The best fit spectral parameters obtained with the power law as well as with the curved power law models are listed in Table~\ref{tab:RRGPD}. 
We find that the majority (17/21) of our remnant candidates are better fitted with a curved power law.
For our remnant candidates, $q$ parameter varies between $-0.49$ to 0.29 with a median value of $-0.19\pm0.05$. 
As expected (9/21) remnant candidates selected from the SPC criterion show systematically higher curvature with 
the $q$ value ranging between $-0.49$ to $-0.23$ with a median value of $-0.32\pm0.03$. 
Two remnant candidates (J021812$-$064334 and J022218$-$032137) appear to show concave spectra ($q$ $>$0) with 150 MHz flux density much higher than the value extrapolated from 325 MHz flux density and 325 MHz $-$ 1.4 GHz spectral index (${\alpha}_{\rm 325}^{\rm 1400}$).
Notably, both these sources have very low surface$-$brightness 
(SB$_{\rm 325~MHz}$ = 39 mJy arcmin$^{-2}$ and 28 mJy arcmin$^{-2}$, respectively), and hence, these 
sources can possibly contain more diffuse low$-$surface$-$brightness emission which is better detected at 150 MHz. 
We find that 4/21 of our remnant candidates are better fitted with a standard power law, and all 
but one of these sources are selected from the USS criterion. 
The USS remnant candidates are likely to be old with the break frequency shifted to the 
low$-$frequency regime (${\nu}_{\rm b}$ $<$325 MHz). 
\par

The $q$ value for remnants is expected to be systematically lower than that for active sources as remnants SEDs tend to be strongly 
curved at higher frequencies due to spectral ageing. Although, we find that the distribution of $q$ parameter for 
active sources in our sample spanning across $-0.93$ to $0.79$ with a median value of $-0.13{\pm}0.01$ is not very different from 
that of the remnant candidates (see Table~\ref{tab:RRGComp}). 
We note that a very low $q$ value also corresponds to inverted SEDs in which curvature seen at lower frequencies is 
due to synchrotron self$-$absorption (SSA). To avoid sources strongly affected by SSA we introduced ${\alpha}_{150}^{325}$ $\leq$ $-0.5$ cutoff. As expected, for the subsample consisting only sources with ${\alpha}_{150}^{325}$ $\leq$ $-0.5$, we find 
that the $q$ distributions of remnant candidates and active sources are widely different with 
only 1.1 $\times$ 10$^{-3}$ probability of null hypothesis to be true in the two-sample KS test.
\begin{table*}
\begin{minipage}{180mm}
\caption{Comparison between remnant candidates and active sources}
\label{tab:RRGComp}
\hskip-2.0cm
\begin{tabular}{ccccccccc}
\hline
Parameter  & \multicolumn{3}{c}{Remnant candidates}  &  \multicolumn{3}{c}{Active sources}  &  \multicolumn{2}{c}{KS test}   \\  \cline{2-4} \cline{5-7} \cline{8-9}
          &  $N_{\rm source}$ & range    &  median  &  $N_{\rm source}$   & range     &  median  &  $D$   & $p-$value  \\ \hline
$S_{\rm 150~MHz}$ (mJy) & 20  & 13.8$-$617.3  & 56.1$\pm$32.8  & 232   & 3.5$-$3053.7 & 67.8$\pm$39.2 & 0.18 & 0.48  \\         
$S_{\rm 325~MHz}$ (mJy) & 21  & 9.1$-$316.2   & 29.0$\pm$14.2  & 246   &  3.4$-$1800.9 & 41.0$\pm$19.3 &  0.26 &  0.11 \\  
${\alpha}_{\rm 150}^{\rm 1400}$ & 20 & $-1.72$$-$$-0.75$ & $-1.10\pm0.05$ & 223 & $-1.32-0.05$  
& $-0.79\pm0.02$  & 0.17  & 0.54  \\  
 $z$   &  19  & 0.14$-$1.89  & 0.65$\pm$0.09 & 160 & 0.059$-$2.66 & 0.84$\pm$0.05 & 0.22 & 0.27 \\  
Size (kpc)   &  19  & 241.6$-$1301 & 469.1$\pm$70.6 &  160 & 75.2$-$1651.1 & 314.3$\pm$22.8 & 0.44 & 1 $\times$ 10$^{-3}$ \\ 
log$L_{\rm 1.4~GHz}$ (W~Hz$^{-1}$) &  17 & 23.70$-$26.73 & 25.03$\pm$0.17 &  154 & 23.42$-$27.39 & 25.47$\pm$0.07 & 0.28 &  0.12 \\
log$L_{\rm 150~MHz}$ (W~Hz$^{-1}$) & 19  & 24.61$-$27.92 & 26.11$\pm$0.19 & 154 & 24.34$-$28.35 & 
26.18$\pm$0.10 & 0.15 & 0.77  \\  
$CP$   & 20  & $>$0.00014~$-$~$>$0.013  & $>$0.0044 &   38 & 0.0017$-$0.305 & 0.025$\pm$0.011      & ...     &   ...    \\  
$SB_{\rm 325~MHz}$ (mJy arcmin$^{-2}$)  & 21  &  22.2$-$264 & 63.4$\pm$14.5 &  246 & 19.20$-$14156.9 & 238.1$\pm$116.5 & 0.41  &  1.6 $\times$ 10$^{-3}$ \\  
 $q$                      &  20  & $-0.49$~$-$~0.29 &  $-0.19\pm0.05$  & 225 & $-0.93$~$-$~0.79 & $-0.13\pm0.01$ & 0.27 & 0.13  \\  
 $q$ (${\alpha}_{150}^{325}~{\leq} -0.5$) &  19  & $-0.49$~$-$~0.29 &  $-0.18\pm0.04$  & 148 & $-0.69$~$-$~0.79 & $-0.08\pm0.01$ & 0.47 & 1.1 $\times$ 10$^{-3}$  \\  
\hline
\end{tabular}
\\
{\bf Notes}$-$ The two sample Kolmogorov$-$Smirnov (KS) test is a non$-$parametric statistical test that examines 
the hypothesis that two samples come from same distribution. $D$ represents the maximum difference between the cumulative 
distributions of two samples and $p-$value is the probability that the null hypothesis, i.e., two samples comes from
same distribution, is true. The error on the median value is taken as 
${\sigma}_{\rm SD}$/$\sqrt{N_{\rm sample}}$ assuming normal distribution for the sample, 
where ${\sigma}_{\rm SD}$ is the standard deviation and $N_{\rm sample}$ is the sample size. 
Occasionally large errors found on the median values are due to the skewed non$-$Gaussian distributions.
\end{minipage}
\end{table*} 	
%
%
\subsection{Surface$-$brightness of remnant candidates}
\label{sec:SB}
Remnants are expected to show low$-$surface$-$brightness emission as the radio emitting
plasma in lobes radiates continuously with no supply of fresh energy. Low$-$surface$-$brightness may also arise 
due to the expansion of radio lobes if they are over$-$pressured at the end of active phase. 
For our remnant candidates, we estimate surface$-$brightness measured 
at 325 MHz (see Table~\ref{tab:RRGSample}). The surface$-$brightness of our remnant candidates is 
defined as the ratio of 325 MHz 
total flux density to the area over which radio emission is detected at $\geq$3$\sigma$ level in the 325 MHz 
GMRT image. 
We note that the surface$-$brightness of remnant candidates should be estimated at the lowest frequency 
owing to the fact that remnants appear brighter at lower frequencies. 
However, due to unavailability of 150 MHz LOFAR images we estimate surface$-$brightness at 325 MHz. 
We find that the 325 MHz surface$-$brightness of our 21 remnant candidates are distributed 
in the range of 
22.2 mJy arcmin$^{-2}$ to 264 mJy arcmin$^{-2}$ with a median value of 63.4$\pm$14.5 mJy arcmin$^{-2}$ (see Table~\ref{tab:RRGSample}). 
The 15/21 remnant candidates identified with the morphological criterion have surface$-$brightness in the 
same range. While, 12/21 remnant candidates identified with the spectral criteria have surface$-$brightness in the range of 29 mJy arcmin$^{-2}$ to 198.6 mJy arcmin$^{-2}$ 
with a median of 64$\pm$20 mJy arcmin$^{-2}$. 
We find that irrespective of selection criteria remnant candidates tend to show low$-$surface$-$brightness. 
It is worth to point out that \cite{Brienza17} found 150 MHz surface$-$brightness in the range of 
10$-$30 mJy arcmin$^{-2}$ for their morphologically selected remnant candidates. 
In general, remnants of very low$-$surface$-$brightness are likely to be old in 
which energy loss has occurred over a much longer period, while the remnants of relatively 
higher surface$-$brightness can be relatively young. 
However, we note that the surface$-$brightness parameter is a frequency and resolution dependent parameter, 
and hence, surface$-$brightness measured with different observations may not be directly comparable.
\subsection{Core prominence of remnant candidates}
\label{sec:CP}
In the literature, core prominence (CP) defined as the ratio of core flux density to total flux density 
has been used as an indicator of the activity status \citep[{\eg}][]{Brienza17}. 
Remnant sources with no detected core can have only upper limits on CP that are expected to be systematically 
lower than the CP values for active sources of similar brightness. 
We determine the CP limits for our remnant candidates and compare their distribution {\it w.r.t.} active sources in our sample. 
Since, compact core is better detected in deep high resolution high$-$frequency observations, while extended radio emission from lobes is better detected in deep low$-$frequency observations, we define 
CP = S$_{\rm core}^{\rm 3.0~GHz}$/S$_{\rm total}^{\rm 150~MHz}$.
The exact frequency at which core flux density is measured may not be very important owing to the fact that radio cores are generally 
expected to have flat spectrum upto higher frequencies \citep{Hardcastle08}. 
We check the detection of core component in our sample sources by using 3.0 GHz VLASS, 1.4 GHz FIRST 
and 1.4 GHz JVLA image cutouts. 
The VLASS images with 3$\sigma$ sensitivity of 0.36 mJy~beam$^{-1}$ and angular resolution of 2$^{\prime\prime}$.5 
are better suited to detect and disentangle core component. However, deep JVLA 
observations with sensitivity (3$\sigma$ = 0.048 mJy beam$^{-1}$) better than that for the 
VLASS and FIRST, are preferred, whenever available. 
We consider the detection of core above 3$\sigma$ level, and task `JMFIT' in the 
Astronomical Image Processing System (AIPS\footnote{http://www.aips.nrao.edu/index.shtml})  
that fits an elliptical Gaussian to the core component, is used whenever core flux density is not 
listed in the catalogues.  
\begin{figure}
\centering
\includegraphics[angle=0,width=8.5cm,trim={0.0cm 6.6cm 0.0cm 7.0cm},clip]{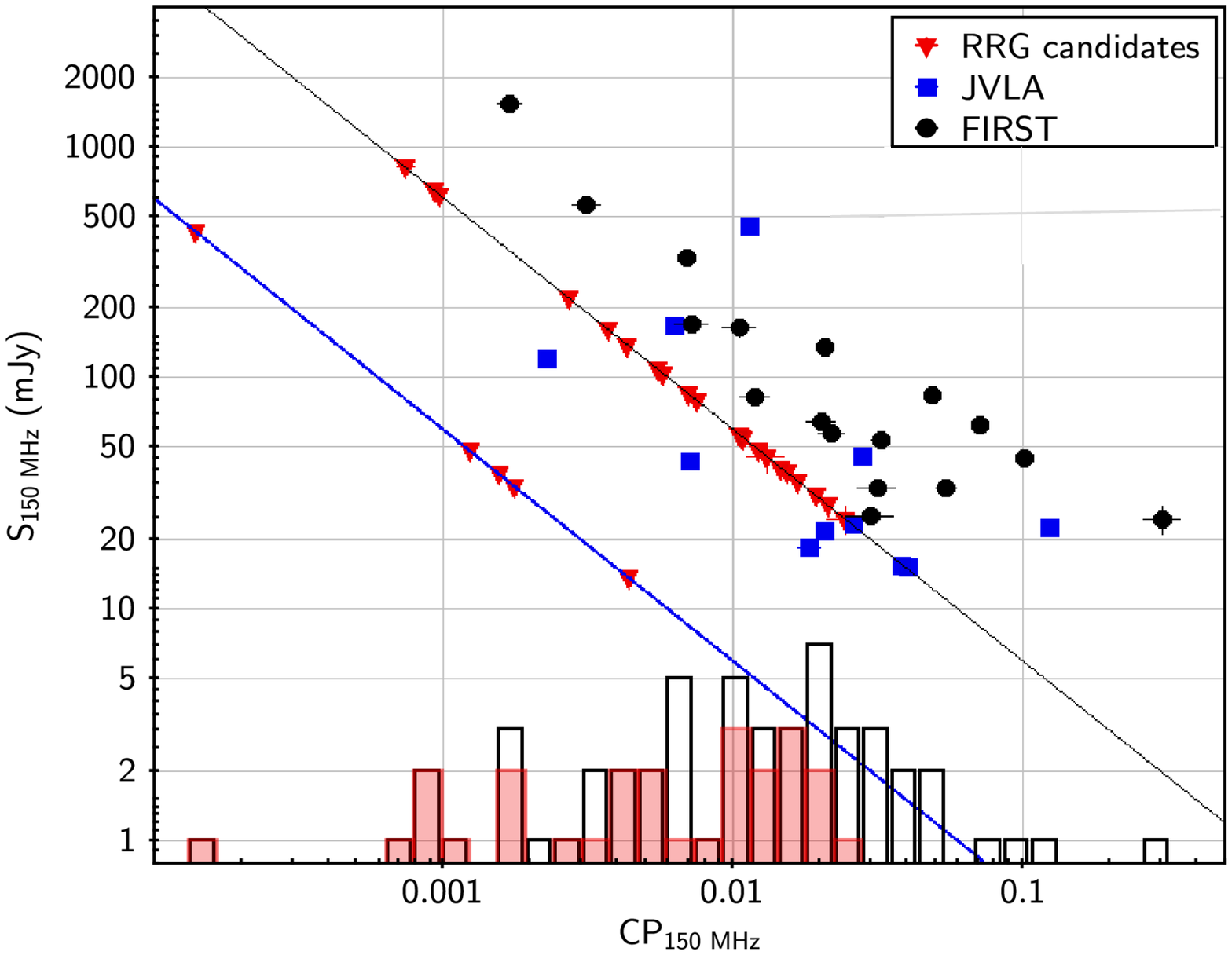}
\caption{150 MHz flux density versus core prominence (CP) parameter. 
For remnant candidates upper limits on CP are placed by using 3$\sigma$ 
flux density limits from 3.0 GHz VLASS or 1.4 GHz JVLA observations. 
The upper limits on CP based on the 3$\sigma$ sensitivity limits of 3.0 GHz VLASS and 1.4 GHz JVLA 
observations are shown in solid black line and dotted blue line, respectively. The histogram of 
CP is shown in the bottom where filled and empty bars represent the distributions of 
remnant candidates and active sources, respectively.} 
\label{fig:CP} 
\end{figure}
For remnant candidates with undetected cores, we place an upper limit on CP by using 3$\sigma$ limit on flux density.  
We note that the VLASS, JVLA and FIRST images of moderate resolutions 
(2$^{\prime\prime}$.5$-$5$^{\prime\prime}$.0) have limitations 
in detecting the core component in many of our sources, particularly in sources with 
relatively small sizes ($<$60$^{\prime\prime}$). 
Deep high$-$frequency radio observations yielding sub$-$arcsec resolution would be ideal to detect the core components 
in such sources. 
In our sample we obtain CP estimates for only 59 sources including upper limits placed on our 21 
remnant candidates. 
\par
Figure~\ref{fig:CP} shows the distributions of CP parameter for the remnant candidates and active sources in our sample. 38 active sources have CP values distributed in the range of 1.7 $\times$ 10$^{-3}$ to 3.05 $\times$ 10$^{-1}$ with a median value of 2.5$\pm$1.1 $\times$ 10$^{-2}$. 
While, 21 remnant candidates have upper limits on CP spanning over 1.4 $\times$ 10$^{-4}$ 
to 1.3 $\times$ 10$^{-2}$ with a median value of 4.4 $\times$ 10$^{-3}$. 
As expected remnant candidates have systematically lower CP limits than the CP values for active sources.  
In principle, remnant sources must have CP = 0 considering the absence of radio core, however, flux limited observations 
can only allow us to place an upper limit. Deep observations can put more stringent limits on CP. 
It is evident that the upper limits on CP derived from the JVLA observations are nearly one order of magnitude deeper than those from the VLASS or FIRST survey (see Figure~\ref{fig:CP}).   
\par 
We find that the CP limits of our remnant candidates are consistent with those for the  remnant candidates reported 
in the literature \citep[{\eg}][]{Mahatma18,Quici21}. 
Figure~\ref{fig:CP} demonstrates that the CP depends on the source flux density such that brighter sources tend to have lower CP in compared to fainter sources. 
The flux density dependence of CP suggests that it is difficult to assign a sharp cutoff limit on the 
CP to segregate active and remnant sources. In fact, powerful lobe$-$dominated FR$-$II radio galaxies can 
have low CP values despite being active. Therefore, we conclude that CP cannot be considered a 
reliable parameter for identifying remnants.
\subsection{Optical host galaxy and redshifts}
\label{sec:optical}
The identification of host galaxies of radio sources is essential to estimate their redshifts 
and luminosities. 
We attempt to identify the host galaxies of our 268 extended radio sources by using the 
HSC$-$SSP imaging survey. 
To identify optical counterparts we opted a method similar to that followed by \cite{Jurlin21}. 
For double$-$lobed radio sources with clear or tentative detection of a core, 
optical counterpart closest to the core position within 5$^{\prime\prime}$.0 of radius is considered. 
Search radius is similar to the synthesized beam$-$sizes of 1.4 GHz FIRST and JVLA observations.  
Although, most of the sources are found to show optical counterparts within 2$^{\prime\prime}$.0. 
For double$-$lobed sources showing attached lobes with unresolved radio core we use flux density weighted centroid position 
to search for their optical counterparts. 
For our remnant sources with no detected radio core and detached radio lobes or 
amorphous morphology we search optical counterpart around the flux density weighted barycentre position that 
lies on the intersection of two lobes axes. 
We caution that the identification of host galaxy of a radio source is not secure, if radio core is not detected. 
\par
With the aforementioned exercise we identified the optical counterparts of only  
179/268 (66.8$\%$) radio sources that have either spectroscopic or photometric redshift estimates.  
Only 79/179 sources have spectroscopic redshifts derived from various spectroscopic surveys  
({\eg}PRIMUS, VIPER, and SDSS), while, remaining 100/179 sources have photometric redshifts from the HSC$-$SSP PDR2 (see~Section~\ref{sec:optical}). 
Among our 21 remnant candidates all but two have redshift estimates. 
For two remnant candidates with no redshift estimate we place an upper limit on the redshift ($z$) $>$1.0 
by comparing them with the H$z$RGs of known redshifts in the S$_{\rm 1.4~GHz}$ versus S$_{\rm 3.6~{\mu}m}$ diagnostic plot \citep[see][]{Singh17}.  
Our 19 remnant candidates have redshifts in the range of 0.14 to 1.89 with a median value of 0.65$\pm$0.09, while 160 active sources having redshifts are distributed over the range of 0.059 
to 2.66 with a median value of 0.84$\pm$0.05. 
The two$-$sample KS test shows that the redshift distributions for our remnant candidates and active sources are 
somewhat different with p$-$value = 0.27 (see~Table~\ref{tab:RRGComp}). 
Thus, on average, our remnant candidates are found at lower redshifts than the active source population.
The systematic difference in redshift distributions can possibly be due to observational bias considering 
the fact that morphological identification of remnant could preferentially pick nearby large$-$size sources.
Although, we find that the redshifts of our remnant candidates tend to be higher than those 
reported in some of the previous studies in the literature. 
For instance, the remnant candidates identified in the Lockman$-$Hole and Herschel$-$ATLAS fields have 
the median redshift values 0.5 and 0.31, respectively \citep[see][]{Jurlin21,Mahatma18}. 
The higher redshifts of our remnant candidates can be attributed to their relatively low radio flux densities 
and the deeper optical data used. We note that a large fraction of our sources only have photometric redshifts, 
and hence, results based on these redshift values should be treated with caution. 
\subsection{Large$-$scale environments}
\label{sec:environment}
It has been suggested that the remnants residing in cluster environments can be detected over much longer 
timescales due to slow or even arrested expansion of lobes in dense intra$-$cluster medium, and hence, one 
can expect a 
higher tendency of remnants to reside in cluster environments  \citep[see][]{Murgia11}. 
We investigate the large$-$scale environments of our remnant candidates by using existing
cluster catalogs built from the optical and X$-$ray surveys in the XMM$-$LSS field. We used the HSC$-$SSP survey cluster catalog that consisted of clusters with richness (N$_{\rm gal}$) $\geq$15 considering all galaxies brighter 
than 24 magnitude in $z-$band \citep{Oguri18}. The clusters detected in the HSC$-$SSP are found to be distributed over 
0.1$<$ $z$ $<$1.1 redshift range with the photometric redshift accuracy of ${\Delta}z$/(1+$z$) $<$0.01. We also used X$-$ray cluster catalog based on the deep {\em XMM$-$Newton} survey 
meant to detect X$-$ray emitting hot gas in intra$-$cluster medium with the flux limit of a 
few times of 10$^{15}$ erg s$^{-1}$ cm$^{-2}$ in 0.5$-$10 keV band within 1$^{\prime}$.0 aperture \citep{Adami18}. 
With follow up optical spectroscopic observations X$-$ray detected clusters were 
found to have redshifts over 0.0 $<$ $z$ $<$ 1.2 with one cluster at $z$ = 2.0. 
\par
We matched our remnant candidates with both the cluster catalogs. 
Within the redshift interval ${\Delta}z$ = 0.1 (a typical photometric redshift error for our remnant candidates),  and a radius corresponding to 1 Mpc, we find that only one J022305$-$041232) of our remnant candidates is 
associated with the cluster which is detected in both optical as well as X$-$ray observations. 
The host of remnant candidate J022305$-$041232 having $z$ = 0.6314 is lying at the projected distance of 187 kpc 
(28$^{\prime\prime}$) from the X$-$ray assigned cluster center. It is worth to point out that the remnant candidate 
(J022305$-$041232) residing in the cluster environment is of relatively small size (249.7 kpc). However, other  
remnant candidates of relatively small sizes (let say $<$350 kpc) are not found to be associated with cluster environments. Considering the redshift upper limit ($z$ $<$1.1) for the existing cluster catalogs, we find that, the fraction of our remnant candidates residing in cluster environments is only 1/15 $\sim$ 6.7$\%$. 
Hence, we conclude that, in general, our remnant candidates reside in less dense environments. 
This result is similar to that found in \cite{Jurlin21} who reported that only a minority (23$\%$) of remnants in 
their sample reside in cluster environments. 
\subsection{Power$-$radio size (P$-$D) plot}
\label{sec:PDPlot}
\begin{figure}[ht]
\centering
\includegraphics[angle=0,width=8.5cm,trim={0.0cm 6.8cm 0.0cm 7.0cm},clip]{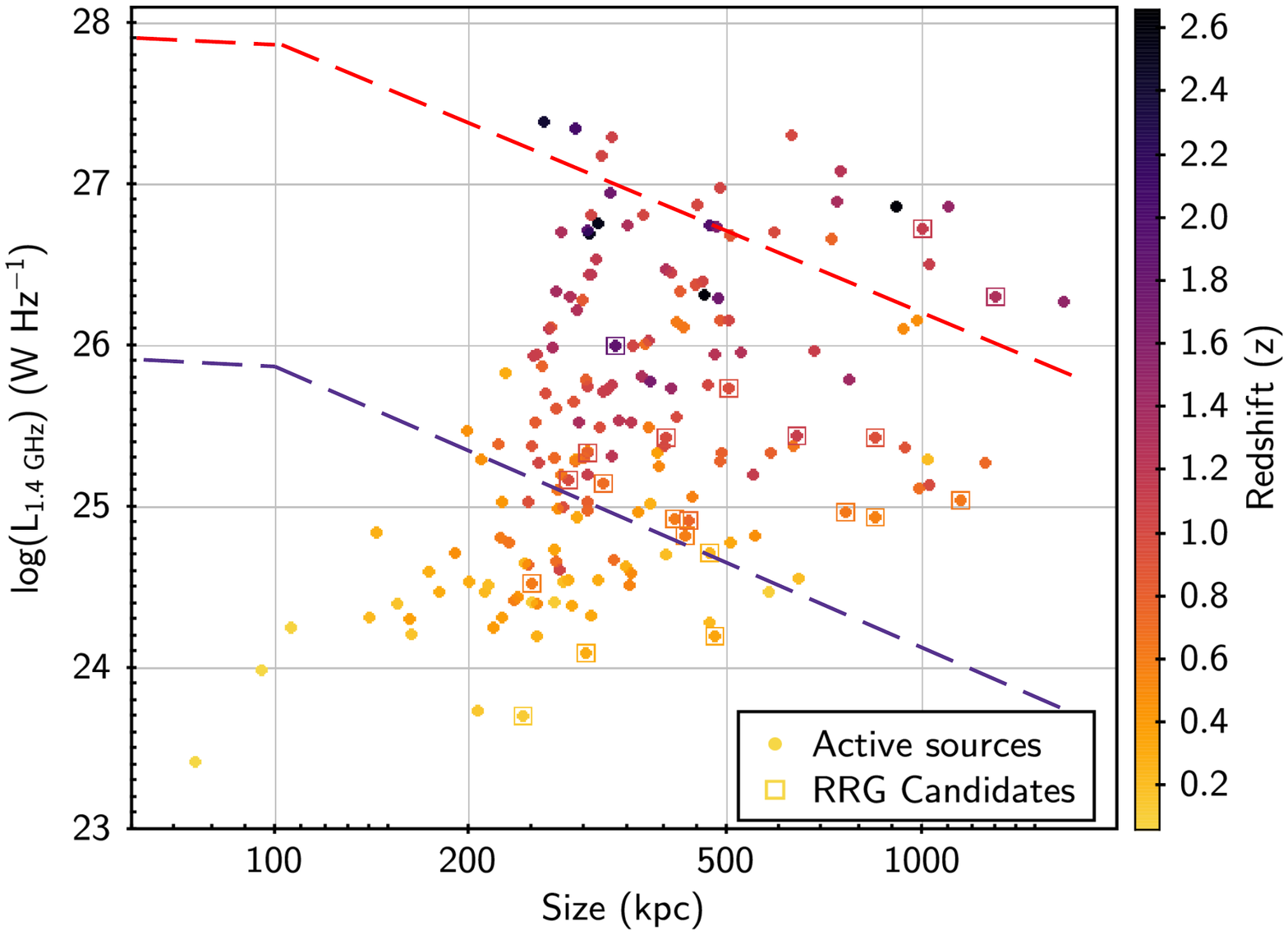}
\caption {The plot of radio size versus 1.4 GHz radio luminosity. Our remnant candidates are marked with square boxes. 
The vertical color bar depicts redshift variation in our sample. The red and blue dashed lines represent 
the evolutionary tracks, as reported in \cite{An12}, for high-power and low-power radio galaxies, respectively.} 
\label{fig:PDPlot} 
\end{figure}

Dynamical evolutionary models predict the growth of radio size as well as radio luminosity 
during the active phase of radio galaxies \citep{An12}. In the remnant phase, 
radio luminosity decreases while radio size can increase, if lobes are over$-$pressured. Therefore, radio power versus radio size diagnostic plot commonly known as the P$-$D plot in the literature, allows us to infer the evolutionary stage of radio galaxies. 
To examine the nature of our remnant candidates we check their positions in the P$-$D plot. 
Figure~\ref{fig:PDPlot} shows 1.4~GHz luminosity versus radio size plot. 
We note that the projected linear size is measured as the end-to-end distance along the direction of maximum elongation in the 325 MHz image. To account for the presence of diffuse extended emission we consider emission upto 3$\sigma$ level. 
The physical radio size is obtained from the linear angular size by using angular diameter distance at the 
corresponding redshift.
\par
We find that, in comparison to active sources, our remnant candidates have systematically larger radio size lying in the range of 241.6 kpc to 1301 kpc with a median value of 469.1$\pm$70.6 kpc. While, active sources  
have radio sizes in the range of 75.2 kpc to 1651.1 kpc with a median value of 314.3$\pm$22.8 kpc 
(see Table~\ref{tab:RRGComp}). The two$-$sample KS test suggests that the radio size 
distributions of our remnant candidates and active sources are different with 
insignificant probability (1 $\times$ 10$^{-3}$) for null hypothesis, that the two sample come 
from the same distribution, to be true. 
We note that despite different radio size distributions for our remnant candidates and active sources, 
there is no sharp segregating value that can be used to differentiate remnant and active sources. 
\par 
The P$-$D plot also demonstrates the comparison of radio luminosities for our remnant candidates 
and active sources. 
We estimate the total radio luminosity in the rest frame by applying a k$-$correction assuming curved power law SEDs
(${L_{\rm 1.4~GHz}}~=~4{\pi}{S_{\rm 1.4~GHz}}~{D}^{2}_{\rm L}~(1 + z)^{-({\alpha}+1)}
~e^{q(ln(1+ z))^{2}}$, 
where $S_{\rm 1.4~GHz}$ is the flux density at 1.4~GHz, $D_{\rm L}$ is the luminosity distance, $q$ is the spectral curvature and 
${\alpha}$ is the spectral index of curved power law). 
There is no remarkable difference between 1.4 GHz radio luminosities of active sources and our remnant 
candidates (see~Figure~\ref{fig:PDPlot}). 
We find that our 17 remnant candidates, with 1.4 GHz flux density and redshift estimates, have 
1.4 GHz luminosities distributed in the range of 5.0 $\times$ 10$^{23}$ W~Hz$^{-1}$ to 5.73 $\times$ 10$^{26}$ W~Hz$^{-1}$ 
with a median value of 1.07$\pm$0.36 $\times$ 10$^{25}$ W~Hz$^{-1}$. While, 154 active sources show 1.4 GHz luminosities 
in the range of 2.63 $\times$ 10$^{23}$ W~Hz$^{-1}$ to 2.45 $\times$ 10$^{27}$ W~Hz$^{-1}$ 
with a median value of 2.95$\pm$0.30 $\times$ 10$^{25}$ W~Hz$^{-1}$. Thus, we find that 1.4 GHz radio 
luminosities of our remnant candidates tend to be systematically little lower than that for active sources. 
The two$-$sample KS test shows that 1.4 GHz radio luminosity distributions of remnant candidates and active sources 
are drawn from the same distribution with a probability of only 0.12. Although, we caution that the small number of remnant candidates 
must be taken into consideration while interpreting the statistical results.   
\par
Further, we also compare 150 MHz luminosities of remnant candidates and active sources noting the fact 
that the low$-$frequency radio observations are more effective 
in capturing diffuse emission of low$-$surface$-$brightness often seen in the evolved and remnant sources. 
The two$-$sample KS test suggests similarity in the 
radio luminosity distributions of remnant candidates and active sources with 77$\%$ probability for them to 
be drawn from same distribution (see Table~\ref{tab:RRGComp}).   
Notably, despite having somewhat different radio size distributions, the 
150 MHz radio luminosity distributions of remnant candidates and active sources are similar. 
This result can be understood, if most of our remnant candidates continue 
to remain as luminous as active sources at 150 MHz even after the cessation of AGN activity. 
The high 150 MHz luminosities of our remnant candidates can only be expected, 
if majority of these are not too old, and have spectral break frequency at $>$150 MHz. 
Indeed, we find that many of our remnant candidates show strong spectral curvature with an apparent break 
at $>$325 MHz (see Section~\ref{sec:SED}).
\par
In the P$-$D plot we draw evolutionary tracks for radio galaxies reported in \cite{An12} (see~Figure~\ref{fig:PDPlot}). 
It is clear that most of our sample sources are large-size radio galaxies similar to FR-I and FR-II type sources 
representing the late phase of evolution. 
We note that our P$-$D plot shows lack of small-size ($<$ 100~kpc) young radio sources due 
to the effect of radio size cutoff limit ($\geq$ 30$^{\prime\prime}$) introduced in our sample. 
We also point out that our P$-$D plot shows a wide diversity in radio sizes of our remnant 
candidates in the range of 241.6 kpc to 1301 kpc.
The radio size of a remnant can depend on various factors such as the 
jet kinetic power during the active phase, longevity of active phase, and the large$-$scale cluster environment 
\citep{Turner15}. In general, small size remnant candidates have relatively low radio luminosities 
(see Figure~\ref{fig:PDPlot}). Thus, small size remnant candidates could have possessed jets of low kinetic 
power during their active phase, if radio luminosity is considered as the proxy for the jet power. 
Since, our small size remnant candidates mostly reside in non$-$cluster 
environments (see Section~\ref{sec:environment}), then a low jet kinetic power and/or a relatively shorter active phase can be the plausible reasons for a relatively small radio size of our remnant candidates. 
\section{Fraction of remnant candidates}
\label{sec:frac}
Determining the fraction of remnants ($f_{\rm rem}$) among the population of radio galaxies can help us to constrain the AGN duty cycle and evolutionary models. 
Using our search for remnant candidates we place an upper limit on $f_{\rm rem}$.
Since we cannot rule out the possibility that some of our remnant candidates may still be active with the presence of a faint radio core falling below the detection limit we get only an upper limit on the number 
and fraction of remnants. 
Our 21 remnant candidates in a sample of 263 extended radio sources above a flux density cutoff limit S$_{\rm 325~MHz}$ $>$8.0 mJy (the completeness limit, see Section~\ref{sec:completeness}), yield an upper limit on $f_{\rm rem}$ to be only 21/263 $<$8.0$\%$.  
In Table~\ref{tab:frac}, we list the fraction of remnants found in our study as well as in other 
previous searches in different deep fields.   
At first glance, the upper limit on $f_{\rm rem}$ $<$8.0$\%$ found in our study appears consistent with that 
reported in some of the previous studies \citep[{\eg}][]{Mahatma18,Quici21}.
Moreover, while comparing $f_{\rm rem}$ reported in different studies one needs to account for the 
differences in sensitivity, resolution, and cutoff limits applied on flux density and angular size. In the following subsections, we discuss the effects of sensitivity, completeness and 
the flux density cutoff limit.      
\begin{table*}
\begin{minipage}{180mm}
\caption{Fraction of remnant sources in different deep fields}
\label{tab:frac}
\hskip-3.0cm
\scalebox{0.9}{
\begin{tabular}{lccccccccccc} \hline 
Field            &  Sky area    &  ${\nu}_{\rm base}$  & rms, ${\nu}_{\rm base}$ & ${\nu}_{\rm high}$ & rms, ${\nu}_{\rm high}$ &      Search   & Flux limit      & Size       &  Fraction    &  Density &   Ref. \\ 
                 & (deg$^{2}$)  &  (MHz)    & (mJy b$^{-1}$) & (GHz)  & (mJy b$^{-1}$)  &  criteria & (mJy) & limit        & of remnants     &   (deg$^{-2}$)    &        \\ \hline     
XMM$-$LSS$-$JVLA   &  5.0  & 325  & 0.15   & 1.4  & 0.016  & M, S, U   & 8.0  &  30$^{\prime\prime}$  &  5/103 $<$5$\%$  & 1.0 & 1  \\
XMM$-$LSS$-$Out   & 7.5   & 325  & 0.15   & 3.0  &  0.12  & M, S, U   & 8.0   &  30$^{\prime\prime}$  & 16/160 $<$10.0$\%$  &  2.1 & 1  \\   
XMM$-$LSS        & 12.5  & 325  & 0.15   & 1.4$-$3.0  & 0.016$-$0.12 & M, S, U & 8.0  &  30$^{\prime\prime}$  &  21/263 $<$8.0$\%$ & 1.7 & 1  \\   
GAMA$-$23       &  8.3  & 216  & 0.9    & 5.5  & 0.024  & M         &  10   &  25$^{\prime\prime}$  &  10/104 $<$10$\%$ & 1.2 & 2  \\
Herschel$-$ATLAS &  140  & 150  & 0.1$-$2.0 & 6.0 & 0.020 & M        &   80    &  40$^{\prime\prime}$  &  11/127 $<$9$\%$ & 0.08 & 3  \\
Lockman$-$Hole   &  35   & 150  & 0.15$-$0.9 & 6.0 & 0.050 & M, S, U  &   40  &  40$^{\prime\prime}$  &  13/158 $<$8$\%$  & 0.37 & 4  \\
ATLBS$-$ESS      &  7.5  & 1400 & 0.072 & 1.4 & 0.072 & M  &  1.0     &  30$^{\prime\prime}$  &  4/119  $<$4$\%$ & 0.53 & 5  \\
\hline
\end{tabular}}
\\
{\bf Notes}$-$ Column 1 : Name of the deep field; column 2: Sky area used for remnant search; column 
3: low$-$frequency at which remnant search is based (${\nu}_{\rm base}$); column 4 : noise$-$rms at 
the base frequency observations (rms, ${\nu}_{\rm base}$); column 5 : high$-$frequency observations used to detect core and compact features (${\nu}_{\rm high}$); column 6 : noise$-$rms of high$-$frequency observations (rms, ${\nu}_{\rm high}$); column 7 : Criteria used to search for remnant candidates (M : morphological criterion, S : spectral curvature criterion, 
U: USS criterion); column 8 : Flux cutoff limit used in the sample; column 9 : Linear angular size 
cutoff applied in the sample; column 10 : Upper limit on the fraction of remnants; column 11 : surface number density  
of remnant candidates; column 12 : References $-$ (1) This work, (2) \citet{Quici21}, (3) \citet{Mahatma18}, 
(4) \citet{Jurlin21}, (5) \citet{Saripalli12}. 
The flux density cutoff is applied at the corresponding base frequency.
\end{minipage}
\end{table*} 
\subsection{Effect of sensitivity and resolution}
\label{sec:sensitivity}
For our remnant candidates, we checked for the absence of radio core by using deep 
1.4 GHz JVLA observations in the XMM$-$LSS$-$JVLA region and 3 GHz VLASS observations in the XMM$-$LSS$-$Out region. 
The comparison of remnant fraction in the XMM$-$LSS$-$JVLA region and in the XMM$-$LSS$-$Out region can easily 
demonstrate the effect of sensitivity.    
In the XMM$-$LSS$-$JVLA region we find only 5 remnant candidates among 103 radio sources, giving an upper limit 
on $f_{\rm rem}$ $<$5$\%$ (see Table~\ref{tab:frac}). 
In the XMM$-$LSS$-$Out region, where 3~GHz VLASS images are used for the core detection, we find a much higher $f_{\rm rem}$ to be 16/165 $<$9.7$\%$. 
The lower remnant fraction in the XMM$-$LSS$-$JVLA region can be explained as the absence of radio core 
is examined at a much fainter level. 
Thus, we demonstrate that deep JVLA observations allow us to place a much tighter constraint 
on the remnant fraction.
We note that, in comparison to the previous searches based on the low$-$frequency radio observations, we obtain the lowest $f_{\rm rem}$ $<$5$\%$ (see Table~\ref{tab:frac}). 
The lowest fraction obtained in our study may be attributed to the lowest flux density cutoff limit 
(S$_{\rm 325~MHz}$ $\geq$8.0 mJy) that causes rapid increase in the number of active sources at the fainter level.    
It is worth mentioning that, in the ATLBS$-$ESS field \cite{Saripalli12} searched for remnants to 
the flux limit of 1.0 mJy at 1.4 GHz 
and found remnant fraction $<$4$\%$. Although, unlike low$-$frequency observations, 1.4 GHz observations 
are likely to miss the detection of sources having low$-$surface$-$brightness emission, 
which in turn can lead to a lower remnant fraction.  
\par 
It is important to note that while comparing $f_{\rm rem}$ we need to account for the uncertainty due to
cosmic variance ({\ie}field-to-field variation due to large-scale structure) and Poisson noise. 
To assess the uncertainty on $f_{\rm rem}$ due to cosmic variance as well as Poisson noise we use Cosmic 
variance calculator\footnote{https://www.ph.unimelb.edu.au/~mtrenti/cvc/CosmicVariance.html} v1.03
based on the method proposed by \cite{Trenti08}. 
The cosmic variance is computed by integrating the two-point correlation function of dark matter halos 
over the survey volume and then multiplying it with the average bias of the source sample which in turn depends on the 
source volume density and average halo occupation fraction.
For our remnant candidates in the XMM-LSS-JVLA region we consider sky-area of 3$^{\circ}$.10 $\times$ 1$^{\circ}$.61, average redshift ($z_{\rm mean}$) of 0.9, average value of ${\Delta}z$ 0.2, 
completeness parameter set to 1.0. 
For a total number of remnant candidates of 5 in the XMM-LSS-JVLA region, we obtain 
cosmic variance 0.29, Poisson uncertainty 0.45 and the total fractional error on the number counts 0.54. 
Therefore, uncertainty in $f_{\rm rem}$ is dominated by Poisson noise.  
In the other deep fields too, uncertainty on $f_{\rm rem}$ is likely to be dominated by the Poisson noise rather than the cosmic variance due to small number of remnant sources detected in the fields. Therefore, we caution that the uncertainties in  the number counts must be accounted while 
comparing $f_{\rm rem}$ in different fields.
\subsection{Completeness}
\label{sec:completeness}
\begin{figure}
\centering
\includegraphics[angle=0,width=8.5cm,trim={0.0cm 7.0cm 0.0cm 6.5cm},clip]{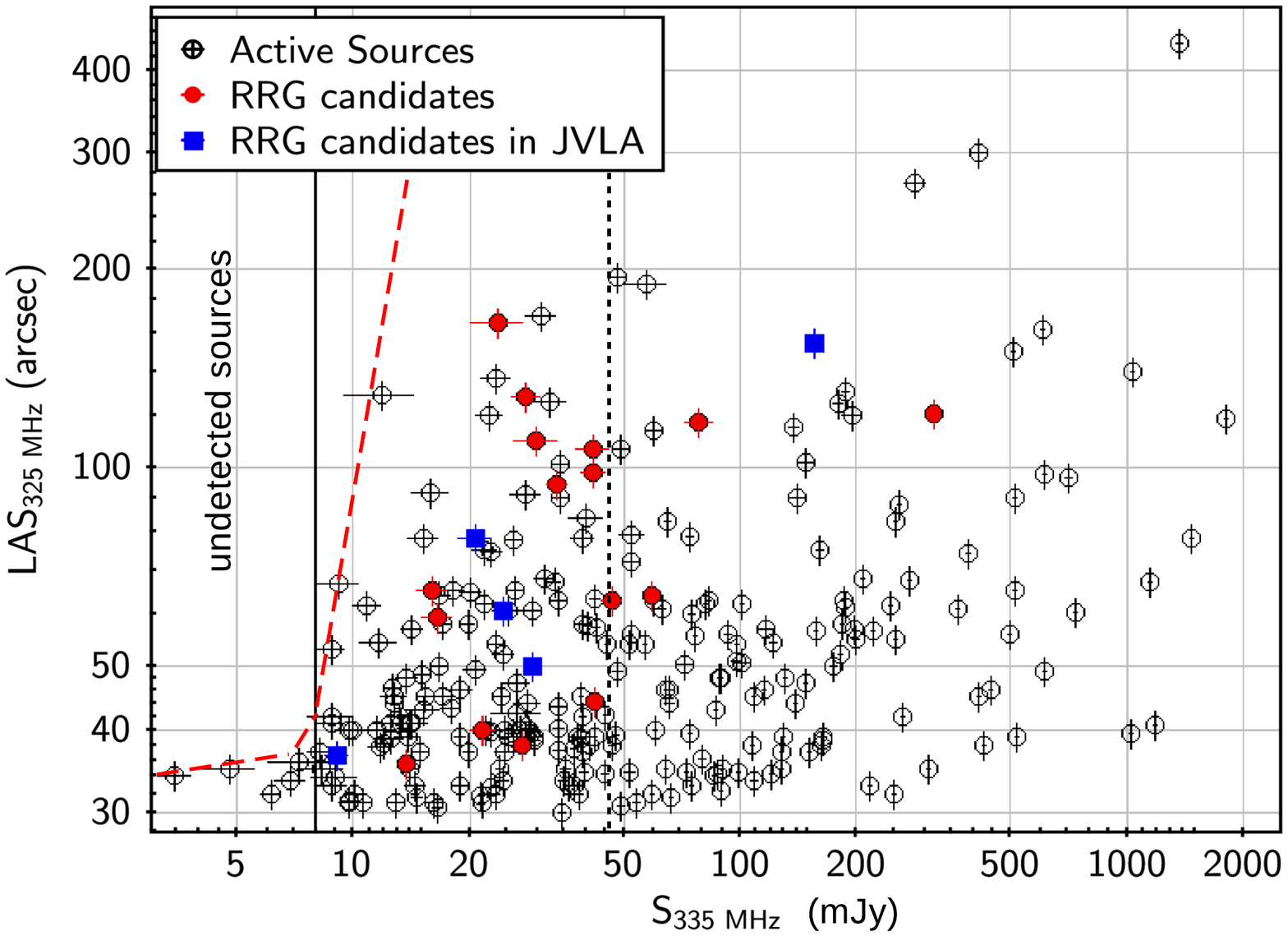}
\caption{Integrated flux density versus linear angular size at 325 MHz.  
Fiducial 5 per cent error bars are placed in the LAS. 
The vertical black line at 8.0 mJy represents a near completeness limit, while dashed curve highlights the limitation of detecting sources of large angular sizes in the fainter regime. The black dotted vertical line represents the flux density cutoff limit considered in \cite{Mahatma18} work.} 
\label{fig:FDVsSize} 
\end{figure}

The completeness of a sample above a flux density limit is an important factor in determining the fraction. 
An incomplete sample lacking the detection of all sources at the fainter end can over$-$estimate the remnant fraction. 
To assess the flux density completeness in our sample, we check the differential source counts {\it w.r.t.} 325 MHz flux density in our sample of 268 sources. 
The ${dN/dS}$ plot as well as 325~MHz flux density distribution show that source counts begin to 
decline after 8.0 mJy. Therefore, our sample can be treated as complete above 8.0 mJy (see Figure~\ref{fig:FDVsSize}).
The completeness flux limit corresponds to 13.7 mJy at 150 MHz, assuming a typical spectral index of -0.7. 
We find 263 out of 268 sources falling above the flux density limit (S$_{\rm 325~MHz}$ $\geq$8.0 mJy). 
Thus, with the use of flux limit cutoff we ensure not to over$-$estimate the remnant fraction. 
\subsection{Remnant candidates at the fainter regime}
\label{sec:fainterregime}
In this subsection we attempt to highlight the importance of searching remnants at lower flux densities.   
From Figure~\ref{fig:FDVsSize} showing a plot of integrated flux density versus LAS measured at 325 MHz, 
it is evident that the remnant fraction depends on the flux density at which they are searched. 
For instance, a higher flux density cutoff limit of 80 mJy at 150 MHz that corresponds to 46.5 mJy at 325 MHz assuming a 
typical spectral index of $-0.7$, considered by \cite{Mahatma18}, would limit us to detect only 5/21 of our remnant candidates. Thus, a significant fraction of remnant candidates (16/21 $\sim$ 76.2$\%$) would be missed by applying a high flux density cutoff limit. \cite{Brienza17} also placed a flux density cutoff limit at S$_{\rm 150~MHz}$ = 40 mJy. 
Therefore, unlike previous studies our study presents the search for remnants in a much fainter flux regime 
reaching down to nearly 13 mJy at 150 MHz.
In Figure~\ref{fig:FDVsSize}, we also highlight the dearth of large$-$size radio sources at fainter flux densities 
which is simply due to observational bias {\ie}faint extended sources remain undetected due to limited surface-brightness achieved 
in our 325 MHz GMRT observations. Due to this bias we find that the remnant candidates of relatively large angular sizes are present at relatively higher flux densities.
Considering the fact that small-size (let say $<$100$^{\prime\prime}$) remnants as well as active radio sources are more abundant, it is unlikely that low fraction of remnants found in our study can be significantly impacted by the surface brightness limitation.  
\par 
The comparison of surface number density of remnant candidates in different flux limited observations can further 
demonstrate the importance of searching them at lower flux densities. 
For instance, in the XMM$-$LSS$-$JVLA region we find only five remnant candidates over an area of 5.0 deg$^{2}$ yielding 
surface number density of 1.0 source deg$^{-2}$. 
In the XMM$-$LSS$-$Out region, the surface number density of remnant candidates is 16/7.5 $\sim$ 2.1 deg$^{-2}$. 
The higher number density in the XMM$-$LSS$-$Out region is owing to the use of the relatively shallow VLASS and FIRST observations for the core detection. 
The remnant number density in the Herschel$-$ATLAS field in only 11/140 $\sim$ 0.08 deg$^{-2}$ 
that can mainly be attributed to the high flux cutoff limit of S$_{\rm 150~MHz}$ = 80 mJy.  
Albeit, deep observations used for the core detection would also bring down the remnant fraction and the number density. 
Interestingly, in the XMM$-$LSS$-$JVLA region we examined the detection of core at similar depth 
and obtain a higher surface number density than that found in the Herschel$-$ATLAS field (see Table~\ref{tab:frac}). 
It is worth to point out that the surface number density of remnant candidates in the XMM$-$LSS$-$JVLA region is 
similar to that found in the GAMA$-$23 field in which the flux density cutoff limit is comparable 
to that for our sample.      

\subsection{Implications of remnant fraction}
\label{sec:implications}
In the literature, there have been attempts to design models that can explain the 
evolution of radio galaxies and can predict the remnant fraction \citep{Brienza17}. 
Recent, evolutionary models predict that the remnant fraction depends inversely to the 
duration of the active phase ($t_{\rm on}$) \citep[see][]{Shabala20}. 
Radio galaxies with longer active phase would result into large$-$size remnants 
that would fade rapidly and yield a lower remnant fraction. 
While, radio galaxies with relatively short active phase would result in small$-$size 
remnants that can be detected for a relatively long time$-$scale in the remnant phase, and hence, 
can show a higher remnant fraction. 
In fact, a high remnant fraction ($>$30$\%$) has been obtained with the models assuming a relatively 
short (30$-$40 Myr) average life$-$time of the active phase \citep[see][]{Brienza17}. 
To model active, remnant and restarted source population \cite{Shabala20} considered $t_{\rm on}$ and jet kinetic power ($Q_{\rm jet}$) as the key parameters for the growth of a radio galaxy and found that a model assuming constant age $t_{\rm on}$ $=$ 300 Myr, and power law distribution of 
log($Q_{\rm jet}$) with an index of $-0.8$ to $-1.2$ predicts the fraction of remnant plus restarted sources 
to be only 2$-$5$\%$. 
While, a somewhat higher fraction (10$\%$) of remnant plus restarted sources can be modelled, if power law 
distributions with an index in the range of $-0.8$ to $-1.2$ are assumed for both $t_{\rm on}$ and $Q_{\rm jet}$. 
Therefore, remnant fraction ($<$5$\%$) obtained in our study seems 
consistent with the model proposed by \cite{Shabala20}. 
\section{Conclusions}
\label{sec:conclusions}
In this paper, we attempt to search and characterize remnant candidates in the XMM$-$LSS field 
using deep 150 MHz LOFAR, 325 MHz GMRT, 1.4 GHz JVLA and 3 GHz VLASS observations. 
These are some of the deepest observations available at the respective frequencies till date. 
We used both morphological as well as spectral criteria to identify the full population of remnant candidates. 
Salient features of our study are outlined as below. 
 \begin{itemize}
\item We identified 21 remnant candidates in a sample of 263 extended radio sources 
(SNR $\geq$10, size $\geq$30$^{\prime\prime}$, and S$_{\rm 325~MHz}$ $\geq$8.0 mJy) 
derived from the 325 MHz GMRT observations over 12.5 deg$^{2}$ in the XMM$-$LSS field. 
\item Majority (15/21 = 71$\%$) of our remnant candidates can be identified by using the 
morphological criterion {\ie}absent core. 
We find that the spectral curvature criterion is useful in identifying remnant candidates of small angular sizes 
($<$60$^{\prime\prime}$) which pose difficulty in deblending core from lobes using images 
of 2$^{\prime\prime}$.5$-$5$^{\prime\prime}$.0 resolution.  
\item Unlike previous studies \citep[{\eg}][]{Brienza17,Mahatma18} our study attempts to search for remnants 
at the faint regime reaching down to 8.0 mJy at 325 MHz that corresponds to 13.7 mJy at 150 MHz assuming 
a typical spectral index of $-0.7$. 
Our remnant candidates have 150 MHz flux densities distributed in the range of 13.8 mJy to 617.3 mJy with a 
median value of 55.1$\pm$39.3 mJy.
\item Our remnant candidates show a wide range of characteristics in terms of spectral index, radio size, 
radio luminosity, and surface$-$brightness. 
In comparison to active sources our remnant candidates 
exhibit steeper spectra (${\alpha}_{\rm 150}^{\rm 1400}$ spans over $-1.71$~$-$~$-0.75$ with a median 
value of $-1.10\pm0.05$), larger radio sizes (in the range of 242 kpc to 1301 kpc
with a median value of 469.1$\pm$70.6 kpc), lower surface$-$brightness (in the range of 22.2 to 264 mJy arcmin$^{-2}$ with a median value of 63.4$\pm$14.5 mJy arcmin$^{-2}$). 
Although, distributions of various parameters show no sharp division between 
remnant candidates and active sources.  
\item Using P$-$D plot we find that our sample sources represent mostly evolved radio galaxies and 
follow evolutionary tracks reported in \cite{An12}. As expected, the radio sizes of our remnant candidates tend 
to be systematically larger than that for active sources, while 1.4~GHz radio luminosities of 
remnant candidates are little lower than that for active sources. Moreover, 
150 MHz radio luminosity distribution of our remnant candidates is similar to that for active sources, 
suggesting that most of our remnant candidates continue to remain 
as luminous as active sources at 150 MHz even after the cessation of AGN activity.
\item Using cluster catalogs based on the deep HSC$-$SSP optical and {\em XMM$-$Newton} X$-$ray surveys we find that 
only one of our remnant candidates is associated with the cluster environment. Therefore, we conclude that, 
in general, remnants reside in non$-$cluster environments which is consistent with the result reported 
in \cite{Jurlin21}.    
\item Deep JVLA observations allow us to place a constraint on the remnant fraction to be (5/103) $<$5$\%$ 
in the XMM$-$LSS$-$JVLA region. The remnant fraction ($<$5$\%$) found in our study is the lowest, compared to previous studies \citep[{\eg}][]{Mahatma18,Jurlin21}.
\item The low fraction of remnants ($<$5$\%$) obtained in our study seems to be consistent with the evolutionary model proposed by \cite{Shabala20} who assumed power law distributions for the active phase life$-$time and jet kinetic power with index $-0.8$ to $-1.2$ to explain 
the observed fraction of remnant plus restarted sources. 
Further, a relatively high population of small$-$size remnants found in our study in consistent with the prediction of 
evolutionary model.
\item Our study shows that the remnant fraction depends on the flux density at which remnants are searched. 
A high flux density cutoff limit would miss a substantial fraction of remnants, and in turn, would change 
the remnant fraction. Thus, our study demonstrates the importance of searching for remnants at the fainter level to 
detect an unexplored population of remnants. 
\end{itemize}
%
%
\begin{acknowledgments}
We thank the anonymous reviewers for useful comments that helped us to improve the manuscript. 
SD, VS and AK acknowledge the support from Physical Research Laboratory, Ahmedabad, funded by the Department 
of Space, Government of India. CHI and YW acknowledge the support of the Department of Atomic Energy, Government 
of India, under project no. 12-R\&D-TFR5.02-0700. We thank the staff of GMRT who have made these 
observations possible. GMRT is run by the National Centre for Radio Astrophysics of the Tata Institute 
of Fundamental Research. The GMRT observations were  carried out under program No. 4404$-$3 supported by 
the Indo$-$French Center for the Promotion of Advanced Research (Centre Franco$-$Indien pour la Promotion 
de la Recherche Avancée).
This paper is based on data collected at the Subaru Telescope and retrieved from the HSC data archive 
system, which is operated by the Subaru Telescope and Astronomy Data Center (ADC) at NAOJ. Data analysis was 
in part carried out with the cooperation of Center for Computational Astrophysics (CfCA), NAOJ. 
%
\end{acknowledgments}
%
%
%
\facilities{GMRT, LOFAR, JVLA, Subaru}

%
\software{TOPCAT \citep{Taylor05}, AIPS \citep{Greisen03}, DS9 \citep{Joye03}, astropy \citep{Astropy18} }
%
%

%
%
%
\bibliography{RRGpaper}{}
\bibliographystyle{aasjournal}
%
%
%
\end{document}